\documentclass{aa}
\usepackage[varg]{txfonts}
\usepackage{natbib}
\usepackage{multirow}
\usepackage{subcaption}
\usepackage{hyperref}
\hypersetup{colorlinks=true, citecolor=blue}

\bibpunct{(}{)}{;}{a}{}{,}

\newcommand{\figref}[1]{Fig.\,\ref{#1}}
\newcommand{\tabref}[1]{Tab.\,\ref{#1}}
\newcommand{\equref}[1]{Eq.\,\ref{#1}}

\newcommand{\micron}{\,$\mu$m}
\newcommand{\citel}[1]{\citeauthor{#1}\,\citeyear{#1}}
\newcommand{\flux}{\,erg\,cm$^{-2}$\,s$^{-1}$}
\newcommand{\lum}{\,erg\,s$^{-1}$}
\newcommand{\NH}{\,cm$^{-2}$}
\newcommand{\kmpersec}{km\,s$^{-1}$}

\makeatletter

\newcommand{\Rnum}[1]{\expandafter\@slowromancap\romannumeral #1@}
\makeatother

\begin{document}

\title{Spectroscopic identification of \textit{INTEGRAL} high-energy sources with VLT/ISAAC\thanks{Based on observations made with ESO Telescopes at the La Silla Paranal Observatory under programme ID 089.D-0181(A).}}
\author{F. Fortin\inst1, S. Chaty\inst1$^{,}$\inst2, A. Coleiro\inst3, J. A. Tomsick\inst4, C. H. R. Nitschelm\inst5}

\institute{
{Laboratoire AIM (UMR 7158 CEA/DRF - CNRS - Université Paris Diderot), Irfu / Département d’Astrophysique, CEA-Saclay, 91191
Gif-sur-Yvette Cedex, France}
\and
{Institut Universitaire de France, 103, boulevard Saint-Michel, 75005 Paris}
\and
{APC, Université Paris Diderot, CNRS/IN2P3, CEA/Irfu, Observatoire de Paris, 10 rue Alice Domon et Léonie Duquet, 75205 Paris Cedex 13, France}
\and
{Space Science Laboratory, 7 Gauss Way, University of California, Berkeley, CA 94720-7450, USA}
\and
{Unidad de Astronom{\'i}a, Universidad de Antofagasta, Avenida Angamos 601, Antofagasta 1270300, Chile}
}

\abstract
{The \textit{INTEGRAL} satellite has been observing the $\gamma$-ray sky for 15 years and has detected over 900 X-ray sources of various nature. However, more than 200 of these sources still lack precise identification.}
{Our goal is to reveal the nature of the high-energy sources detected by \textit{INTEGRAL}. In particular, we want to improve the census of X-ray binaries.}
{Photometry and spectroscopy were performed in July 2012 on 14 \textit{INTEGRAL} sources in near-infrared at the Very Large Telescope on the European Southern Observatory-UT3 telescope equipped with the ISAAC spectrograph. We used K$_s$ images reaching to a depth of magnitude 18.5 to look for unique counterparts to high-energy detections to check for both extended sources and photometric variability. The analysis of near-infrared spectral features allows us {to constrain} the nature of these X-ray sources by comparing them to {stellar} spectra atlases.}
{We present photometric and/or spectroscopic data for 14 sources (IGR J00465$-$4005, IGR J10447$-$6027, IGR J12489$-$6243, IGR J13020$-$6359, IGR J13186$-$6257, IGR J15293$-$5609, IGR J17200$-$3116, IGR J17404$-$3655, IGR J17586$-$2129, IGR J17597$-$2201, IGR J18457+0244, IGR J18532+0416, IGR J19308$+$0530, and IGR J19378$-$0617). We conclude that 5 of these are active galactic nuclei, 5 are cataclysmic variables, 2 are low- or intermediate-mass X-ray binaries, and 2 are Be high-mass X-ray binaries.}
{}


\titlerunning{Identification of \textit{INTEGRAL} sources with VLT/ISAAC.}
\authorrunning{Fortin et al.}
\maketitle

\section{Introduction}\label{sect:intro}
The \textit{INTEGRAL} satellite has been observing the high-energy sky between 15 keV and 10 MeV for 15 years. However, the nature of the sources detected at high energies is often uncertain and requires further observations at low energies in optical and near-infrared (nIR) wavelengths. This is why a significant fraction ($\sim$\,20\%) of the \textit{INTEGRAL} detections still need better constraints to have a robust identification. According to the catalogue of \textit{INTEGRAL} detections published by \cite{bird_ibis_2016}, it is expected that a majority of the unknown sources are active galactic nuclei (AGN) and high-energy binaries; the latter are binary star systems that host an accreting compact object.

High-energy binaries fall into three main categories depending on the compact object and the mass of the companion star: cataclysmic variables (CV), low-mass X-ray binaries (LMXB), and high-mass X-ray binaries (HMXB). While LMXBs host either a neutron star (NS) or a black hole (BH),  CVs host a white dwarf\,; both have a low-mass companion star (typically M\,$\leq$\,1\,M$_\sun$). High-energy radiation is released through accretion of matter from the companion star overflowing its Roche lobe. An accretion disc can form around the compact object and feed it gradually, which often leads to transient behaviour. The less common intermediate-mass X-ray binaries (IMXBs) have a companion of mass between 1 and 10\,M$_\sun$ and the accretion process is similar to that of LMXBs. For the sake of consistency with the literature and especially \cite{bird_ibis_2016}, we group IMXBs with LMXBs.     

The HMXBs host a massive star (typically M\,$\geq$\,10\,M$_\sun$), around which orbits either a NS or a BH. Two main subcategories exist among HMXBs, based on the evolutionary phase of the companion star. In Be-types (BeHMXB), the secondary is a fast-rotating main-sequence O/B star that sheds matter from its equator as a consequence of high centrifugal force. A decretion disc thus forms around the companion star. The accretion of matter occurs when the compact object passes through that disc. Supergiant binaries \mbox{(sgHMXB)} host an evolved O/B supergiant star that feeds a compact object with matter through an intense stellar wind, driven by its tremendous luminosity. The sensitivity of \textit{INTEGRAL} at higher energies made it possible to differentiate two new subclasses of sgHMXBs, as reviewed in \cite{chaty_optical/infrared_2013}. Obscured HMXBs are intrinsically absorbed ($N_H > 10^{23}$\,cm$^{-2}$), while supergiant fast X-ray transients (SFXTs) show short and intense bursts of high-energy radiations.

\begin{table*}[ht!]
\centering
\small
\caption{Positions and uncertainty of the X-ray detections of the IGR sample.\label{tab:xpos}}
\begin{tabular}{lllrrl}
\hline\hline\\[-1.5ex]
Source & RA J2000 & Dec J2000 & \textit{l} & \textit{b} & Uncertainty at 90\%\\
       & (X-ray) & (X-ray) & (\degr)           & (\degr)           & (\arcsec) \\[1.5ex]
\hline\\[-1.5ex]
IGR J00465$-$4005  & 00:46:20.71 & $-$40:05:47.3  & 307.2621 & $-$76.9884 & 4\farcs26 (\textit{Swift})$\,^a$ \\
IGR J10447$-$6027  & 10:44:51.89 & $-$60:25:12.0  & 287.9185 & 1.2915     & 0\farcs65 (\textit{Chandra})$\,^b$ \\
IGR J12489$-$6243  & 12:48:46.44 & $-$62:37:43.1  & 302.6257 & 0.2417     & 0\farcs64 (\textit{Chandra}) $\,^c$\\
IGR J13020$-$6359  & 13:01:58.72 & $-$63:58:08.7  & 304.0885 & $-$1.1212  & 0\farcs39 (\textit{XMM})   $\,^d$ \\
IGR J13186$-$6257  & 13:18:25.08 & $-$62:58:15.5  & 305.9915 & $-$0.2599  & 0\farcs64 (\textit{Chandra}) $\,^e$ \\
IGR J15293$-$5609  & 15:29:29.37 & $-$56:12:13.3  & 323.6587 & 0.1712     & 0\farcs64 (\textit{Chandra}) $\,^c$   \\ 
IGR J17200$-$3116  & 17:20:05.92 & $-$31:16:59.4  & 355.0221 & 3.3472     & 0\farcs64 (\textit{Chandra}) $\,^f$  \\  
IGR J17404$-$3655  & 17:40:26.86 & $-$36:55:37.4  & 352.6259 & $-$3.2725  & 0\farcs64 (\textit{Chandra}) $\,^e$  \\  
IGR J17586$-$2129  & 17:58:34.56 & $-$21:23:21.6  & 7.9862   & 1.3265     & 0\farcs64 (\textit{Chandra}) $\,^e$ \\ 
IGR J17597$-$2201  & 17:59:45.52 & $-$22:01:39.17 & 7.5696   & 0.7704     & 0\farcs60 (\textit{Chandra}) $\,^g$  \\  
IGR J18457$+$0244  & 18:45:40.38 & $+$02:42:09.2  & 34.6819  & 2.5135     & 2\farcs16 (\textit{XMM-Newton})   $\,^d$    \\  
IGR J18532$+$0416  & 18:53:15.91 & $+$04:17:48.26 & 36.9651  & 1.5519     & 1\farcs14 (\textit{XMM-Newton})  $\,^d$    \\   
IGR J19308$+$0530  & 19:30:50.77 & $+$05:30:58.09 & 42.3807  & $-$6.1852  & 0\farcs6~~  (\textit{Chandra})$\,^g$   \\
IGR J19378$-$0617  & 19:37:33.1  & $-$06:13:04    & 32.5905  & $-$13.0737 & 3\farcs5~~  (\textit{Swift})  $\,^h$   \\
\hline

\end{tabular}

\begin{minipage}{.77\textwidth}
\begin{tiny}
$^a$\cite{landi_swift/xrt_2010-1}, $\,^b$\cite{fiocchi_five_2010}, $\,^c$\cite{tomsick_localizing_2012}, $\,^d$\cite{rosen_XMM-Newton_2016}, $\,^e$\cite{tomsick_chandra_2009}, $\,^f$\cite{tomsick_chandra_2008}, $\,^g$\cite{ratti_chandra_2010}, $\,^h$\cite{rodriguez_swift_2008}
\end{tiny}
\end{minipage}

\end{table*}

Precisely identifying high-energy sources requires further observations, for which the nIR domain is well adapted. Firstly, many \textit{INTEGRAL} sources (later called IGRs) lie near the Galactic plane, where optical radiation is absorbed by dust while infrared is not. Secondly, most of a binary's nIR emission comes from the companion star or the accretion disc, which is ideal to identify their nature by constraining its spectral type.

In this paper, we present a sample of 14 IGR sources (\tabref{tab:xpos}) for which we performed nIR photometry and/or spectroscopy. We aim to confirm a unique nIR counterpart for each of these IGRs and provide additional constraints on their nature such as the spectral type of companion stars in X-ray binaries. The IBIS instrument  on board \textit{INTEGRAL} has a wide field of view, but does not have enough spatial resolution to associate accurately an optical/nIR counterpart to the high-energy detections. Precise X-ray localization is thus given by either \textit{Chandra}, \textit{XMM-Newton,} or \textit{Swift} telescopes. Section \ref{sect:observations} describes\ nIR photometric and spectroscopic observations by the European Southern Observatory (ESO), along with data reduction processes. Section \ref{sect:results} compiles all the previously published results on the sources along with the new results of our nIR observations. In Section \ref{sect:discuss} we discuss the results and outcomes of these observations before concluding in Section \ref{sect:conclusion}.

\section{Observations}\label{sect:observations}
The observations were performed in 2012 (P.I. S. Chaty) on a sample of 14 \textit{INTEGRAL} sources (programme ID 089.D-0181(A), see \tabref{tab:xpos}). Both nIR photometry and spectroscopy were performed at ESO in Chile on the 8 m Very Large Telescope Unit 3 Nasmyth A (VLT/UT3) equipped with the ISAAC instrument. {The atmospheric conditions at the time of the observations were satisfactory and delivered an average seeing of 0\farcs7 in the K$_s$ band.

\subsection{Photometry}
Near-infrared images were obtained through a K$_{s}$ standard filter (1.98 -- 2.35\micron). The field of view for individual images was $2\farcm5\times2\farcm5$ with a pixel scale of 0.148\arcsec. For each source, five frames were taken. Each frame was {acquired} with a random spatial offset (10\arcsec\,in average) that follows the jitter acquisition method standardly used in nIR observations.

Data reduction was performed with standard Image Reduction and Analysis Facility (IRAF\footnote{IRAF is distributed by the National Optical Astronomy Observatories, which are operated by the Association of Universities for Research in Astronomy, Inc., under cooperative agreement with the National Science Foundation.}) procedures. After {dark subtraction and} flat-field correction, the sky value was estimated and subtracted through the median of the five jittered images taken on each source. The images were then aligned based on precise astrometry and averaged to optimize the signal-to-noise ratio (S/N).

Aperture photometry was performed with the \textit{IRAF.apphot.qphot} tool to derive the apparent K$_s$ magnitudes. {The counts were compatible with the linearity regime of the detector.} The flux was integrated in a circle around the source (typically of radius 1\farcs2) and corrected for background estimated in an annulus {with inner and outer radii of 2\arcsec\,and 3\arcsec\,, respectively}. The size of the integration circle and background annulus were {individually} chosen to minimize sky contribution and pollution from nearby stars. Eight {photometric} standard stars (\tabref{tab:stdstars}) were used to compute the {average} zero-point of photometry at the date of acquisition ($Z_p=0.972\pm0.056$). To correct for extinction, we used the value\footnote{\href{http://www.eso.org/sci/facilities/paranal/decommissioned/isaac/tools/imaging\_standards.html\#Extinction}{http://www.eso.org/sci/facilities/paranal/decommissioned/isaac/tools/ imaging\_standards.html\#Extinction}} estimated for Paranal by ESO\,: $\kappa_{K_s}=0.07$\,mag\,airmass$^{-1}$. The magnitudes we derived are available in \tabref{tab:log}, along with the log of our observations.

\begin{table}[h!]
\caption{List of standard stars used to calibrate the photometry.\label{tab:stdstars}}
\centering
\begin{tabular}{lll}
\hline\hline\\[-1.5ex]
Source         & RA J2000     & Dec J2000      \\
\hline\\[-1.5ex]
\textbf{S677D} & 23:23:34.4  & $-$15:21:04.21 \\
\textbf{S294D} & 00:33:15.19 & $-$39:24:05.54 \\
\textbf{S279F} & 17:48:22.33 & $-$45:25:39.94 \\
\textbf{S273E} & 14:56:51.45 & $-$44:49:10.6  \\
\textbf{S234E} & 20:31:20.24 & $-$49:38:59.21 \\
\textbf{S071D} & 18:28:09.1  & $-$69:25:59.74 \\
\textbf{L547}  & 18:51:15.47 & $-$04:15:53.82 \\
\textbf{FS29}  & 21:52:25.44 & $+$02:23:21.3  \\
\hline
\end{tabular}
\end{table}

\begin{table*}[h]
\centering
\small
\caption{Log of our photometric and spectroscopic acquisitions.\label{tab:log}}
\begin{tabular}{ccccllcl}
\hline\hline\\[-1.5ex]
Source & RA J2000 & Dec J2000  & Observation date & \multicolumn{2}{c}{Exp. time} & Airmass & K$_{s}$ mag.  \\
       & (nIR)    & (nIR)      & (UTC)            & Phot.        & Spec.          &         &     \\[1ex]
\hline\\[-1.5ex]
IGR J00465$-$4005 & {00:46:20.681} & {$-$40:05:49.26}  & 2012-07-12T08:27:07.61 & 100\,s   & 960\,s & 1.090 & 14.820 $\pm$ 0.059 \\
IGR J10447$-$6027 & {10:44:51.925} & {$-$60:25:11.78}  & 2012-07-12T23:38:32.45 & 100\,s   & 960\,s & 1.666 & 14.100 $\pm$ 0.060 \\
IGR J12489$-$6243 & {12:48:46.422} & {$-$62:37:42.53}  & 2012-07-14T00:05:57.84 & 100\,s   & 480\,s & 1.399 & 14.724 $\pm$ 0.060 \\
IGR J13020$-$6359 & {13:01:58.723} & {$-$63:58:08.88}  & 2012-07-13T00:52:19.40 & 29.5\,s  & 480\,s & 1.476 & 11.373 $\pm$ 0.058 \\
IGR J13186$-$6257 & {13:18:25.041} & {$-$62:58:15.66}  & 2012-07-12T00:28:30.21 & 100\,s   & 480\,s & 1.377 & 13.080 $\pm$ 0.058 \\
IGR J15293$-$5609 & {15:29:29.394} & {$-$56:12:13.42}  & 2012-07-12T01:52:14.30 & ...      & 480\,s & 1.202 & ...                \\
IGR J17200$-$3116 & {17:20:05.920} & {$-$31:16:59.62}  & 2012-07-13T06:07:36.45 & 100\,s   & 960\,s & 1.405 & 12.202 $\pm$ 0.058 \\
IGR J17404$-$3655 & {17:40:26.862} & {$-$36:55:37.39}  & 2012-07-14T02:42:28.86 & ...      & 960\,s & 1.025 & ...                \\
IGR J17586$-$2129 & {17:58:34.558} & {$-$21:23:21.55}  & 2012-07-12T03:36:55.12 & ...      & 480\,s & 1.006 & ...                \\
IGR J17597$-$2201 & {17:59:45.518} & {$-$22:01:39.48}  & 2012-07-12T03:52:08.04 & 100\,s   & 960\,s & 1.019 & 13.091 $\pm$ 0.058 \\
IGR J18457$+$0244 & {18:45:40.388} & {$+$02:42:08.88}  & 2012-07-12T05:51:29.41 & 100\,s   & 960\,s & 1.302 & 14.6~~~~~$\pm$ 0.2 \\
IGR J18532$+$0416 & {18:53:16.028} & {$+$04:17:48.24}  & 2012-07-13T07:28:22.53 & 100\,s   & 960\,s & 1.970 & 13.968 $\pm$ 0.060 \\
IGR J19308$+$0530 & {19:30:50.756} & {$+$05:30:58.12}  & 2012-07-12T08:07:51.83 & ...      & 480\,s & 1.863 & ...                \\
IGR J19378$-$0617 & {19:37:33.029} & {$-$06:13:04.76}  & 2012-07-13T08:06:57.13 & ...      & 480\,s & 1.582 & ...                \\
\hline
\end{tabular}
\end{table*}

\subsection{Spectroscopy}
Long-slit spectroscopy was acquired with ISAAC in short wavelength spectroscopy-low resolution mode (SWS-LR). The 0\farcs6 slit was used to obtain a spectral resolution of $R=750$ in the K band (1.8 -- 2.5\,$\mu$m) providing a dispersion of 7.14\,\AA\,per pixel. The full width of narrow OH lines from sky emission were measured to be 26$\pm$1\,\AA\, at 22\,000\,\AA, which is compatible with the theoretical instrumental resolution of $R=750$. For each source, eight spectral frames were taken. Each spectra was acquired with a slight spatial offset ($\sim$\,30\arcsec) along the slit, following the ESO nodding procedure.

Data reduction was also performed with standard IRAF tools. Each spectrum was corrected by dark and flat frames. The overall sky value was estimated with the median of the eight spectra and then subtracted. Each spectrum was isolated via \textit{IRAF.apall} package with respect to its local background. The trace of the spectra was fitted with a seventh order Tchebyshev polynomial. The eight extracted spectra were then combined through a median to correct for cosmic rays.

Atmospheric absorption was corrected on each reduced spectrum using \textit{Molecfit} \citep{kausch_molecfit:_2015,smette_molecfit:_2015}. This radiation transfer tool fits the atmospheric features based on meteorological parameters on the date acquisitions were performed. As the spectral types of the observed stars were unknown, telluric correction via standard stars would have added artefacts due to spectral differences between standard and observed stars. However, the lower detector response towards 20\,000\,\AA\, leads to a poor fit of the underlying CO$_2$ telluric feature and introduces a residual artefact on almost all of our spectra.

A first wavelength solution was derived using the argon and xenon lamp spectra provided by the standard calibration procedure in ESO for ISAAC. The task \textit{IRAF.identify} allowed us to derive a solution with a RMS of 0.15\AA. However, after reviewing OH sky lines for individual sources, we noticed that the wavelengths were shifted by a constant that spans from 10 to 50\,\AA\,depending on the source. We thus derived an individual wavelength zero-point correction for each source using eight prominent OH lines.

\subsection{Astrometry\label{sect:astrometry}}
Good astrometric calibration is crucial to find nIR counterparts to high-energy detections. While images acquired with ISAAC for photometry also came with astrometric calibration, it was not precise enough. We refined the astrometry solution with the help of GAIA (Graphical Astronomic Image Analysis) by matching the positions of the sources in each field of view with data from 2MASS Point Source Catalogue (PSC) and/or Gaia DR1. Sources for which only spectra were acquired have noticeably more uncertainty in their astrometric calibration, since the calibration was performed on a single exposure that is taken before every batch of spectra for slit positioning. The RMS for each astrometric calibration is given in \tabref{tab:summary}

\section{Results}\label{sect:results}
Observational data obtained with VLT/ISAAC are used for several purposes. First, the photometric data allowed us to produce deep nIR images around the high-energy positions of the IGR sources. This was a way to check for blended or fainter sources and provide unambiguous counterparts for IGRs. Second, the good spatial resolution of ISAAC allowed us to distinguish extended sources, thus greatly helping the identification process. Third, accretion-driven sources are prone to photometric variability, which is verified by comparing our results to nIR photometry from the literature, mostly found in the 2MASS PSC. The analysis of spectral features allowed us to identify extragalactic sources from the redshift of recognizable emission lines such as hydrogen 2.166\micron\,Brackett (7--4) (Br$\gamma$). As for the Galactic sources, we compared the features in their spectrum to spectral atlases (\citeauthor{kleinmann_spectra_1986}\,\citeyear{kleinmann_spectra_1986}, \citeauthor{hanson_spectral_1996}\,\citeyear{hanson_spectral_1996}\,\&\,\citeyear{hanson_medium_2005}, \citeauthor{ramirez_luminosity_1997}\,\citeyear{ramirez_luminosity_1997}, \citeauthor{lenorzer_atlas_2002}\,\citeyear{lenorzer_atlas_2002}, \citeauthor{harrison_detection_2004}\,\citeyear{harrison_detection_2004}) to derive their spectral type.

\subsection*{Distance estimate of X-ray binaries}
After deriving the spectral type of the companion star and its K$_s$ magnitude, we can estimate its distance using the hydrogen column density found in the literature, along with typical effective temperatures and stellar radii ranges from the spectral type. For our calculations, we use tabulated data from \textit{Allen's Astrophysical Quantities} \citep{cox_allens_2000} linking spectral type to effective temperature and radius.

Let m$^*$ and M$^*$ be the apparent and absolute magnitudes of the considered stars, d the distance in parsec, and A the extinction. Then the distance modulus is\,
\begin{equation}
m^* - M^* = 5\,log(d) - 5 + A
.\end{equation}

If we set L$_K^*$ as the luminosity of the stars in the K band, we also have\,
\begin{equation}
M^*_K - M_{K,\odot} = -2.5\,log\left(\frac{L_K^*}{L_{K,\odot}}\right)
.\end{equation}

Such luminosity is proportional to the surface area of emission and the spectral emissivity in the K band\,, that is
\begin{equation}
L \propto R^2\,B_{\nu}(T_{eff},2.2\,\mu m)
.\end{equation}

We can then isolate the distance and write\,
\begin{equation}
d(pc) = \left(\frac{R^*}{R_{\odot}}\right)\,\left(\frac{B_{\nu}^*}{B_{\nu,\odot}}\right)^{\frac{1}{2}} \times 10^{0.2\,(m^*_K - A_K - M_{K,\odot} + 5)}
.\end{equation}

We estimate A$_K$ with the following formula: A$_V$ = N$_H$/2.21$\times$10$^{21}$\NH\, \citep{guver_Relation_2009} and A$_K$ = 0.09$\times A_V$ \citep{whittet_book_2003}. We note that the distance estimate we obtain is rather rough, since small changes in temperature or radius affect it drastically, especially when considering giant stars or hot B-type stars. Moreover, we do not take into account the cases in which the binary has an accretion disc that participates in the nIR flux, such that our estimates are a lower limit on the distance.

\subsection{Active galactic nuclei}

\textbf{\subsubsection*{IGR J00465$-$4005}}

First discovered by \cite{bird_4th_2010} with the \textit{IBIS} gamma-ray imager on board \textit{INTEGRAL}, this source coincides with the highly absorbed galaxy ESP 39607 at redshift $z=0.201$ present in the NASA Extragalactic Database (NED). \cite{monet_usno-b_2003} reported optical magnitudes $R=17.3$ and $B=17.8,$ while the infrared magnitudes given in 2MASS \citep{skrutskie_two_2006}\,are as follows: $J=16.379\pm0.129$, $H=15.844\pm0.164$ and $K=14.845\pm0.133$. Various parameters were then derived in \cite{landi_swift/xrt_2010} by fitting \textit{Swift/XRT} data with a double power law. The intrinsic absorption is $N_H\sim2.4\times10^{23}$\NH, the photon index $\Gamma\sim2.5,$ and the 2--10\,keV flux is approximately $1.2\times10^{-12}$\flux. \cite{masetti_unveiling_2010} provided optical spectroscopy in which the authors detect features redshifted at z=0.201 from a Seyfert 2 galaxy.

Our imaging data show a single point-like source in the \textit{Swift} 4\farcs26 error circle in an otherwise clear field of view (\figref{fig:fc:00465}). This allowed us to attribute a nIR counterpart and derive a magnitude of K$_s=14.820\pm 0.059$. This value is compatible with 2MASS J00462068$-$4005491 (0\farcs1 apart from one another). There is no significant difference from the K magnitude provided by 2MASS.

The overall smooth nIR spectrum presents a single, recognizable feature that we attribute to a redshifted hydrogen Paschen (7--4) (Pa$\alpha$ 1.875\micron) emission line (\figref{fig:spec:00465}) at $z=0.202\pm 0.002$. The Br$\gamma$ line is shifted out of our spectral range, therefore the redshift was only estimated on the Pa$\alpha$ line.

We thus confirm the extragalactic nature of IGR J00465$-$4005 and estimate its redshift to be $z=0.202\pm0.002$. At this distance, the 2--10\,keV X-ray luminosity is $9.3\times10^{43}$\lum. Along with the high column density, this is compatible with a Seyfert 2 galaxy as previously suggested.

\textbf{\subsubsection*{IGR J10447$-$6027}}

Originally discovered by \cite{leyder_hard_2008} when \textit{INTEGRAL} was observing the region of Eta Carina, it was first associated with a young stellar object (YSO). This hypothesis was rejected by \cite{fiocchi_five_2010} with \textit{Chandra} observations that allowed these authors to isolate a single point-like source of X-rays. Fiocchi et al. used an absorbed power law with an extra 0.2\,keV thermal component to fit the X-ray data. The 0.2--10\,keV flux is 1.7$\pm$0.3$\times10^{-12}$\flux\, and the photon index is $\Gamma = 1.0\substack{+0.3\\-0.6}$. The column density  is thought to be very high ($N_H\sim2\times10^{23}$\NH), which is consistent with the fact that no optical counterpart is present in the USNO-B1.0 catalogue, which indicates a rather red source. A nIR counterpart was then found in 2MASS (2MASS J10445192$-$6025115, $J=15.308$, $H=14.967\pm0.103,$ and $K=13.977\pm0.098$). 

The 0\farcs65 \textit{Chandra} error circle allowed us to isolate a single nIR counterpart in ISAAC data (\figref{fig:fc:10447}) with a magnitude of $K_s=14.100\pm0.060,$ which is consistent with the value for 2MASS J10445192$-$6025115; this source is however 0\farcs3 apart from our candidate, probably because of the presence of another nearby 2MASS source. The spectrum shows two prominent emission lines that are consistent with redshifted hydrogen Br$\gamma$ and Pa$\alpha$ at $z=0.047\pm0.001$ (\figref{fig:spec:10447}).

We then conclude that IGR J10447$-$6027 is an AGN at $z=0.047\pm0.001$. The 0.2--10\,keV X-ray luminosity at this distance is $8.4\times10^{42}$\lum. This X-ray luminosity and the high column density are compatible with a Seyfert 2 galaxy.

\textbf{\subsubsection*{IGR J18457$+$0244}}

Detected for the first time by \cite{bird_4th_2010}, follow-up observations by \textit{Swift} \citep{landi_swift/xrt_2010} and \textit{XMM-Newton} \citep{bodaghee_xmm-newton_2012} allowed the latter authors to pinpoint a single 2MASS counterpart (2MASS J18454039$+$0242088, $J=16.244, H=15.274\pm0.136, K=14.643\pm0.122$), even though this counterpart is only outside the 2.5\arcsec \textit{XMM-Newton} 90\% confidence radius. The study led by \cite{bodaghee_xmm-newton_2012} reveals a possible $4.4$\,ks oscillation and a slightly redshifted iron K$\alpha$ line although both significances are low. It is suggested that this source is most likely an AGN, but needs further observations to reject totally the hypothesis of it being an absorbed X-ray pulsar because of potential periodic modulation at high energies. Refined \textit{XMM-Newton} astrometry provides better positioning (see \tabref{tab:xpos}) for this source with an error circle of 2\farcs16 \citep{rosen_XMM-Newton_2016}.

At this position we find a single nIR counterpart that corresponds to 2MASS J18454039$+$0242088 (\figref{fig:fc:18457}). Our deep K-band images reveal an axisymetric extended source that has a bright centre and dimmer wings of dimensions 2\farcs5$\times$6\farcs5 (\figref{fig:contour:18457}). The centre of the extended source lies less than 0\farcs1 away from the 2MASS position.
The K$_s$ magnitude is $14.6\pm0.2,$ which is compatible with no variability from the 2MASS value, and the K-band spectrum does not reveal any particular feature (\figref{fig:spec:18457}).

A smooth spectrum suggests a synchrotron origin, perhaps due to IGR J18457+0244 seen flaring. Adding the fact that the K$_s$ image reveals an extended source, we suggest this source is the counterpart of the high-energy detections and is very likely to be an AGN.

\textbf{\subsubsection*{IGR J18532$+$0416}}

The detection of this source was first reported in \cite{bird_4th_2010} and then localized with \textit{Swift-XRT} \citep{fiocchi_swift_2011}. Follow-up observations with \textit{XMM-Newton} \citep{bodaghee_xmm-newton_2012} allowed them to further constrain the position of the source and associate it with 2MASS J18531602$+$0417481 ($J=16.488, H=15.15\pm0.093, K=13.864\pm0.072$); this is the only source compatible with the position uncertainty of $2\farcs5$. \cite{bodaghee_xmm-newton_2012} fitted the \textit{XMM-Newton} spectrum by a power law with an additional Gaussian to account for a K$_{\alpha}$ line. The spectrum is compatible with a photon index $\Gamma=1.4\pm0.1$ and a column density $N_H=1.98\pm0.08\times10^{22}$\NH. The X-ray flux at 0.2--12\,keV is 1.46$\times 10^{-12}$\flux. A feature near 6\,keV is likely to be a redshifted K$\alpha$ line at $z=0.051$, suggesting an AGN although the S/N is poor. However, a faint periodic signal at 1408\,s might be present in the light curve, which if confirmed would rather suggest that this source is an HMXB with an accreting NS. A new astrometric solution is available in 3XMM-DR7 (\tabref{tab:xpos}) and gives an updated position along with a 90\% error circle of 1\farcs14.

The ISAAC image reveals that no nIR source falls in the new \textit{XMM-Newton} error circle (\figref{fig:fc:18532}). We find that the previous counterpart 2MASS J18531602$+$0417481 (source \#1) is an extended source that is very likely a galaxy (\figref{fig:contour:18532}). However, it lies 1\farcs7 away from the new \textit{XMM-Newton} position. The K$_s$=13.968$\pm$0.060 is compatible with the 2MASS value. The nIR image also reveals a very faint source (\#2, K$_s$>19) 1\farcs1 away from the \textit{XMM-Newton} position. Spectroscopy was performed on \#1 and reveals no particular feature (\figref{fig:spec:18532}). We note that extending the 90\% error circle to a 3$\sigma$ circle of 1\farcs6 for the \textit{XMM-Newton} position and 0\farcs24 for the 2MASS object (\#1), both candidates are viable counterparts.

IGR J18532+0416 is thought to be either an AGN or a HMXB. If we assume candidate \#1 is the counterpart, it would then be an AGN. The fact that we spatially resolve the galaxy is consistent with the suggested redshift of $0.051$ derived in X-rays. If we assume instead that candidate \#2 is the counterpart, we would identify it as an HMXB. Its faint magnitude associated with the column density and a fiducial BV companion star would place this source between 64 and 80\,kpc, right outside the Milky Way. This makes candidate \#2 very unlikely to be the counterpart to IGR J18532+0416, thus we conclude that candidate \#1 is its counterpart and that it is an AGN.

Our nIR imaging thus favours IGR J18532+0416 being an extragalactic source and more precisely an AGN at redshift $z=0.051$ as derived from X-ray observations. Its 0.2--12\,keV X-ray luminosity would be about $1.6\times10^{43}$\lum\,at this distance.

\textbf{\subsubsection*{IGR J19378$-$0617}}

The first detection by \textit{INTEGRAL} was reported by \cite{molkov_integral/ibis_2004}, before a slightly adjusted position is given by \cite{bird_3rd_2007}. The authors already suggested the detection of a Seyfert 1 AGN. Further \textit{Swift} observations lead to a better positioning of the source and an extended nIR counterpart falls inside the $3\farcs5$ error circle (2MASX J19373299-0613046, $J=12.673\pm0.018, H=11.598\pm0.083, K=10.721\pm0.038$). This source matches with SS 422/1H 1934-063 given in the catalogue from \cite{molkov_integral/ibis_2004}. It is identified as a Seyfert 1.5 galaxy at $z=0.011$, and is already known as a radio (NVSS J193733-061304) and X-ray source (1RXS J193732.8-061305). \cite{rodriguez_swift_2008} fit a \textit{Swift} spectrum using a power law and derived a column density $N_H=0.15\pm0.05\times10^{22}$\NH\, and a photon index $\Gamma=2.5\pm0.2$, indicating a soft and not intrinsically absorbed source.

The K$_s$ image shows a rather bright, isolated source that lies inside the \textit{Swift} error circle (\figref{fig:fc:19378}). Associated with 2MASS J19373301$-$0613047, our source lies about 0\farcs25 away; the slight separation could be due to our lack of a good quality photometry image. The ISAAC K-band spectrum of IGR J19378$-$0617 shows a single emission line (\figref{fig:spec:19378}) attributed to a redshifted hydrogen Br$\gamma$ at $z=0.011\pm0.001$, which is consistent with the results of the previous studies.
We confirm the extragalactic nature of IGR J19378$-$0617, an AGN located at $z=0.011\pm0.001$.

\bigbreak
\subsection{Cataclysmic variables}

\textbf{\subsubsection*{IGR J12489$-$6243}}

This source was first detected by \textit{INTEGRAL/IBIS} as reported by \cite{bird_4th_2010}. Better spatial localization was obtained with \textit{Chandra} \citep{tomsick_localizing_2012}. However, two X-ray candidates were found near the 4\farcm1 \textit{INTEGRAL} error circle: one inside (1\farcm4) and the other right outside (5\farcm42). But because the closest source is too soft, the latter is more likely to be the actual counterpart. The authors fit the \textit{Chandra} spectrum of IGR J12489$-$6243a with a power law, which indeed indicates a very hard photon index of $\Gamma=-0.83\substack{+0.76\\-0.56}$ and a column density lower than $1.2\times 10^{22}$\NH. The 2--10\,keV flux is 5.3$\times10^{-13}$\flux. This strongly supports a Galactic origin for IGR J12489$-$6243a and is likely to be either a CV or an HMXB.

In our 2012 ISAAC observations on IGR J12489$-$6243a, we find a point-like source at the position given by \textit{Chandra,} {which centroid lies inside} the 90\% error circle (\figref{fig:fc:12489}). This is the only counterpart we find in the ISAAC image. {The only catalogued counterpart is CXOU J124846.4-623743: the closest 2MASS source is located more than 3\arcsec away.} The K$_s$ magnitude is $14.724\pm0.060$. The K-band spectrum reveals three emission lines: \ion{He}{I} 2.058\micron, \ion{He}{I} 2.162\micron,\,and Br$\gamma$ (\figref{fig:spec:12489}). The S/N does not allow us to identify any other features. According to \cite{hanson_spectral_1996}, few OB stars show both \ion{He}{I} 2.058\micron\,and Br$\gamma$ in emission and when they do, they also present hydrogen's Pfund series in emission. The fact that IGR J12489$-$6243 does not have any Pfund emission leads us to think it is rather a K--M main-sequence star, although it does not show the CO absorption series.

Given the hardness of X-ray emissions and the variability detected by \textit{INTEGRAL}, this source could very well be a cataclysmic binary, that is a low-mass main-sequence star orbiting a white dwarf. There is a caveat however in the fact that the nIR counterpart does not show the series of absorption lines due to $^{12}$CO that is often seen in K--M stars. A possible explanation could be that the companion star depleted its atmosphere into the white dwarf (see Sect.\,\ref{discussion:cv}) or that the nIR emission is dominated by the accretion disc.
We suggest a distance of 250--730\,pc for a K--M main-sequence companion star. That would correspond to a 2--10\,keV luminosity between $4-30\times10^{30}$\lum, which is compatible with an accreting CV.

\textbf{\subsubsection*{IGR J15293$-$5609}}

\cite{tomsick_localizing_2012} used a \textit{Chandra} spectrum of IGR J15293$-$5609 to fit a power law and derive a column density $N_H=3.4\substack{+2.8\\-2.3}\times10^{21}$\NH\, along with a photon index $\Gamma$=2.4$\substack{+0.6\\-0.5}$, which favours a Galactic origin. The authors fit the spectral energy distribution (SED) from infrared to optical with no IR excess and derived an effective temperature between 4\,200 and 7\,000\,K. According to the parallax given in the Sydney Observatory Galactic Survey \citep{fresneau_vizier_2007}, this source is located  at $1.56\pm0.12$\,kpc. It was thus possible to estimate the radius of the star to be between 12.0 and 16.4\,R$_\sun$, such that \cite{tomsick_localizing_2012} suggested this is a binary hosting an early KIII giant companion. Given the distance, the authors suggested the 0.3--10\,keV luminosity of $1.4\substack{+1.0\\-0.4}\times10^{32}$\lum\,is consistent with the accretor being a white dwarf.

During the 2012 ISAAC run we only performed nIR spectroscopy for this source since its SED was already well known. A single source lies inside the 0\farcs64 \textit{Chandra} circle and its position (\figref{fig:fc:15293}) is consistent with 2MASS J15292939$-$5612133 {($J=9.620\pm0.026, H=8.962\pm0.022, K=8.747\pm0.024$, located 0\farcs1 away)}. The K-band spectrum shows the $^{12}$CO and $^{13}$CO series in absorption; these features are only seen in cooler G, K, and M stars (\figref{fig:spec:15293}). The lack of hydrogen Br$_{\gamma}$ rules out a class G or early K star. However, the presence of  the absorption lines \ion{Na}{i} 2.21\micron\,and \ion{Ca}{i} 2.26\micron\,in close proportions to those of CO leads the identification towards a main-sequence star, rather than an evolved red giant. In particular, the equivalent width ratio of CO(2-0) to both Na 2.21\micron\,and Ca 2.26\micron\,as determined in \cite{ramirez_luminosity_1997} is a strong indicator of the luminosity class; we derive a ratio of CO/(Na+Ca) of $-0.25\substack{+0.09\\-0.08}$, which is perfectly consistent with a late-K main-sequence star.

There a discrepancy between spectroscopic and both photometric and parallax measurements; the only agreement is on the K spectral class, corresponding to an effective temperature below 5\,000\,K.

Assuming a KV secondary from nIR spectroscopy, the distance to the system would be between 45 and 85\,pc and would instead yield a 0.3--10\,keV X-ray luminosity of $1-4\times10^{29}$\lum: this would be on the low end of low-luminosity CVs \citep{reis_X-ray_2013}.

Assuming a 1.56\,kpc distance, the bright K$_s$ magnitude can only be explained by a giant companion star. That would mean our spectrum shows an abnormal quantity of metals (Na, Ca, Mg) compared to the reference KIII star with solar or subsolar metallicity. In \cite{ivanov_Medium-Resolution_2004}, several spectra of K3III stars of increasing metallicity (-1.7 to 0.5\,dex) are also shown to have slightly increasing Na 2.21\micron\,to CO(2$-$0) ratio. A KIII star of metallicity higher than 0.5\,dex may match our spectroscopic observations. However, we lack reference spectra to confirm such a claim. Although in theory, several studies such as that led by \cite{stehle_chemical_1999} have predicted metal-enriched secondaries in CVs due to the exchange of metal-rich nova ejecta.

Both high energy and nIR suggest that IGR J15293$-$5609 is a CV. However, there is an uncertainty on the secondary star. A KV star, highly compatible with our ISAAC nIR spectroscopy, would indicate a close, low-luminosity CV. A KIII star would fit the previous distance estimate and be part of a CV of typical X-ray luminosity, but the secondary would then be abnormally metal rich.

\textbf{\subsubsection*{IGR J17200$-$3116}}

\cite{masetti_unveiling_2006} reported an optical counterpart to the \textit{INTEGRAL} detection \citep{walter_14_2004} thanks to a refined position given by \textit{Chandra}. The presence of an H$\alpha$ emission line and the overall reddened spectrum along with a column density of $N_{H}=1.9\substack{+0.9\\-0.5}\times10^{22}$\NH\,\citep{tomsick_chandra_2008} would be typical of \textit{INTEGRAL} absorbed HMXB systems. However, \cite{esposito_X-ray_2014} used \textit{XMM-Newton} data to fit a power law with an extra 1.2\,keV black body to derive a photon index of 0.81$\pm$0.06, a 1--10\,keV flux of 2.74$\times10^{-11}$\flux, and a column density N$_H$=1.31$\pm$0.07\NH. The authors also detected a 326\,s X-ray pulse period. The pulse period is slow for a typical NS inside an LMXB, while  the hard photon index is compatible with both HMXB and CV.

We find a single, bright counterpart just inside the \textit{Chandra} error circle in the ISAAC field of view (\figref{fig:fc:17200}). Its K$_{s}$ magnitude is $12.202\pm0.058$. Our source lies less than 0\farcs1 away from 2MASS J17200591$-$3116596 ($J=13.581\pm0.056, H=12.334\pm0.057, K=11.983\pm0.043$). Our K$_s$ measurement is slightly dimmer than expected from 2MASS (0.176 mag in K$_s$), which would indicate that this source is variable. The K-band spectrum of this source reveals the $^{12}$CO absorption series after 2.3\micron\,(\figref{fig:spec:17200}), which is typical of a low-mass star \citep{kleinmann_spectra_1986}. Since the CO absorption is well visible up to the 6$^{th}$ series, we suggest this is a K star rather than an M star. Two helium emission lines are present, along with a \ion{Mg}{I} absorption line. The low intensity of the \ion{Mg}{I} line and the absence of other lines such as \ion{Na}{I} and \ion{Ca}{I} favours an evolved KIII star rather than a main-sequence star.

The hard photon index, slow pulse period, and KIII companion star are all compatible with a low-mass star orbiting around a white dwarf. We suggest IGR J17200$-$3116 is a symbiotic CV. Assuming a KIII star, the distance would be between 4--8\,kpc. That would yield a 1--10\,keV X-ray luminosity between $0.7-2.8\times10^{35}$\lum. We note that \cite{orio_two_2007} found symbiotic stars with similar luminosities.

\textbf{\subsubsection*{IGR J17404$-$3655}}

The first \textit{INTEGRAL} detection was reported in \cite{bird_3rd_2007} and a \textit{Swift} position was reported later in \cite{landi_swift/xrt_2008} along with the fit of the \textit{Swift} spectrum with an unabsorbed power law ($\Gamma$=0.24, 2--10\,keV flux of 1.2$\times$10$^{-11}$\flux). Further optical observations led \cite{masetti_unveiling_2009} to identify this source as an LMXB. The authors estimated its distance to be 9.1\,kpc and thus its luminosity to be 1.2$\times10^{35}$\lum. However, the \textit{Chandra} spectrum presented in \cite{tomsick_chandra_2009} is compatible with a photon index $\Gamma=-0.30\substack{+0.30\\-0.24}$ and is too hard for a typical LMXB; the authors suggested it is more likely to be caused by an HMXB hosting a highly magnetized NS.

A single point-like source lies within the \textit{Chandra} error circle (\figref{fig:fc:17404}). {It lies less than 0\farcs1 away from 2MASS J17402685$-$3655374 ($J=15.340\pm0.094,H=14.564,K=14.194$).} ISAAC nIR spectrum shows the two emission lines of \ion{He}{I} 2.058\micron\,and Br$\gamma$, along with a faint \ion{C}{IV} emission line. This could correspond to a hot B-type star as shown in \cite{hanson_spectral_1996}. However, such stars show Br$\gamma$ and Pfund emission series after 2.3\micron\,in comparable strengths. This is not the case for IGR J17404$-$3655. We do not detect any Pfund line (\figref{fig:spec:17404}), while the Br$\gamma$ line is rather strong; that is stronger than the line we detect in IGR J13186$-$6257, which also shows Pfund emission (see \figref{fig:spec:13186}).

A study of CVs in the nIR domain by \cite{harrison_detection_2004} revealed that their spectra do not always show the absorption series of $^{12}$CO that we usually see in cooler, less massive stars and that is still present the Br$\gamma$ and \ion{He}{i} lines mentioned above. In particular, the spectrum of IGR J17404$-$3655 resembles that of a K3--5V star orbiting a white dwarf. Thus, according to K-band spectroscopy, we would rather identify this source as a CV with a K3--5V companion star.

The hard high-energy spectrum along with the aforementioned nIR spectral features are compatible with IGR J17404$-$3655 being a CV with a K3--5V companion. This would represent a distance between 530 and 700\,pc, for a 2--10\,keV X-ray luminosity of 4--7$\times10^{32}$\lum; these values are compatible with a CV.

\textbf{\subsubsection*{IGR J17586$-$2129}}

Discovered by \textit{INTEGRAL} \citep{bird_second_2006} and further observed by \textit{Chandra} \citep{tomsick_chandra_2009}, a power-law fit indicates a highly absorbed source ($N_H=9-22\times10^{22}$\NH) with a photon index of $\Gamma=0.23\substack{+0.59\\-0.54}$. It coincides with a deeply reddened 2MASS source (2MASS J17583455$-$2123215, $J=11.380\pm0.040, H=9.530\pm0.034, K=8.437\pm0.027$) with $I-K_s\sim7$. The high absorption may indicate that it lies very far away, probably several kpc. Given a fiducial distance of 5\,kpc, \cite{tomsick_chandra_2009} considered its 0.3--10\,keV luminosity of $3\times10^{34}$\lum\,to be too bright for a CV, and its hardness makes it more likely to be an HMXB rather than an AGN.

The high K$_s$ {magnitude} and good localization of this source allow us to easily isolate it in the ISAAC K-band image (\figref{fig:fc:17586}). {The source lies less than 0\farcs1 away from 2MASS J17583455$-$2123215.} We can clearly see the $^{12}$CO and $^{13}$CO absorption series after 2.3\micron. Although they are not as well defined as in the spectrum of IGR J17200$-$3116 (\figref{fig:spec:17586}), we see the six CO absorption series and {thus} suggest it is a K star. The spectrum shows a faint \ion{Na}{I} absorption line. If this was a main-sequence star, the Na line should be more prominent and we should also detect \ion{Ca}{I} and \ion{Mg}{I} absorption lines. We suggest this spectrum is compatible with an evolved KIII star.

The high column density directly translates into A$_V$ = 40--100. The bright K magnitude combined with the type KIII star hypothesis would yield a distance between 130 and 300\,pc for the lowest column density estimate (N$_H \simeq 9\times 10^{22}\,cm^{-2}$). These distances give X-ray luminosities that are on the low end for CVs. This scenario is very unlikely since such high absorption may not only entirely be due to interstellar medium, but also to intrinsic absorption from material near the accretor. If we then consider instead the average extinction A$_V$=6.48 in a 5\arcmin\, radius around the source \citep{schlafly_Measuring_2011}, we derive a distance of 1.1\,kpc and a 0.3--10\,keV luminosity of $3\times10^{33}$\lum; these values are more consistent with classical values for CVs.

Along with the lack of calcium and sodium in absorption, the nIR spectrum suggests a KIII star that disagrees with the HMXB hypothesis. Even though this source has high intrinsic absorption that could suggest an LMXB, the hard photon index points towards a CV. We suggest this source is more likely to be a symbiotic CV with a KIII companion.

\bigbreak
\subsection{Low- or intermediate-mass X-ray binaries}

\textbf{\subsubsection*{IGR J17597$-$2201}}

Reported by \cite{lutovinov_IGR17597-2201_2003}, this \textit{INTEGRAL} source is first localized with XMM-Newton \citep{walter_XMM-Newton_2006}. It is subject to type I X-ray bursts detected by \textit{JEM-X} \citep{brandt_detection_2007}.\, This suggests the presence of a NS inside an LMXB. The 4\arcsec\,\textit{XMM-Newton} error circle allowed \cite{chaty_multi-wavelength_2008} to identify several counterpart candidates with NTT/EMMI. Among these, the candidate identified as \#1 has optical and infrared magnitudes that are consistent with an LMXB nature. Further \textit{Chandra} observations by \cite{ratti_chandra_2010} allowed these authors to isolate a single counterpart in \textit{i'} band (NTT/EMMI), which corresponds to candidate \#1 in \cite{chaty_multi-wavelength_2008}. The average X-ray flux at 0.2--12\,keV taken from 3XMM DR7 is $2.4\times10^{-12}$\flux.

A single source is present inside the \textit{Chandra} 0\farcs64 error circle on the ISAAC image (\figref{fig:fc:17597}), which corresponds to the aforementioned candidate. Its K$_s$ magnitude is 13.091$\pm$0.058. There are no catalogued optical/infrared sources for this star. While its spectrum has an overall low signal, features such as the Br$\gamma$ and $^{12}$CO series are visible in absorption. Weaker absorption lines (\ion{Na}{i} and \ion{Ca}{i}) are possibly present with the caveat of low S/N (\figref{fig:spec:17597}). According to \cite{kleinmann_spectra_1986}, the presence of both Br$\gamma$ and $^{12}$CO series in absorption indicates a giant star. Based on the strength ratio of these features (EW$_{Br\gamma}/$EW$_{CO(2-0)}\sim 0.4$), we suggest this star is between a G8 and K0 type.

Using the average extinction A$_V$=16.99 in a radius of 5\arcmin\,around the source \citep{schlafly_Measuring_2011}, we derive a distance of 5.6\,kpc, which corresponds to an average 0.2--12\,keV X-ray luminosity of $1.1\times10^{36}$\lum. This is a typical value for an LMXB.
Both X-ray and nIR data agree that IGR J17597$-$2201 is an LMXB with a G8--K0III companion star.

\textbf{\subsubsection*{IGR J19308$+$0530}}

First reported in \cite{bird_second_2006}, follow-up observations with \textit{Swift} \citep{rodriguez_swift_2008} and \textit{Chandra} \citep{ratti_chandra_2010} allowed the latter to localize the position of the X-ray source with an accuracy of $0\farcs64$ and identify the nIR counterpart (2MASS J19305075$+$0530582, $J=9.617\pm0.032, H=9.245\pm0.023, K=9.130\pm0.023$). This star is very bright and is reported to be of spectral type F8 \citep{mccuskey_stellar_1949}. The \textit{Swift} spectrum \citep{rodriguez_swift_2008} is dominated by a soft black-body component at 0.2\,keV with N$_H$<1.5$\times$10$^{21}$\NH, corresponding to a 2--10\,keV flux of 3.3$\times$10$^{-13}$\flux. This is expected in neutron star LMXBs and IMXBs in a state of quiescence \citep{jonker_optical_2004}. A study from \cite{ratti_igr_2013} provides phase-resolved optical spectroscopy of this source. They measured an orbital period of 14.7\,h and a mass ratio q = 1.78$\pm0.04$, for a secondary of type F4V. The authors suggested the primary is most likely a white dwarf, but do not entirely rule out a NS. Its distance would be between 300--450\,pc and its 2--10\,keV X-ray luminosity between 0.5--4$\times10^{30}$\lum\,\citep{ratti_chandra_2010}.

The ISAAC field of view on this position reveals an isolated bright source (\figref{fig:fc:19308}) situated at the position of 2MASS J19305075$+$0530582. K-band spectroscopy shows a prominent Br$\gamma$ line in absorption (\figref{fig:spec:19308}) along with faint absorption lines from metals (\ion{Na}{i}, \ion{Ca}{I}). The signal of the spectrum does not allow us to constrain well enough the strength ratio of the metals to Br$\gamma$\,; as such, our spectrum is compatible with types F8--G0V--III according to \cite{kleinmann_spectra_1986}.
Our data is thus compatible with an intermediate-mass star and we confirm the identification from \cite{ratti_igr_2013} of F4V, thus IGR J19308$+$0530 is an IMXB.

\bigbreak
\subsection{High-mass X-ray binaries}

\textbf{\subsubsection*{IGR J13020$-$6359}}

\cite{bird_second_2006,bird_3rd_2007} first identified this source as  an HMXB based on its proximity with  2RXP J130159.6-635806, which is an accreting pulsar according to  \cite{chernyakova_discovery_2005}. The authors also mentioned a noticeable variability and derive a column density of $2.48\times10^{22}$\NH. A follow-up study with \textit{Swift/XRT} by \cite{rodriguez_swift_2009} shows that the spectral characteristics of IGR J13020$-$6359 are compatible with this source being an accreting pulsar. Assuming a distance between 4--7\,kpc \citep{chernyakova_discovery_2005}  its 2--10\,keV luminosity is about 8--9$\times10^{34}$\lum\,, which is typical of high-mass stars accreting on pulsars \citep{bodaghee_description_2007}.

The ISAAC imaging reveals there is a single source within the 0\farcs39 \textit{XMM-Newton} error circle (\figref{fig:fc:13020}). This source is associated with a rather bright 2MASS object (2MASS J13015871$-$6358089{, $J=12.962\pm1.339, H=12.047\pm0.031, K=11.346\pm0.088$, 0.1\arcsec\, away}). We measure a magnitude $K_s=11.373\pm0.058$, showing no significant variability from the 2MASS value. Its nIR spectrum shows features leading towards a high-mass main-sequence star (\figref{fig:spec:13020}): first the presence of the two emission lines \ion{He}{I} 2.058\micron\,and Br$\gamma$\,and second hydrogen's distinctive Pfund series (25--5 to 17--5) after 2.350\micron. According to the spectral atlas given by \cite{hanson_spectral_1996}, the presence of both \ion{He}{I} and Br$\gamma$ in emission (the latter being more intense than \ion{He}{I}) is only seen in BVe stars. Also, the Pfund series in emission is reported in \cite{lenorzer_atlas_2002} to be present in B0--6Ve stars, which further confirms that this source is an early-type star.

The distance was previously estimated in \cite{chernyakova_discovery_2005} with the rough spectral type and magnitude of the companion star (B-type star at 4--7\,kpc for T$_{eff}\sim$10\,000\,K and 6--10 solar radii). Our data suggest it might be closer, between 0.8 and 2.3\,kpc, using the column density derived in X-rays, temperatures between 12\,000 and 25\,000\,K, and a radius of 2--3.6 R$_{\odot}$, which are typical values for B0--6V stars.

We conclude 2MASS J13015871$-$6358089 is the counterpart to the X-ray emission and is of spectral type B0--6Ve. This agrees with the high-energy characteristics of IGR J13020$-$6359, which we suggest is an HMXB with an accreting pulsar and a B0--6Ve companion.

\textbf{\subsubsection*{IGR J13186$-$6257}}\label{sect:13186}

This source is first reported in \cite{landi_swift/xrt_2008} and later associated with CXOU J131825.0-625815 \citep{tomsick_chandra_2009}\,; the position given by \textit{Chandra} allowed Tomsick et al. to find the IR counterpart to be 2MASS J13182505$-$6258156 ($J=13.581$, $H=12.689$, $K_s=12.842\pm0.050$). The authors derive a column density of $1.8\substack{+6.6\\-1.3}\times10^{23}$\NH\,although the local value could be from $2\times10^{22}$ to $8.4\times10^{23}$\NH, which resembles the characteristics of many \textit{INTEGRAL} HMXBs.

A single counterpart is easily found among the ISAAC field of view on the edge of the 0\farcs64 \textit{Chandra} position (\figref{fig:fc:13186}). This counterpart is compatible with 2MASS J13182505$-$6258156, which is 0\farcs1 away. We obtain a K$_s$ magnitude of $13.080\pm0.058$, which is 0.219\,mag dimmer than expected from 2MASS, indicating a potentially variable source. 

The nIR spectrum is very similar to that of IGR J13020$-$6359. However the signal is not as good.\, \ion{He}{I} and Br$\gamma$ are visible in emission, but the Pfund series is not as prominent and suffers from low S/N (\figref{fig:spec:13186}). To confirm its presence, we used a slightly different reduction method that favours signal at longer wavelengths.

The first reduction process treats each of the eight raw spectra individually before combining them. It is especially adapted for bright stars as there is no issue when extracting each spectra from the background. However, faint stars do not seem to benefit from this method since it is harder to {estimate the trace over the entire range} on a raw, single spectrum. The individual trace fitting is thus not as reliable.
The second reduction process first combines the raw spectra to optimize the S/N. The fit of the {trace} gives better results at longer wavelengths, but sacrifices S/N at shorter wavelengths.

Using the second reduction method on IGR J13186$-$6257 reveals that the Pfund series in emission is definitely present. This is the only source in the sample that gave noticeably different results when comparing the two reduction processes. In particular, applying the second method to IGR J12489$-$6243 and IGR J17404$-$3655, both of which present \ion{He}{I} and Br$\gamma$ in emission, did not reveal any additional feature either Pfund in emission or $^{12/13}$CO in absorption.

IGR J13186$-$6257 is thus very likely to be a B0--6Ve star orbiting a compact object, together forming a BeHMXB. If we consider the uncertainty on the local value of column density, this heavily absorbed binary could be situated from 0.7 to 5.4\,kpc.

\bigbreak

\section{Discussion}\label{sect:discuss}

In this section we discuss the overall results of the nIR observations in comparison to the high-energy identifications and previous distributions of object types among the whole catalogue of \textit{INTEGRAL} sources.

\subsection{AGN candidates identified through imaging}\label{sect:agn}
Near-infrared imaging on IGR J18457+0244 and IGR J18532+0416 (Fig. \ref{fig:contour:agn}) shows extended sources inside (or adjacent to) the high-energy error circles. These two sources do not present any particular features in their nIR spectrum, as opposed to the other AGNs presented in this paper. Even though their signal is rather low, hydrogen emission lines --if present-- should still be detectable, as we easily detect hydrogen in IGR J00465$-$4005, an AGN of similar K$_s$ luminosity. Since they are spatially resolved, their redshift is too low to shift the Br$\gamma$ line out of our spectral range, which is why we are confident about the actual absence of features in their spectrum. We thus suggest that this source is associated with synchrotron emission, coming from flaring AGNs.

\subsection{Abundance anomalies in cataclysmic variables}\label{discussion:cv}
Near-infrared spectroscopy of IGR 12489-6243 and IGR J17404$-$3655 reveals both helium (\ion{He}{I} 2.058\micron) and Br$\gamma$ emission lines similar to OB secondaries in HMXB systems. However, the absence of the Pfund series in emission (see Sect.\,\ref{sect:13186}) leads us to doubt the massive star hypothesis, even though these sources do not show CO absorption as G- or K-type stars would.

The deciding factor comes from nIR spectroscopy on a sample of 12 CVs \citep{harrison_detection_2004} that show abundance anomalies. In particular, some of these sources lack CO absorption and have both \ion{He}{i} 2.058\micron\,and Br$\gamma$ emission, which is very similar to our results on the IGRs mentioned above.

Such spectra could be explained by rather old CVs in which the G/K-type secondary has had time to deplete its atmosphere (were CO absorption usually takes place) into the white dwarf. Helium and hydrogen emission lines would then come from the heated accretion disc rather than from the secondary star itself.

This relies on the fact that while single stars are rather well constrained in terms of element abundances and spectral features, the accreting phase in a binary system can drastically change these characteristics.

\begin{figure}[h]
\includegraphics[width=.49\textwidth]{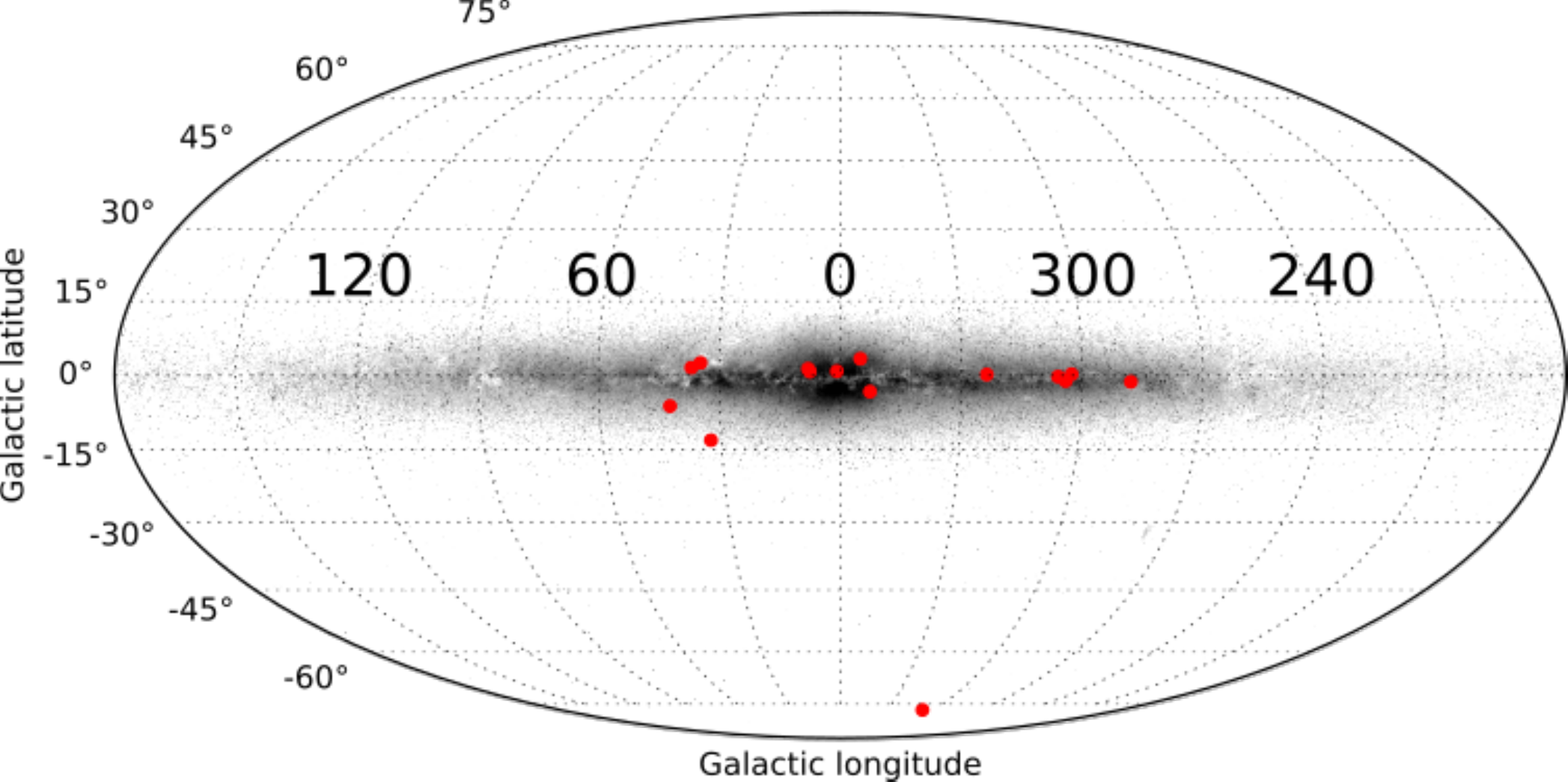}
\caption{Spatial distribution of the \textit{INTEGRAL} sources identified in this paper.\label{fig:galdistrib}}
\end{figure}

\subsection{Source distribution}\label{sect:distrib}

In \cite{bird_ibis_2016} is given the list of all the \textit{INTEGRAL} detections so far, along with their nature when available. These sources are either AGNs (39\%), L/IMXBs (14\%), HMXBs (12\%), or CVs (6\%). For a total of 939 objects, 23\% were still unidentified. Among our 14 sources, 13 are mentioned in \cite{bird_ibis_2016}, while the last only appears in \cite{bird_second_2006} because the detection threshold was modified. Their distribution in the sky is shown in Fig. \ref{fig:galdistrib}.

The two sets of data appear to be rather different from one another (see \figref{fig:sourcedistrib}); however, they have a large gap in the total number of sources. To properly estimate how significant the differences are, we used the following method: First, we assume that the distribution in \cite{bird_ibis_2016} is representative of all the \textit{INTEGRAL} sources (both identified and unidentified). Then, we build a statistical model by drawing 14 random sources multiple times (typically 10$^6$) following probabilities that are derived from our hypothesis (\equref{eq:proba}) for each type of source,
\begin{equation}\label{eq:proba}
p_{type} = \frac{N_{type}}{N_{total}}
.\end{equation}

The statistical model is thus made of the average $\overline{n}_{mod}$ and the standard deviation $\sigma_{mod}$ of the 10$^6$ draws, for each type of source. We finally perform a chi-square test (\equref{eq:chisquare}) to compare our observed set of data $n_{obs}$ to the statistical model built from the set of data in \cite{bird_ibis_2016},

\begin{equation}\label{eq:chisquare}
\centering
\chi^{2} = \sum\limits_{type} \frac{\left(n_{obs,type} - \overline{n}_{mod,type} \right)^{2}}{\sigma^2_{mod,type}}
.\end{equation}

The test returns $\chi^{2}=15.98$ and with 3 degrees of freedom the corresponding \textit{p}-value (the probability that a sample from the model would present such a large difference) is 0.001. Since the threshold \textit{p}-value for statistical significance is usually set to 0.05, this means our data is significantly different from the model. \figref{fig:stat:full} shows that while our number of LMXBs and HMXBs are statistically compatible with the results in \cite{bird_ibis_2016}, we have a shortcoming of AGNs and CVs are over-represented.

The lower number of AGNs can be explained by the fact that our study focussed on sources that are located within $\pm$15\degr\,from the Galactic plane (except for one), while the sources in \cite{bird_ibis_2016} are scattered over the whole sky. We thus built another model, using only the sources that lie in the Galactic plane for both sets of data. This now concerns 412 sources from \cite{bird_ibis_2016} and 13 from our study.

The chi-square test using the new model returns $\chi^{2}=12.56$, and the associated \textit{p}-value is 0.002 with 2 degrees of freedom since we added the Galactic plane constraint. While we now have a compatible number of AGNs, the high number of CVs in our data is still significantly different from what is expected (see \figref{fig:stat:GP}).

There could be different explanations as for why we identified so many CVs. The initial hypothesis concerning the representativity of the \cite{bird_ibis_2016} catalogue could be wrong, which means the remaining >200 unidentified \textit{INTEGRAL} sources would contain a larger  percentage of CVs. This would be an interesting result since \textit{INTEGRAL} discovered obscured sgHMXBs thanks to its high energy range and good sensitivity. Therefore, the aforementioned CVs would look like HMXBs at high energies and would most likely be intermediate polars, in which the white dwarf is highly magnetized, and present a harder high-energy spectrum than regular CVs. We note that four of our CVs were first suggested to be potential HMXBs based on high-energy data, and this confirms once again the necessity of multiwavelength studies to identify binary systems.

\begin{figure}[h]
\includegraphics[width=.5\textwidth]{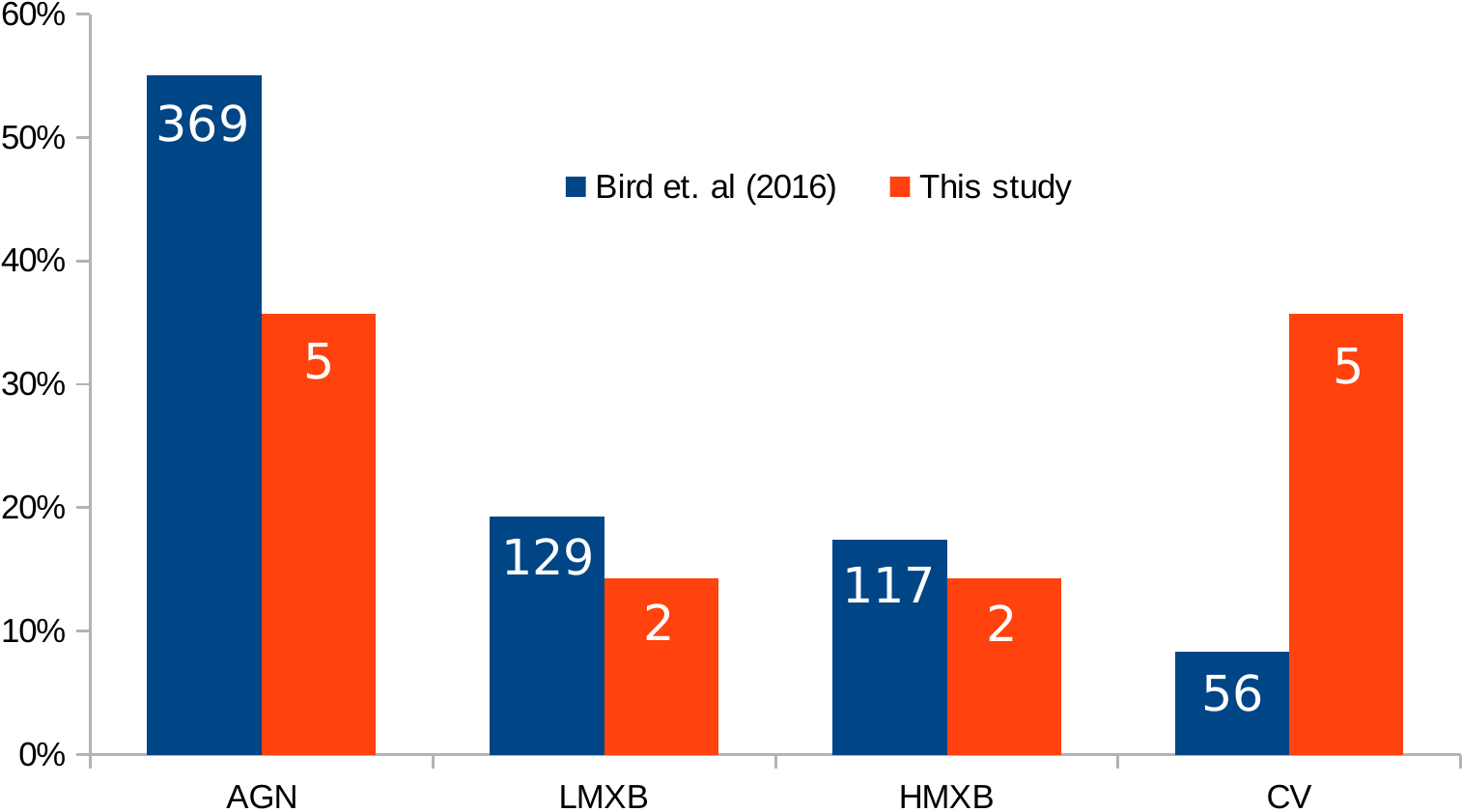}
\caption{Distribution of the 14 sources presented in this paper, alongside the 671 identified IGR sources presented in \cite{bird_ibis_2016}. The left axis is normalized to the fraction (\%) of the sources with respect to the total number of sources in their study, while the actual number of sources are indicated in white. \label{fig:sourcedistrib}}
\end{figure}

\begin{figure}
\begin{subfigure}{.5\textwidth}
\includegraphics[width=\textwidth]{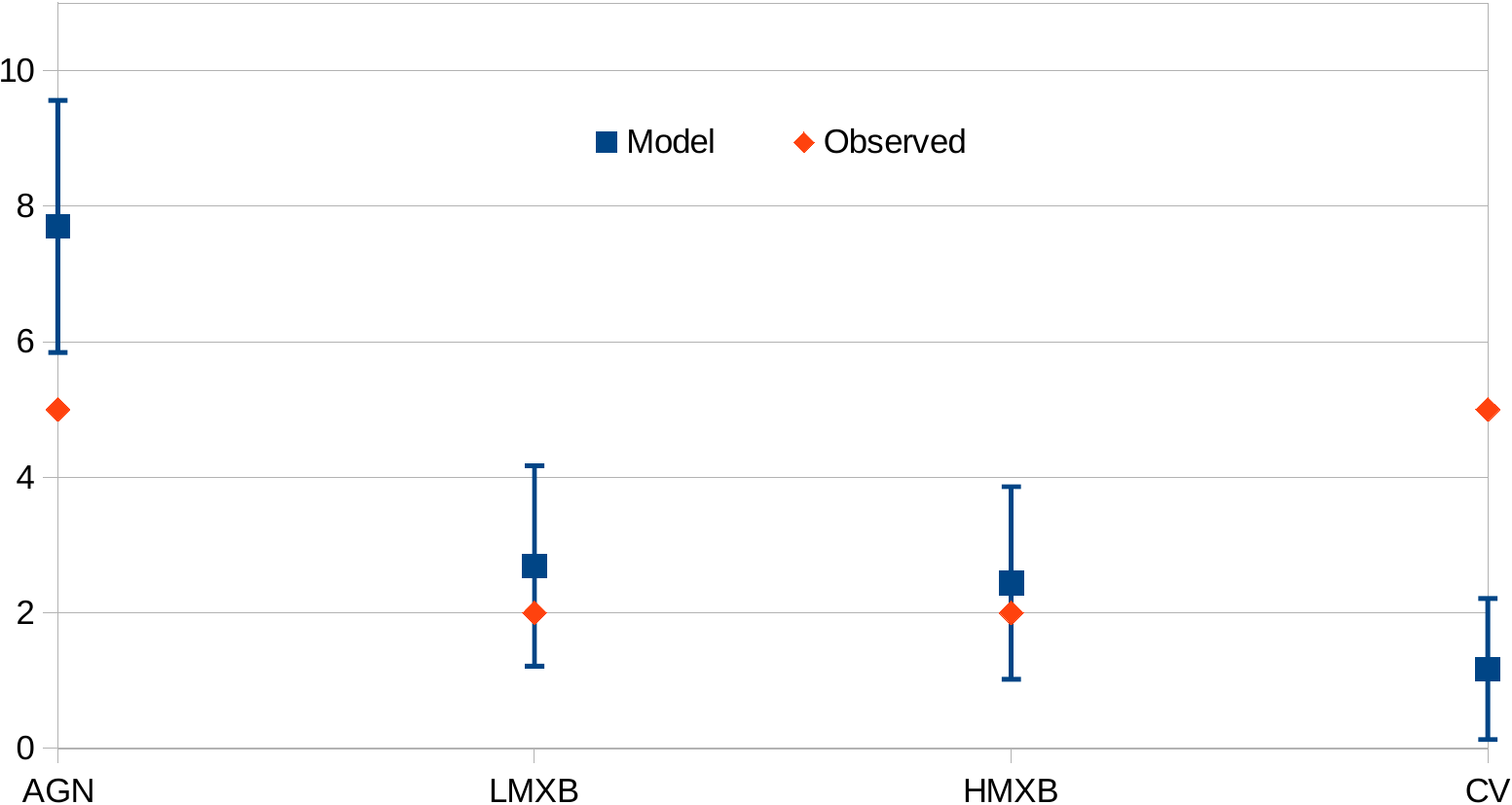}
\caption{}
\end{subfigure}

\begin{subfigure}{.5\textwidth}
\includegraphics[width=\textwidth]{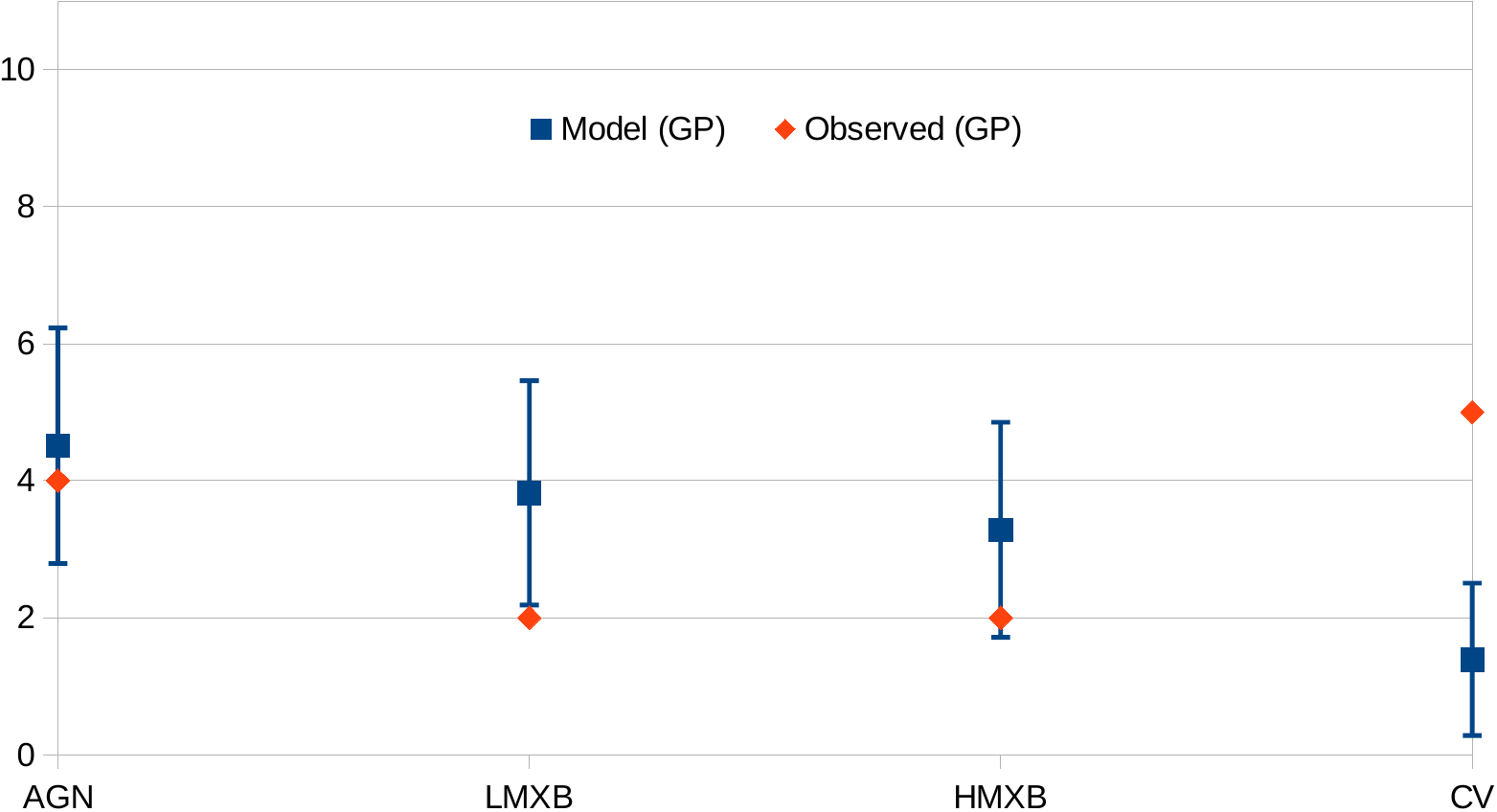}
\caption{}
\end{subfigure}

\caption{Statistical models derived from the data in \cite{bird_ibis_2016} vs. our results. Panel a: Comparison between the whole-sky model and our data. Panel b: Comparison between the Galactic plane model (GP) and our data, both restricted to sources whithin $\pm$15\degr from the GP. }
\end{figure}

\subsection{Current census of high-energy sources}\label{sect:census}
We give an inventory of the various sources detected by \textit{INTEGRAL} at high energies (\tabref{tab:census}). In addition to the previous studies (\citel{liu_catalogue_2006}\,\&\,\citeyear{liu_catalogue_2007}, \citel{coleiro_infrared_2013}, \citel{bird_ibis_2016}), we take into account the results of the current paper. All these sources were cross-correlated since some are common to two or more catalogues. The total number of confirmed AGNs detected by \textit{INTEGRAL} is 373, and represents about 31\% of the IGR sources in the Galactic plane. There are currently 217 known LMXBs, 60 CVs, and 167 HMXBs. Among the latter, we can differentiate 70 BeHMXBs and 35 sgHMXBs 10 of which are SFXTs. 

\begin{table}[h!]
\caption{Census of high-energy sources.\label{tab:census}}
\centering
\begin{tabular}{lrr}
\hline\hline\\[-1.5ex]
Sources       & Census & (\%) \\[1ex]
\hline\\[-1.5ex]
\textbf{AGN}  & 373    & 45.7\% \\[1ex]
\textbf{LMXB} & 217    & 26.6\% \\[1ex]
\textbf{HMXB} & 167    & 20.4\% \\
Be            & 70     & 8.6\% \\
sg(SFXT)      & 35(10) & 4.3 (1.2)\% \\[1ex]
\textbf{CV}   & 60     & 7.3\% \\
\hline

\end{tabular}
\end{table}

The low statistics on the high-energy sources, and in particular on X-ray binaries, comes from the rather long and difficult identification process, for which multiwavelength observations are needed. The most crucial step is obtaining a good X-ray localization of the sources (typically <\,1\arcsec\,in the Galactic plane, <\,2\arcsec\,in less crowded areas). Among the 288 IGR sources for which the nature is either uncertain or completely unknown, 251 (87\%) have a localization that is less accurate than 2\arcsec, which in most cases do not  find a single optical/nIR counterpart before the position is better constrained.

\section{Conclusions}\label{sect:conclusion}
We presented nIR observations of 14 \textit{INTEGRAL} sources with the VLT/ISAAC instrument. The photometric and spectroscopic data allowed us to pinpoint nIR counterparts to the high-energy detections and identify or better constrain the nature of the sources. Among these sources, there are 5 AGNs, 5 CVs, 2 BeHMXBs, and 2 I/LMXBs. While the proportions between types are not fully consistent with those published in \cite{bird_ibis_2016}, we still expect that the remaining unidentified \textit{INTEGRAL} sources contain a significant amount of AGNs, X-ray binaries, and CVs. This could be a great resource for the two latter, since the current census of binaries is not so high and would benefit from having more candidates with a well-constrained nature. In turn, this will help to perform population studies, derive accurate classifications, and answer more general questions on stellar evolution in binaries in the context of stellar merging endpoints and the detection of gravitational waves.

\section*{Acknowledgements}

We thank the anonymous referee for careful reading and valuable input, which helped to improve this paper. We also thank Federico Garcia for his suggestions regarding several of the sources presented in this paper. This work was supported by the Centre National d'Etudes Spatiales (CNES). It is based on observations obtained with MINE: the Multiwavelength INTEGRAL NEtwork. This research has made use of the IGR Sources page maintained by J.\,Rodriguez \& A.\,Bodaghee (\href{http://irfu.cea.fr/Sap/IGR-Sources}{http://irfu.cea.fr/Sap/IGR-Sources})\,; data products from the Two Micron All Sky Survey, which is a joint project of the University of Massachusetts and the Infrared Processing and Analysis Center/California Institute of Technology, funded by the National Aeronautics and Space Administration and the National Science Foundation\,; data obtained from the 3XMM \textit{XMM-Newton} serendipitous source catalogue compiled by the ten institutes of the \textit{XMM-Newton} Survey Science Centre selected by ESA\,; the SIMBAD database and VizieR catalogue access tool, operated at CDS, Strasbourg, France\,; and NASA’s Astrophysics Data System Bibliographic Services, operated by the Smithsonian Astrophysical Observatory under NASA Cooperative Agreement NNX16AC86A.

\begin{table*}[h]

\centering
\caption{Summary of the identifications derived in this study with VLT/ISAAC nIR data. The uncertainty (in arcseconds) given on the centroid of the source comes from the RMS of the astrometric calibration (see \ref{sect:astrometry}).\label{tab:summary}}

\begin{tabular}{lllrll}
\hline\hline\\[-1.5ex]
Source & RAJ2000 & DEJ2000 & Unc.      & Previous identification & Our identification (this paper) \\
       & (nIR)   & (nIR)   & (\arcsec) & (comment)               & (spectral type / comment)       \\[1ex]
\hline\\[-1.5ex]
IGR J00465$-$4005 & {00:46:20.681} & {$-$40:05:49.26} & {0.060} & AGN Sey 2 (z=0.201) & AGN (Sey 2, z=0.202$\pm0.002$) \\
IGR J10447$-$6027 & {10:44:51.925} & {$-$60:25:11.78} & {0.080} & ?                   & AGN (Sey 2, z=0.047$\pm0.001$) \\
IGR J12489$-$6243 & {12:48:46.422} & {$-$62:37:42.53} & {0.053} & CV / HMXB ?         & CV (K/M companion, peculiar) \\
IGR J13020$-$6359 & {13:01:58.723} & {$-$63:58:08.88} & {0.164} & HMXB (NS)           & BeHMXB (B0--6Ve companion) \\
IGR J13186$-$6257 & {13:18:25.041} & {$-$62:58:15.66} & {0.072} & HMXB ?              & BeHMXB (B0--6Ve companion) \\
IGR J15293$-$5609 & {15:29:29.394} & {$-$56:12:13.42} & {0.136} & CV (K-type giant ?) & CV (K5V--III companion) \\
IGR J17200$-$3116 & {17:20:05.920} & {$-$31:16:59.62} & {0.056} & HMXB                & Symbiotic CV (KIII companion) \\
IGR J17404$-$3655 & {17:40:26.862} & {$-$36:55:37.39} & {0.125} & HMXB (NS) ?         & CV (K3--5V companion, peculiar) \\
IGR J17586$-$2129 & {17:58:34.558} & {$-$21:23:21.55} & {0.092} & HMXB ?              & Symbiotic CV (KIII companion) \\
IGR J17597$-$2201 & {17:59:45.518} & {$-$22:01:39.48} & {0.110} & LMXB (NS) ?         & LMXB (G8--K0III companion) \\
IGR J18457$+$0244 & {18:45:40.388} & {$+$02:42:08.88} & {0.043} & Pulsar / AGN ?      & AGN \\
IGR J18532$+$0416 & {18:53:16.028} & {$+$04:17:48.24} & {0.037} & HMXB / AGN ?        & AGN ($z=0.051$) \\
IGR J19308$+$0530 & {19:30:50.756} & {$+$05:30:58.12} & {0.252} & IMXB (F4V companion)& IMXB (F8-G0V-III companion) \\
IGR J19378$-$0617 & {19:37:33.029} & {$-$06:13:04.76} & {0.204} & Sey1.5 (z=0.011)    & AGN (Sey1.5, z=0.011$\pm0.001$) \\

\hline
\end{tabular}
\end{table*}

\begin{footnotesize}
\bibliographystyle{aa}
\bibliography{./references.bib}
\end{footnotesize}

\twocolumn
\begin{figure}[h!]
\caption{Contours around the nIR counterparts of the AGNs identified through imaging. The fields of view are 14$\arcsec\times14\arcsec$; north is up and east is left. \label{fig:contour:agn}}
\begin{subfigure}{.24\textwidth}
\includegraphics[width=\textwidth]{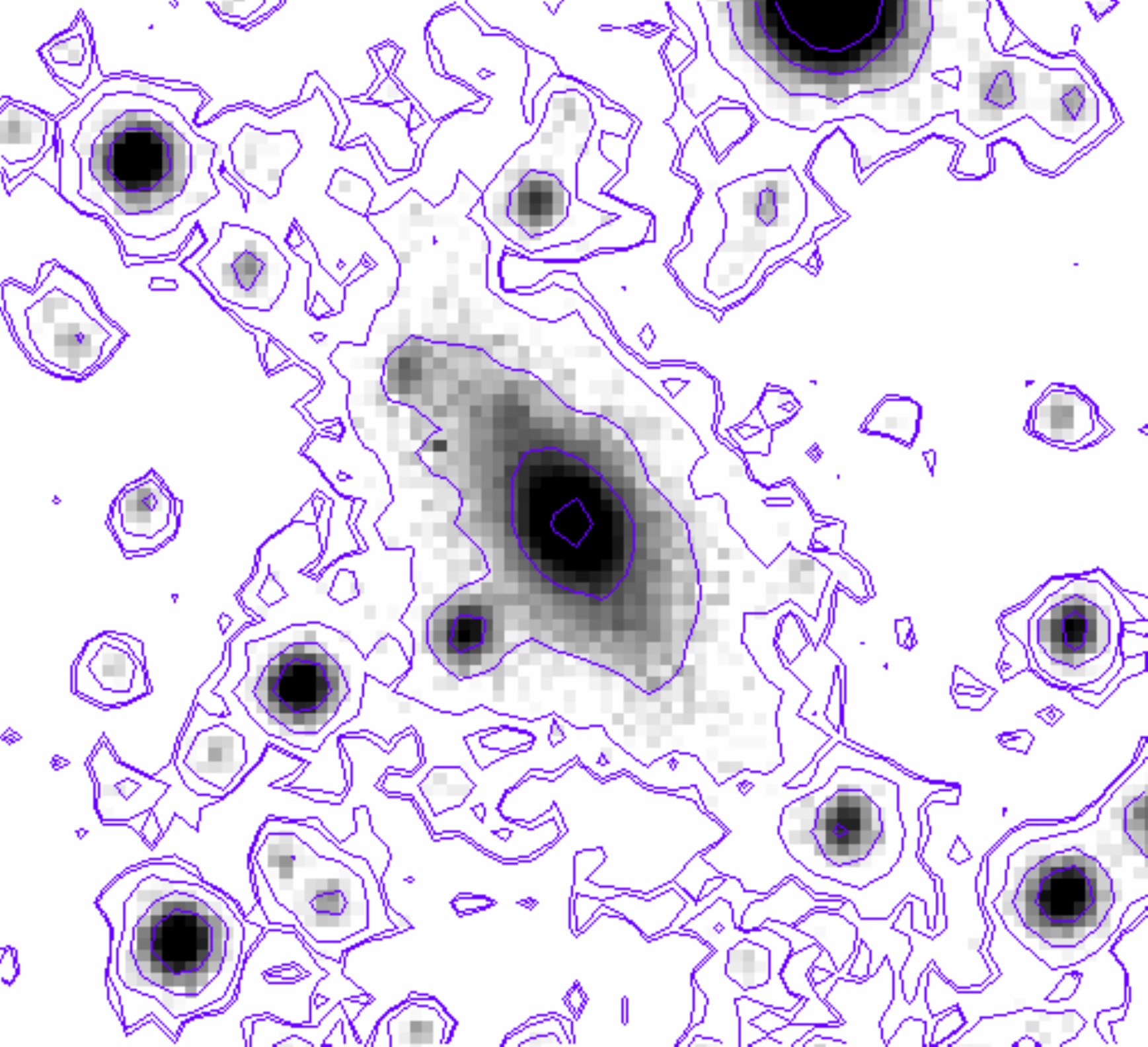}
\caption{IGR J18457$+$0244\label{fig:contour:18457}}
\end{subfigure}
\hfill
\begin{subfigure}{.24\textwidth}
\includegraphics[width=\textwidth]{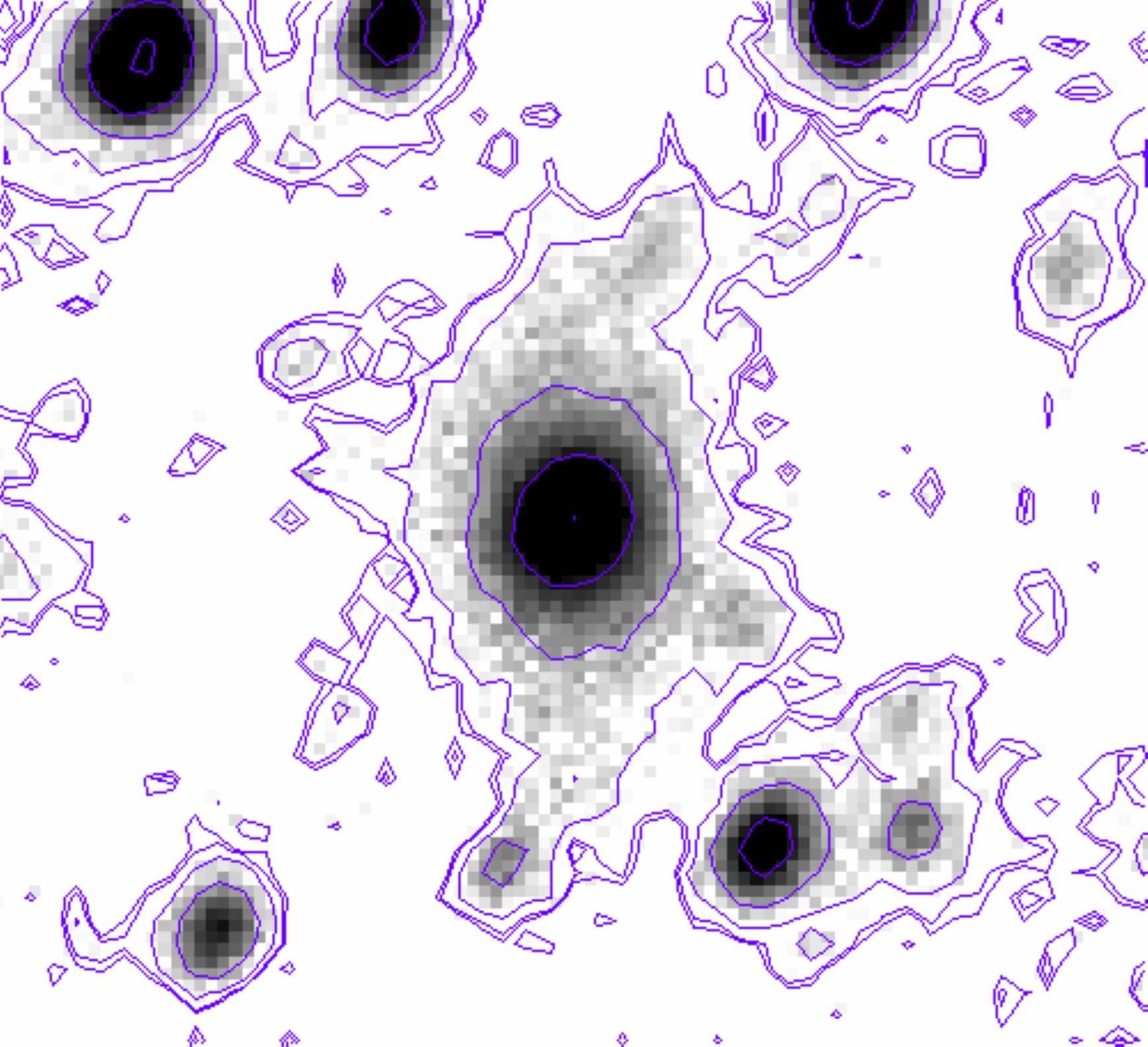}
\caption{IGR J18532$+$0416\label{fig:contour:18532}}
\end{subfigure}
\end{figure}

\begin{table}[hb!]

\caption{Results of spectroscopy on IGR sources. {All features are fitted with a Gaussian with relation to the local background. We provide} the rest wavelength $\lambda$, fitted wavelength $\lambda_{fit}$, FWHM, equivalent width (EW; a negative value indicates an emission line), {and the heliocentric velocity for each identified line in binary systems.}\label{tab:lines:1}}
\centering
\footnotesize

\begin{tabular}{rcccrr}

\hline\hline\\[-1.5ex]
Line & $\lambda$   & $\lambda_{fit}$  & FWHM  & EW   &V$_r$    \\
     & (\AA)       & (\AA)            & (\AA) & (\AA) & (\kmpersec) \\
\hline\\[-1.5ex]

\multicolumn{6}{c}{\textbf{IGR J00465$-$4005}} \\
Pa~$\alpha$ & 18745.9 & 22540.1 & 54.4 & -80.1 \\
\hline\\[-1.5ex]

\multicolumn{6}{c}{\textbf{IGR J10447$-$6027}} \\
Pa~$\alpha$ & 18745.9 & 19615.4 & 42.5 & -148.8 & \\
Br~$\gamma$ & 21655.1 & 22668.6 & 33.3 & -10.0 & \\
\hline\\[-1.5ex]

\multicolumn{6}{c}{\textbf{IGR J12489$-$6243}} \\
\ion{He}{I} & 20581.4 & 20584.4 & 33.3 & -8.9 & 27 \\ 
\ion{He}{I} & 21617.1 & 21621.1 & 40.5 & -6.4 &  38 \\
Br~$\gamma$  & 21655.1 & 21664.8 & 40.0 & -12.0 & 117 \\
\hline\\[-1.5ex]

\multicolumn{6}{c}{\textbf{IGR J13020$-$6359}} \\
\ion{He}{I} & 20581.4 & 20587.0 & 46.3 & -8.8 & 66 \\
Br~$\gamma$  & 21655.1 & 21660.8 & 34.8 & -7.4 & 63 \\
Pf(25-5) & 23737.5 & 23743.9 & 28.5 & -3.7 & 65 \\
Pf(24-5) & 23821.5 & 23831.4 & 45.6 & -4.1 & 109 \\
Pf(23-5) & 23918.5 & 23928.5 & 29.7 & -5.7 & 109 \\
Pf(22-5) & 24028.4 & 24038.1 & 34.1 & -6.7 & 105 \\
Pf(21-5) & 24157.4 & 24167.2 & 32.4 & -7.1 & 106 \\
Pf(20-5) & 24307.4 & 24312.3 & 29.2 & -8.3 & 44 \\
Pf(19-5) & 24483.3 & 24489.3 & 32.7 & -10.1 & 57 \\
Pf(18-5) & 24693.3 & 24701.3 & 38.9 & -14.7 & 77 \\
Pf(17-5) & 24946.2 & 24961.5 & 32.9 & -16.6 & 168 \\
\hline\\[-1.5ex]

\multicolumn{6}{c}{\textbf{IGR J13186$-$6257}} \\
\ion{He}{I} & 20581.4 & 20571.5 & 38.1 & -4.7 & -161 \\
Br~$\gamma$  & 21655.1 & 21656.3 & 21.8 & -2.7 & 0 \\
Pf(22-5) & 24028.4 & 24053.0 & 32.0 & -2.0 &  289 \\
Pf(21-5) & 24157.4 & 24177.0 & 21.6 & -2.0 &  226 \\
Pf(20-5) & 24307.4 & 24322.6 & 32.7 & -3.6 &  170 \\
Pf(19-5) & 24483.3 & 24510.4 & 41.7 & -5.7 &  314 \\
Pf(18-5) & 24693.3 & 24712.5 & 28.8 & -5.9 &  216 \\
Pf(17-5) & 24946.2 & 24966.9 & 32.5 & -8.0 &  232 \\
\hline\\[-1.5ex]

\end{tabular}

\end{table}

\begin{table}
\caption{Results of spectroscopy on IGR sources (continued from Table \ref{tab:lines:1}).}\label{tab:lines:2}
\centering
\small
\begin{tabular}{rcccrr}

\hline\hline\\[-1.5ex]
Line & $\lambda$   & $\lambda_{fit}$  & FWHM  & EW    &V$_r$   \\
     & (\AA)       & (\AA)            & (\AA) & (\AA) & (\kmpersec) \\
\hline\\[-1.5ex]

\multicolumn{6}{c}{\textbf{IGR J15293$-$5609}} \\
\ion{Na}{I}  & 22056.4 & 22055.0 & 34.4 & 2.3 &  -36 \\
\ion{Ca}{I} & 22607.9 & 22612.3 & 30.8 & 1.9 &  41 \\
\ion{Mg}{I} & 22807.7 & 22809.5 & 43.5 & 1.9 &  7 \\
CO (2-0) & 22928.7 & 22940.6 & 46.3 & 5.3 & 139 \\
CO (3-1) & 23220.7 & 23229.3  & 37.4 & 4.9 & 94 \\
$^{13}$CO (2-0) & 23441.6 & 23441.8 & 22.0 & 0.4 &  -14 \\
CO (4-2) & 23528.6 & 23528.0 & 45.1 & 4.6 &  -25 \\
CO (5-3) & 23822.5 & 23840.6 & 46.2 & 6.1 &  211 \\
$^{13}$CO (4-2) & 24030.4 & 24040.0 & 25.8 & 1.2 &  103 \\
CO (6-4) & 24135.4 & 24144.5 & 30.4 & 4.0 &  96 \\
CO (7-5) & 24454.3 & 24461.4 & 27.2 & 3.4 &  70 \\
\hline\\[-1.5ex]

\multicolumn{6}{c}{\textbf{IGR J17200$-$3116}} \\
\ion{He}{I} & 20581.4 & 20576.1 & 21.4 & -1.9 &  -91 \\
\ion{He}{I} & 21120.2 & 21120.7 & 19.6 & -0.7 &  -7 \\
CO (2-0) & 22928.7 & 22956.3 & 69.6 & 11.8 &  347 \\
CO (3-1) & 23220.7 & 23240.3 & 42.4 & 7.7 &  239 \\
$^{13}$CO (2-0) & 23441.6 & 23463.1 & 38.8 & 3.4 &  261 \\
CO (4-2) & 23528.6 & 23537.1 & 65.4 & 10.2 &  94 \\
$^{13}$CO (3-1) & 23732.5 & 23747.1 & 56.0 & 4.8 &  170 \\
CO (5-3) & 23822.5 & 23847.6 & 55.1 & 9.2 &  302 \\
$^{13}$CO (4-2) & 24030.4 & 24049.8 & 26.5 & 2.8 &  228 \\
CO (6-4) & 24135.4 & 24163.0 & 57.8 & 10.7 &  329 \\
$^{13}$CO (5-3) & 24334.4 & 24345.9 & 23.0 & 4.5 &  128 \\
CO (7-5) & 24454.3 & 24480.4 & 63.1 & 13.0 & 306 \\
\hline\\[-1.5ex]

\multicolumn{6}{c}{\textbf{IGR J17404$-$3655}} \\
\ion{He}{I} & 20581.4 & 20583.1 & 44.7 & -12.1 &  23 \\
\ion{C}{IV}  & 20774.3 & 20776.2 & 20.4 & -1.7 &  15 \\
Br~$\gamma$  & 21655.1 & 21658.2 & 55.4 & -20.9 &  31 \\
\hline\\[-1.5ex]

\multicolumn{6}{c}{\textbf{IGR J17586$-$2129}} \\
\ion{Na}{I}  & 22056.4 & 22058.5 & 16.3 & 1.0 &  18 \\
CO (2-0) & 22928.7 & 22948.2 & 40.2 & 5.7 &  244 \\
CO (3-1) & 23220.7 & 23239.3 & 45.9 & 6.1 &  229 \\
CO (4-2) & 23528.6 & 23536.5 & 52.2 & 4.8 & 90 \\
$^{13}$CO (3-1) & 23732.5 & 23747.7 & 66.5 &  1968 & 181 \\
CO (5-3) & 23822.5 & 23845.1 & 58.5 & 5.3 &  273 \\
$^{13}$CO (4-2) & 24030.4 & 24059.2 & 40.9 &  1615 & 348 \\
CO (6-4) & 24135.4 & 24154.6 & 66.5 & 4.6 &  227 \\
\hline\\[-1.5ex]

\multicolumn{6}{c}{\textbf{IGR J17597$-$2201}} \\
Br~$\gamma$ & 21655.1 & 21666.0 & 36.1 & 1.8 &  140 \\
\ion{Na}{I} & 22056.4 & 22072.0 & 21.7 & 0.8 &  201 \\
\ion{Ca}{I} & 22607.9 & 22614.7 & 20.7 & 0.7 &  79 \\
CO (2-0) & 22928.7 & 22958.2 & 45.3 & 4.1 &  375 \\
CO (3-1) & 23220.7 & 23248.6 & 38.9 & 4.9 &  349 \\
CO (4-2) & 23528.6 & 23540.3 & 29.2 & 2.9 &  138 \\
CO (5-3) & 23822.5 & 23852.7 & 48.0 & 5.3 &  369 \\
CO (6-4) & 24135.4 & 24163.1 & 30.8 & 3.2 &  333 \\
\hline\\[-1.5ex]

\multicolumn{6}{c}{\textbf{IGR J19308$+$0530}} \\
Br~$\gamma$  & 21655.1 & 21656.6 & 57.5 & 4.1 &  23 \\
\ion{Na}{I} & 22056.4 & 22052.7 & 34.3 & 0.6 & -48 \\
\ion{Ca}{I} & 22607.9 & 22610.2 & 19.6 & 0.5 &  32 \\
\hline\\[-1.5ex]

\multicolumn{6}{c}{\textbf{IGR J19378$-$0617}} \\
Br~${\gamma}$ & 21656.1 & 21883.0  & 45.6 & -5.7 & \\
\hline\\[-1.5ex]

\end{tabular}
\end{table}

\onecolumn
\begin{figure}[h!]
\caption{K-band spectra of the \textit{INTEGRAL} sources identified as AGN. Spurious features from residual artefacts are labelled in red.\label{fig:AGN}}

\begin{subfigure}{\textwidth}
\begin{minipage}{.49\textwidth}
\includegraphics[width = \textwidth]{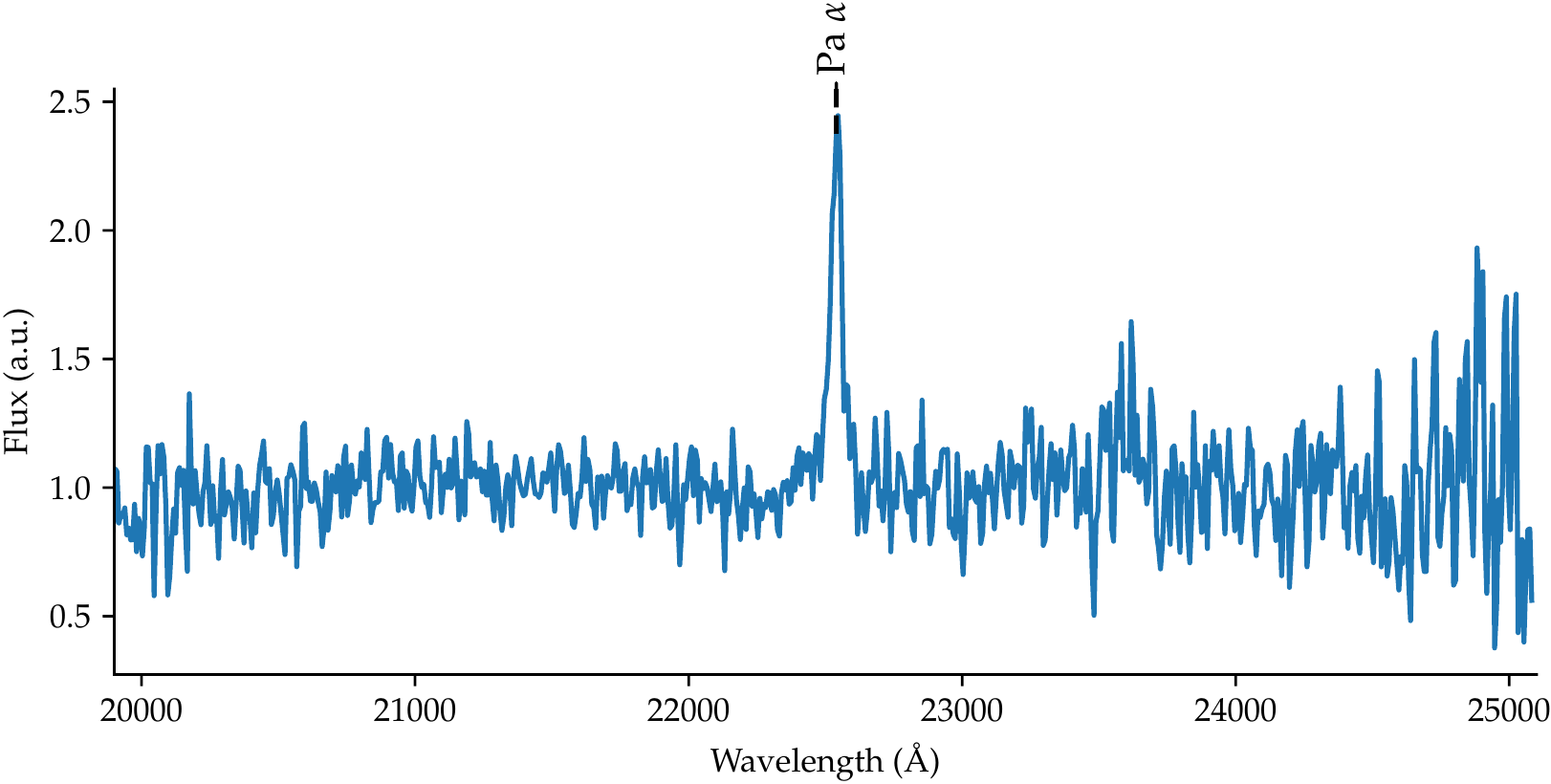}
\caption{IGR J00465$-$4005\label{fig:spec:00465}}
\end{minipage}
\hfill
\begin{minipage}{.49\textwidth}
\includegraphics[width = \textwidth]{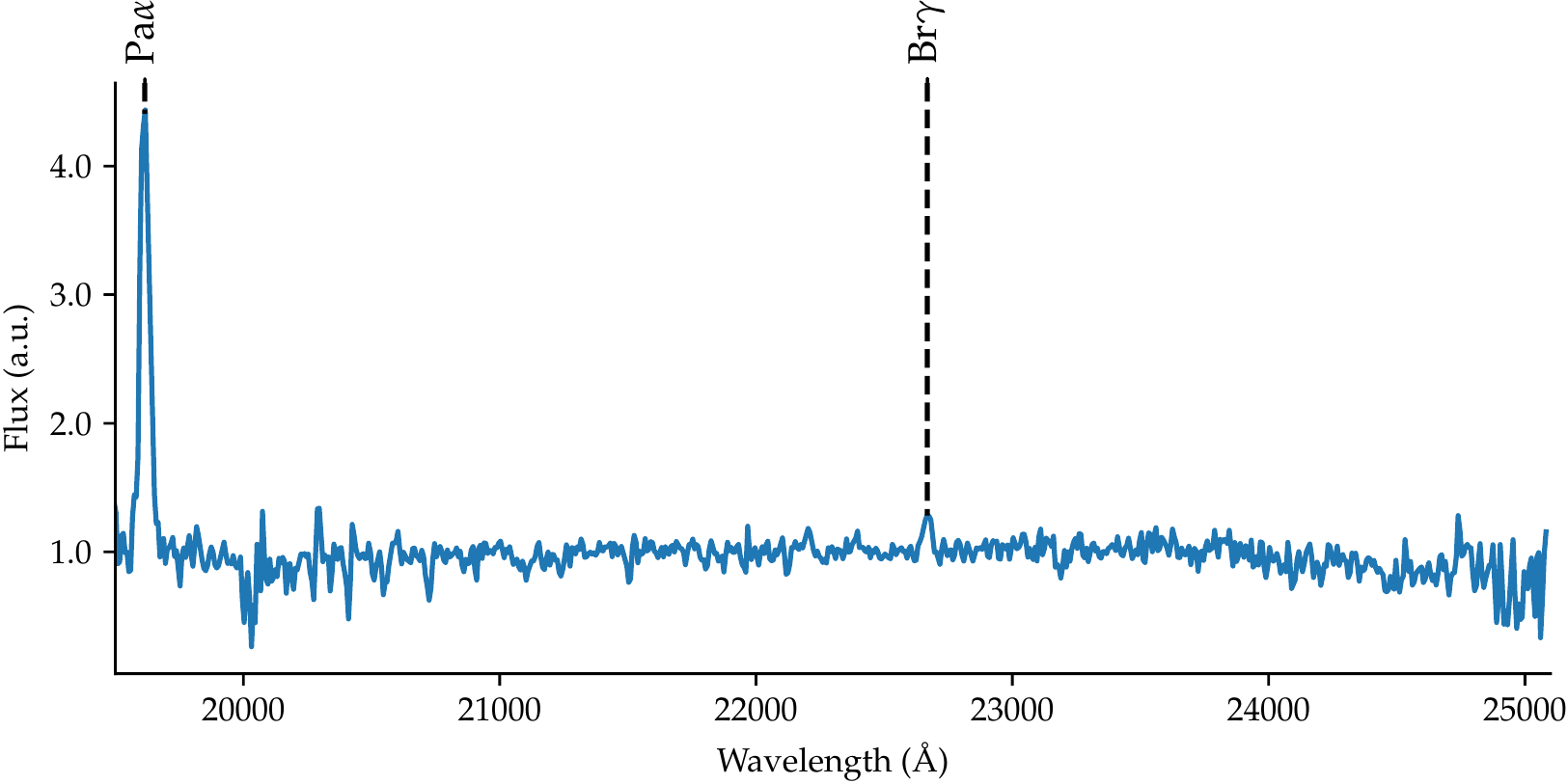}
\caption{IGR J10447$-$6027\label{fig:spec:10447}}
\end{minipage}
\end{subfigure}

\begin{subfigure}{\textwidth}
\begin{minipage}{.49\textwidth}
\includegraphics[width = \textwidth]{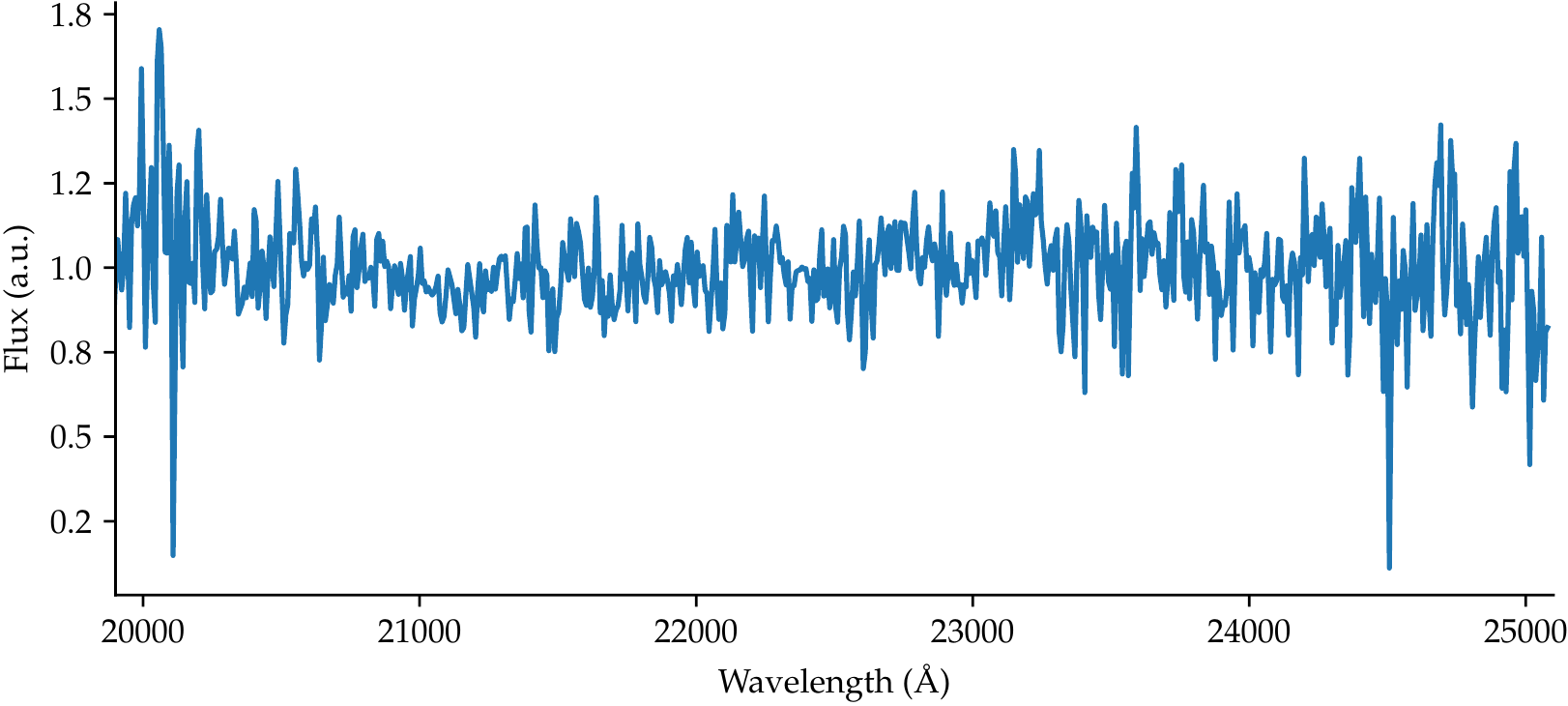}
\caption{IGR J18457$+$0244\label{fig:spec:18457}}
\end{minipage}
\hfill
\begin{minipage}{.49\textwidth}
\includegraphics[width = \textwidth]{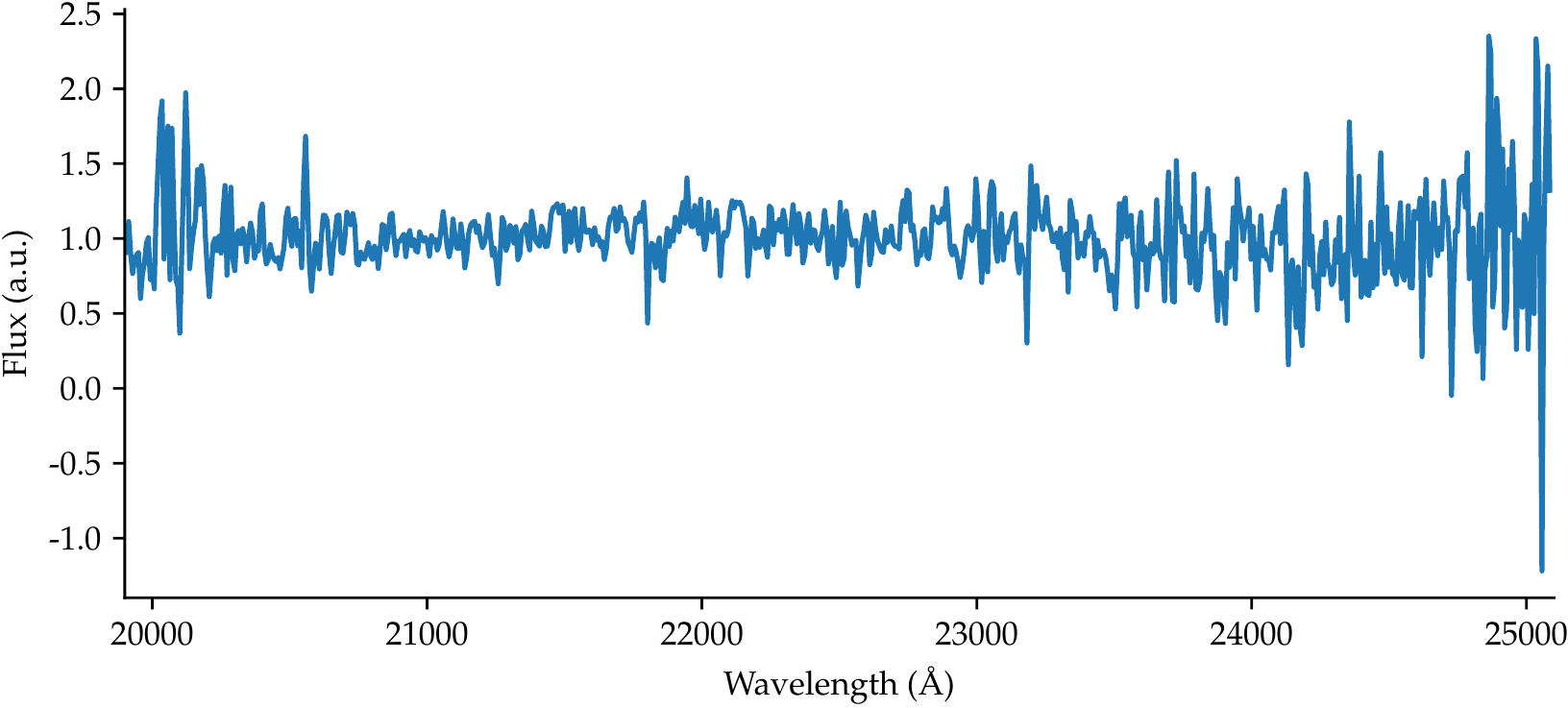}
\caption{IGR J18532$+$0416\label{fig:spec:18532}}
\end{minipage}
\end{subfigure}

\begin{subfigure}{.49\textwidth}
\includegraphics[width = \textwidth]{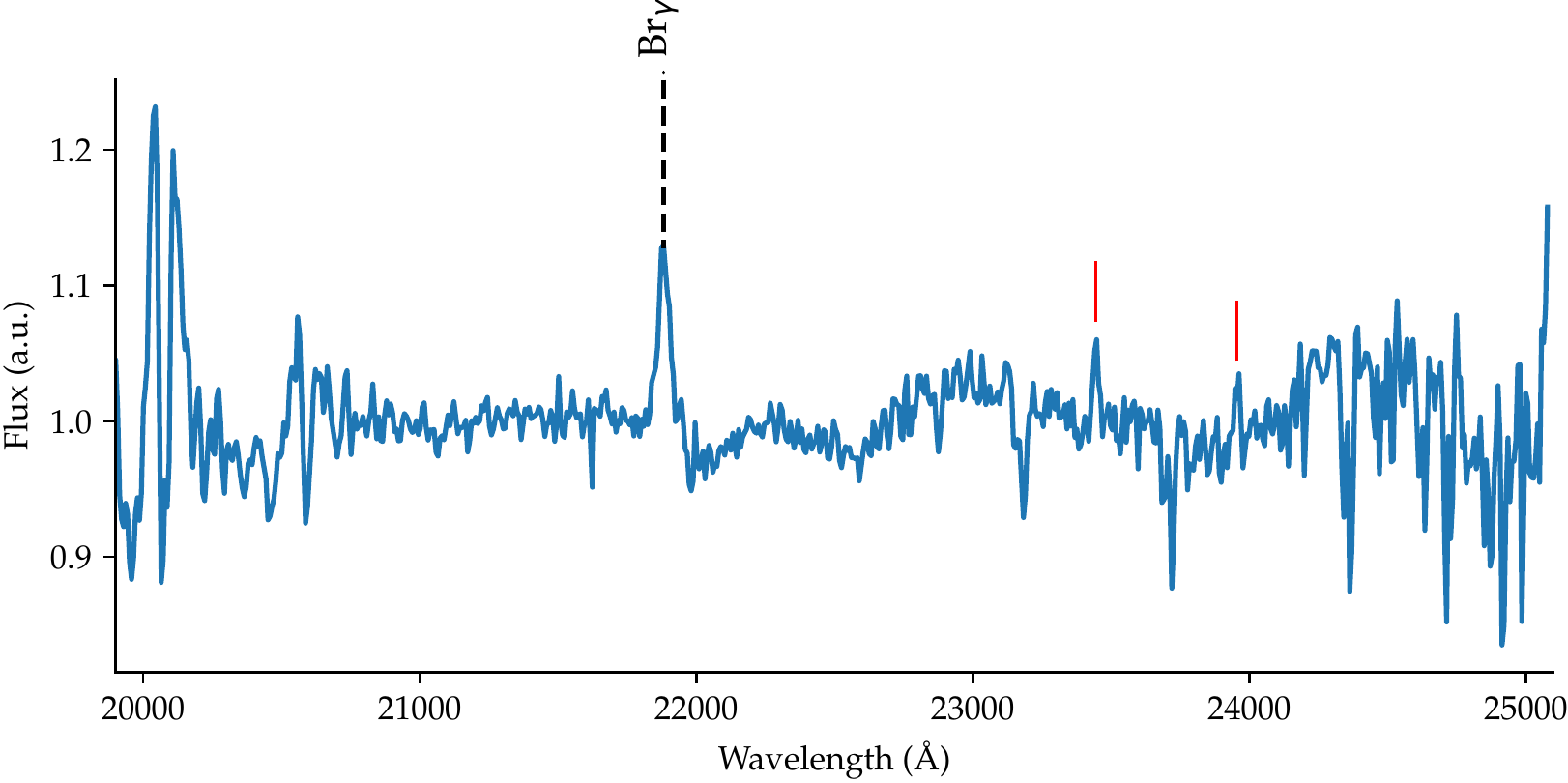}
\caption{IGR J19378$-$0617\label{fig:spec:19378}}
\end{subfigure}
\hfill

\end{figure}

\begin{figure}[h!]
\caption{K-band spectra of the \textit{INTEGRAL} sources identified as HMXBs.\label{fig:HMXB}}

\begin{subfigure}{\textwidth}
\begin{minipage}{.49\textwidth}
\includegraphics[width=\textwidth]{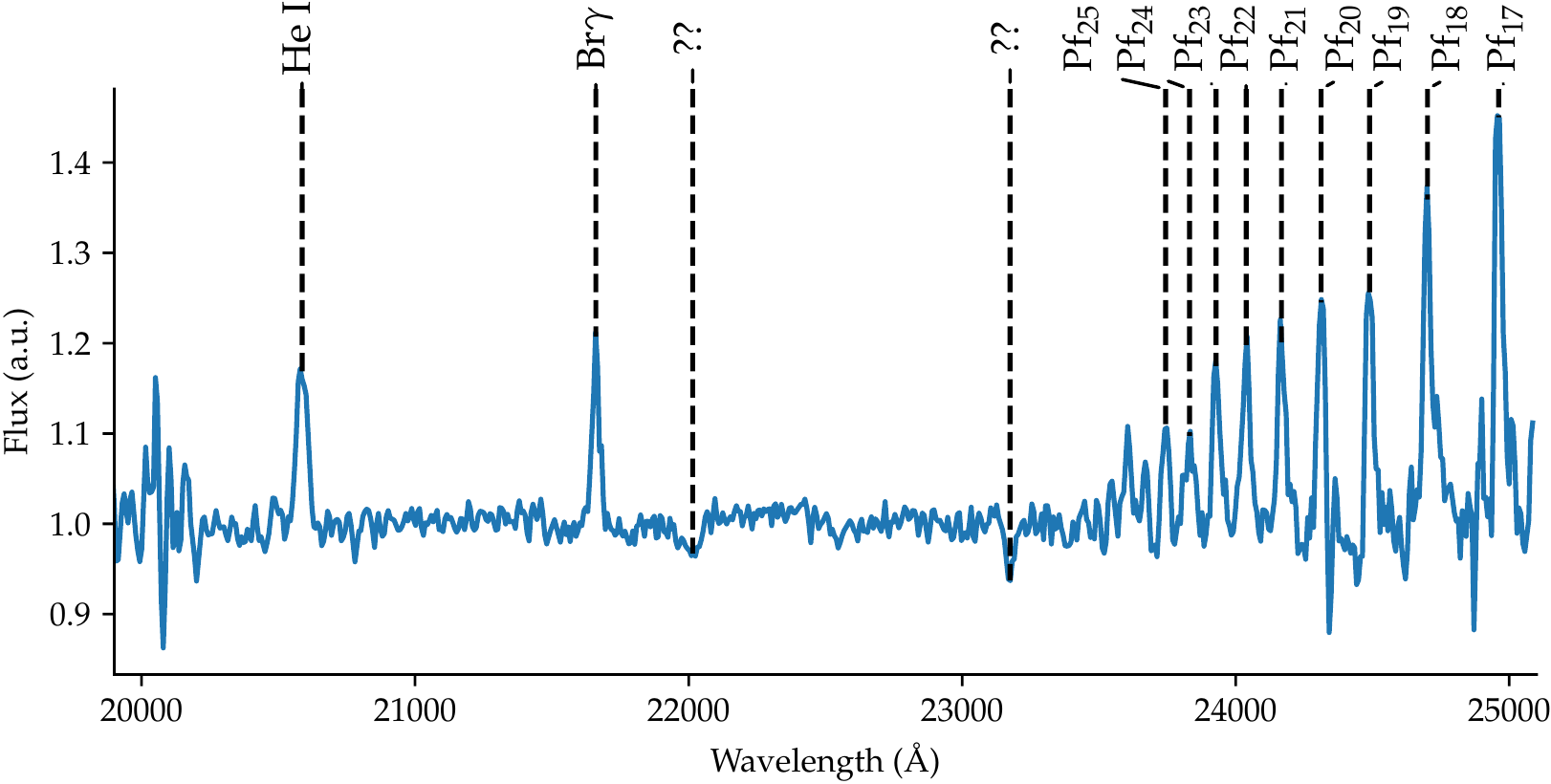}
\caption{IGR J13020$-$6359\label{fig:spec:13020}}
\end{minipage}
\hfill
\begin{minipage}{.49\textwidth}
\includegraphics[width=\textwidth]{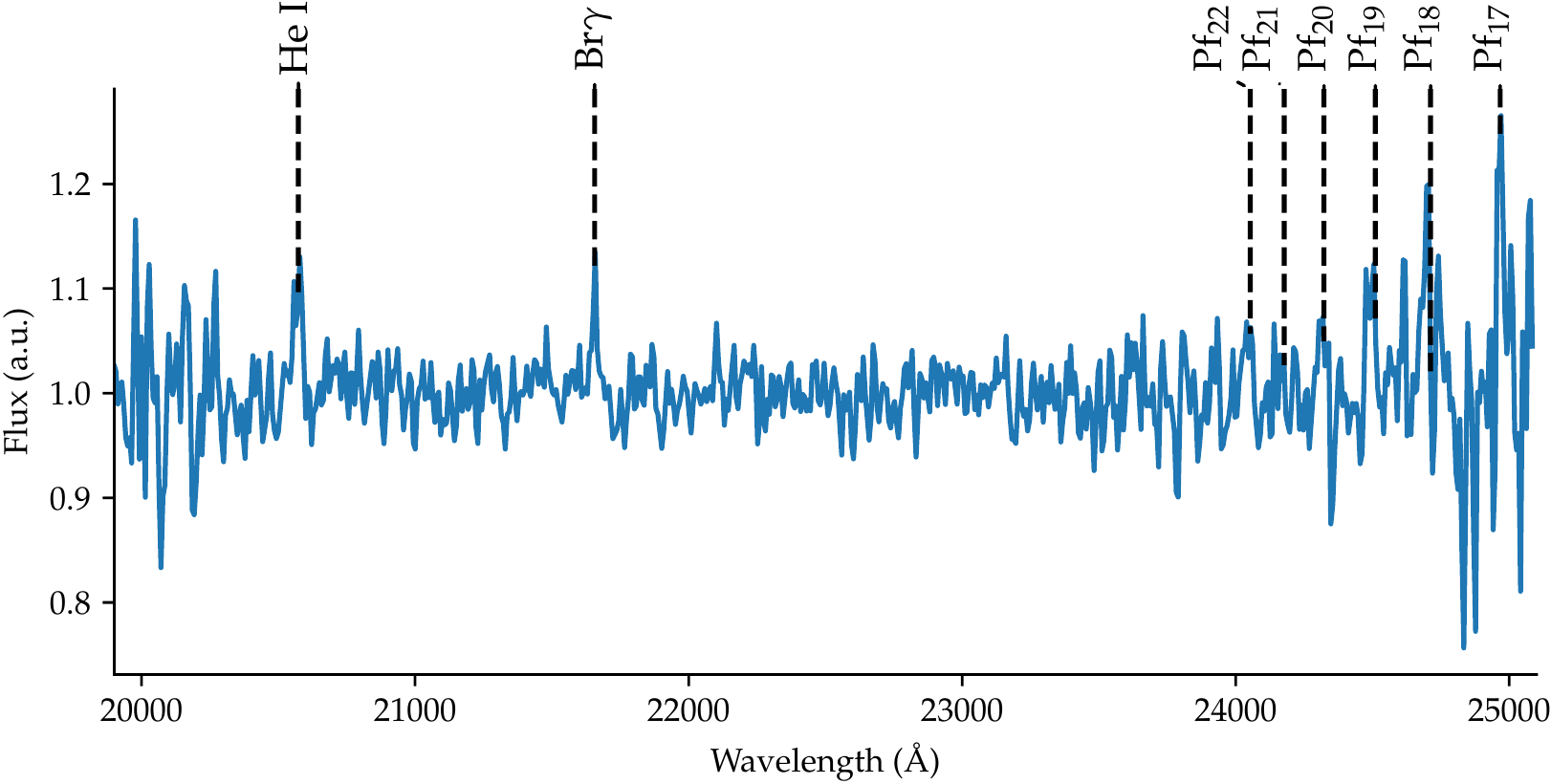}
\caption{IGR J13186$-$6257\label{fig:spec:13186}}
\end{minipage}
\end{subfigure}

\end{figure}

\begin{figure*}[h!]
\caption{K-band spectra of the \textit{INTEGRAL} sources identified as CVs.\label{fig:CV}}

\begin{subfigure}{\textwidth}
\begin{minipage}{.49\textwidth}
\includegraphics[width=\textwidth]{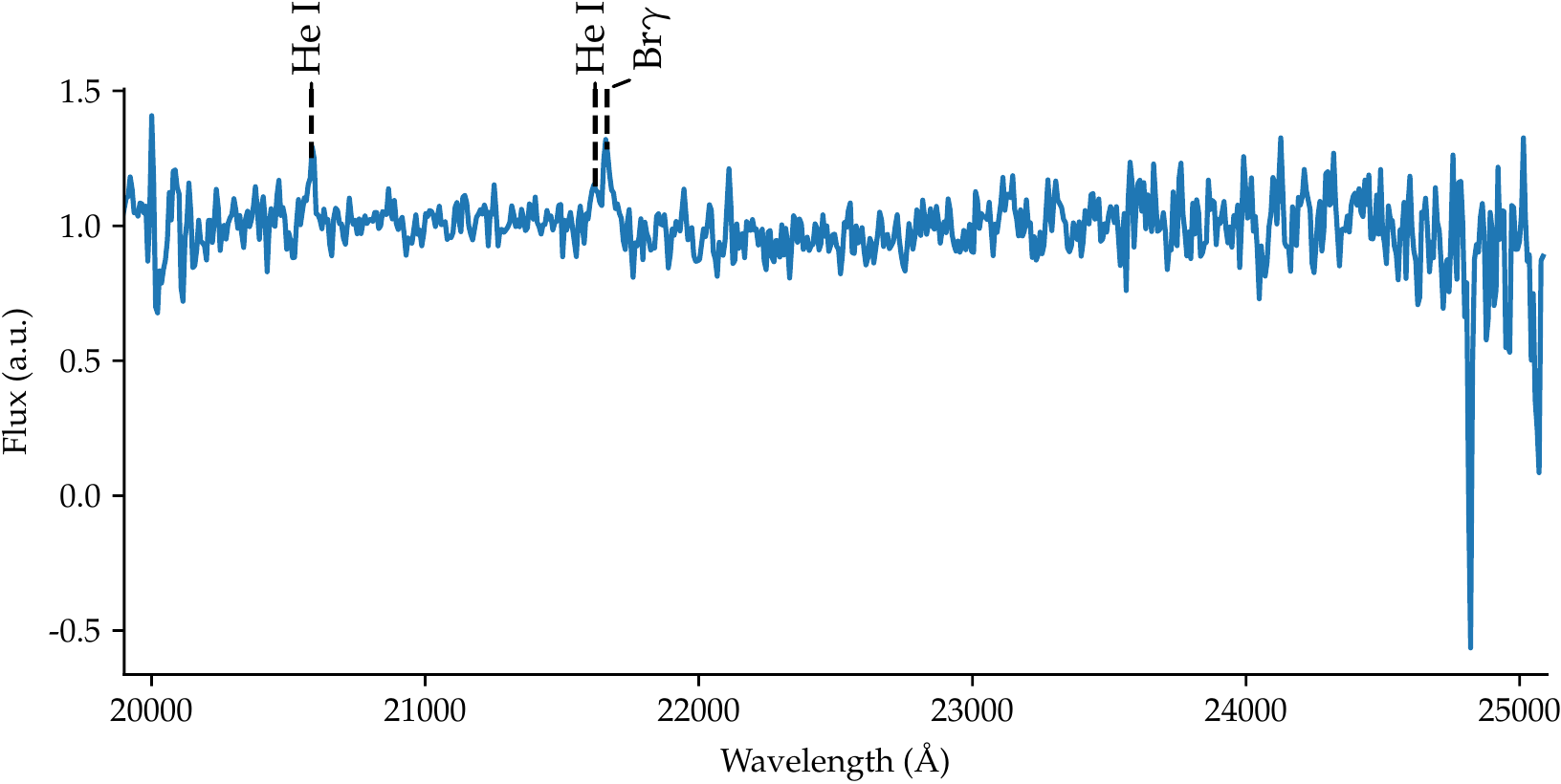}
\caption{IGR 12489$-$6243\label{fig:spec:12489}}
\end{minipage}
\hfill
\begin{minipage}{.49\textwidth}
\includegraphics[width=\textwidth]{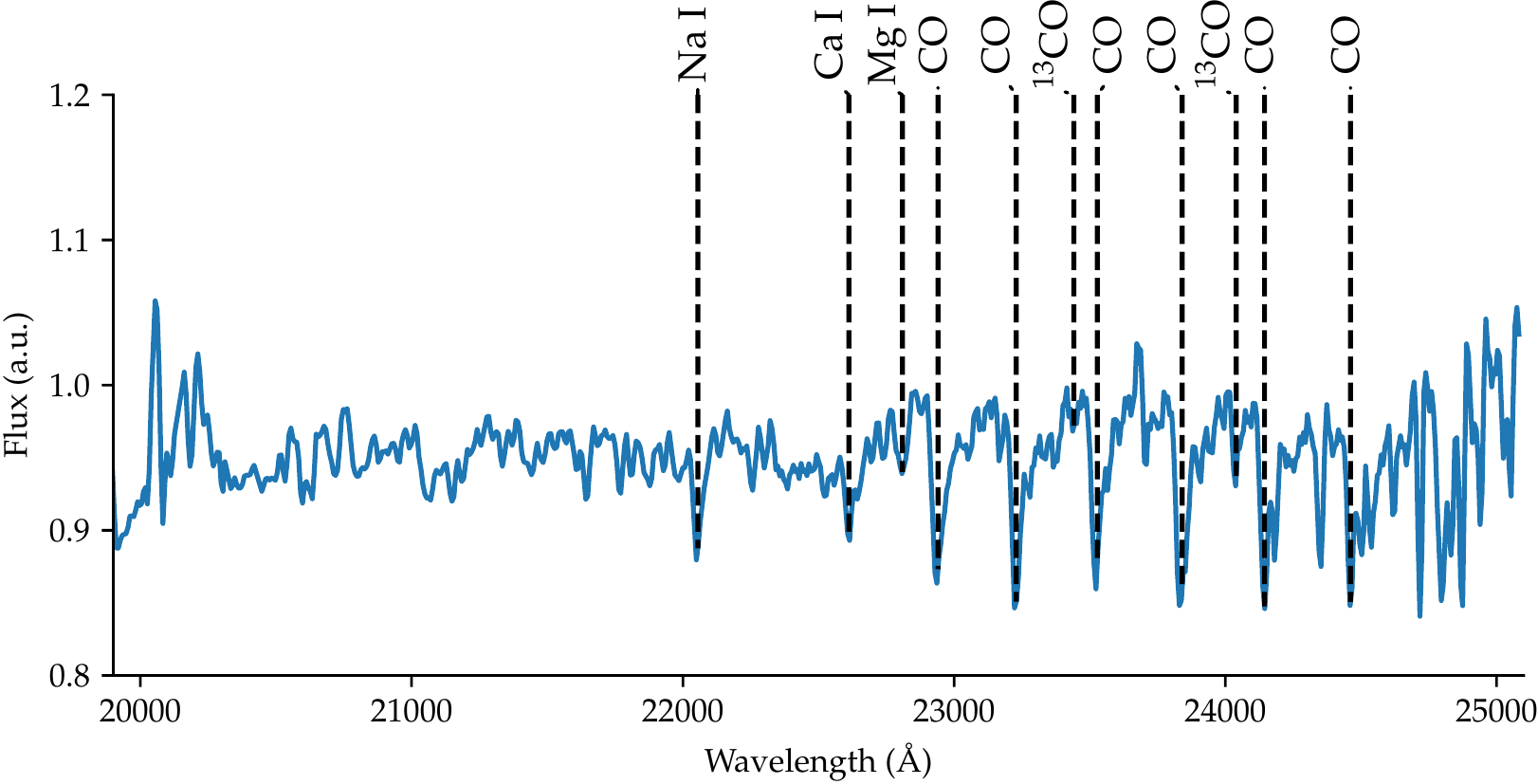}
\caption{IGR J15293$-$5609\label{fig:spec:15293}}
\end{minipage}
\end{subfigure}

\begin{subfigure}{\textwidth}
\begin{minipage}{.49\textwidth}
\includegraphics[width=\textwidth]{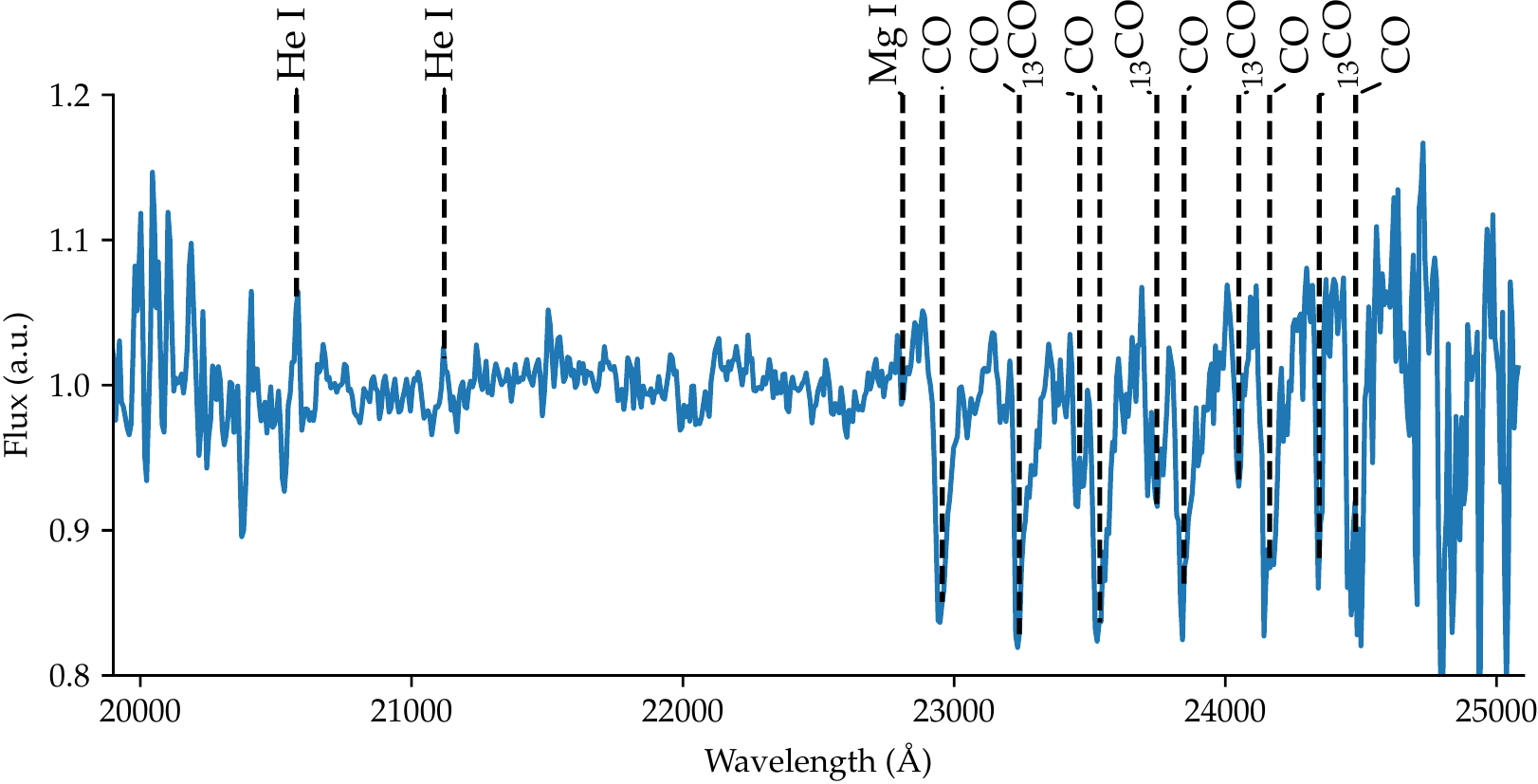}
\caption{IGR J17200$-$3116\label{fig:spec:17200}}
\end{minipage}
\hfill
\begin{minipage}{.49\textwidth}
\includegraphics[width=\textwidth]{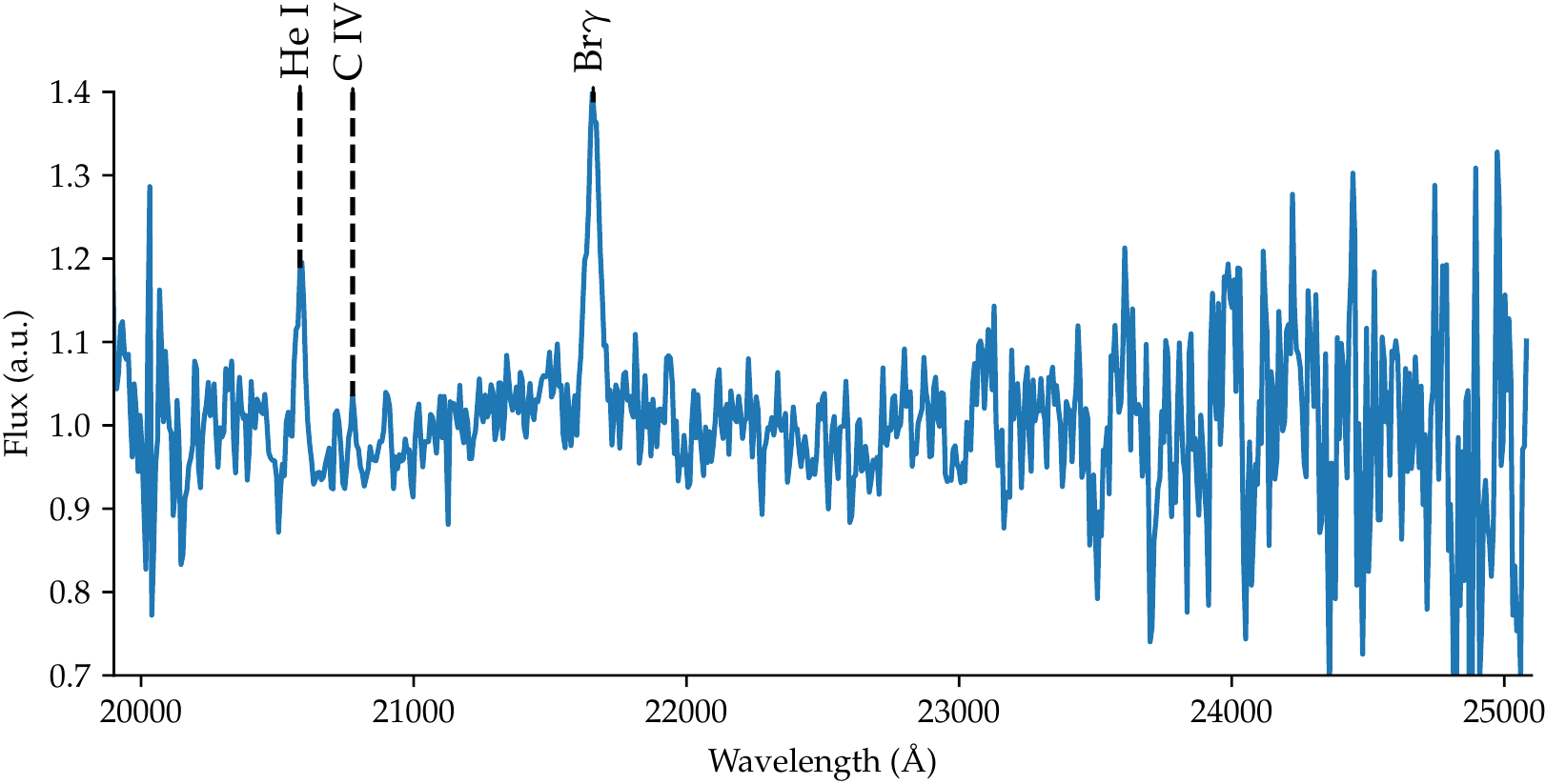}
\caption{IGR J17404$-$3655\label{fig:spec:17404}}
\end{minipage}
\end{subfigure}

\begin{subfigure}{.49\textwidth}
\includegraphics[width=\textwidth]{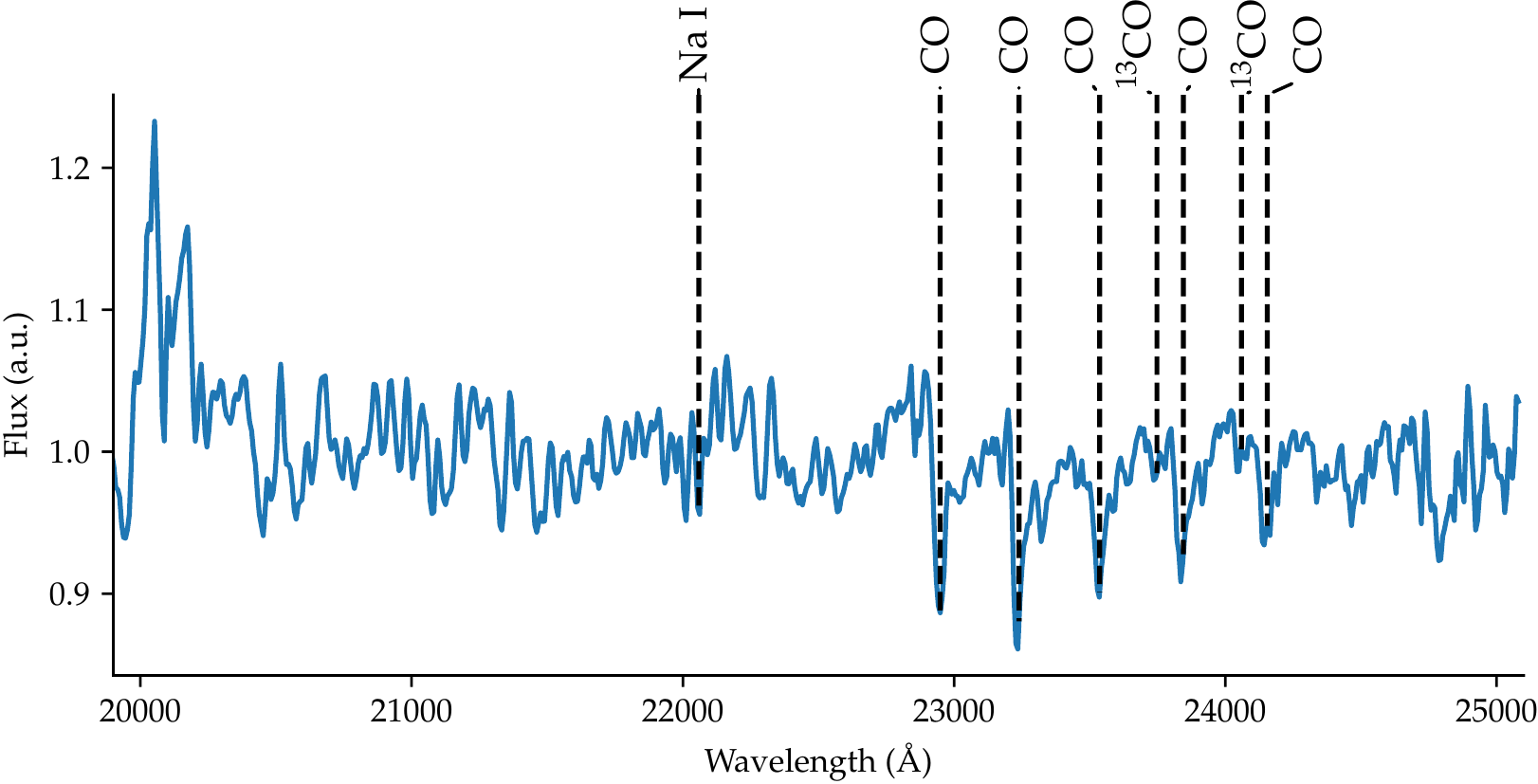}
\caption{IGR J17586$-$2129\label{fig:spec:17586}}
\end{subfigure}
\hfill

\end{figure*}

\begin{figure*}[h!]
\caption{K-band spectra of the \textit{INTEGRAL} sources identified as LMXB/IMXBs.  Spurious features from residual artefacts are labelled in red.\label{fig:LMXB}}

\begin{subfigure}{\textwidth}
\begin{minipage}{.49\textwidth}
\includegraphics[width=\textwidth]{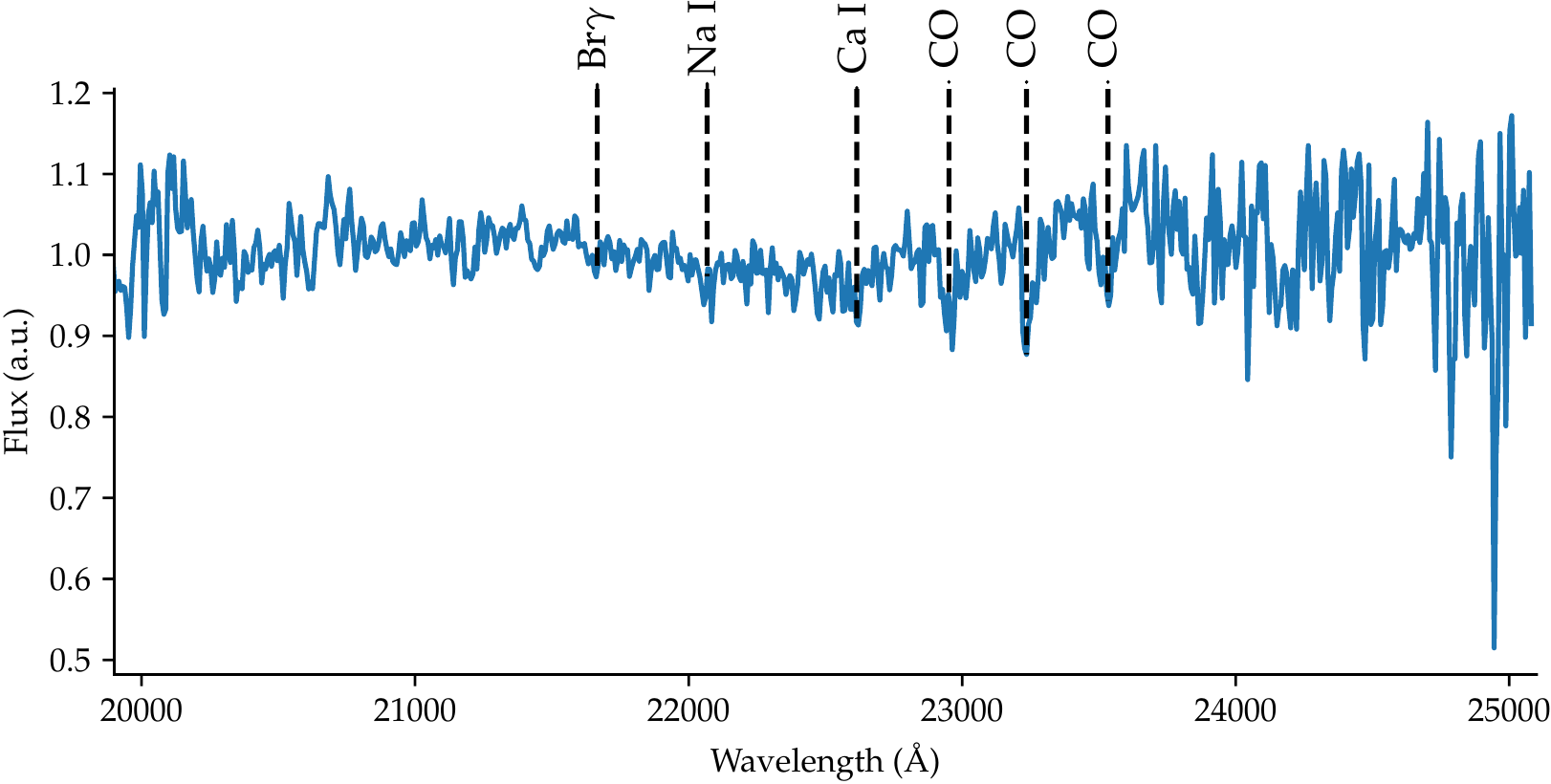}
\caption{IGR J17597$-$2201\label{fig:spec:17597}}
\end{minipage}
\hfill
\begin{minipage}{.49\textwidth}
\includegraphics[width=\textwidth]{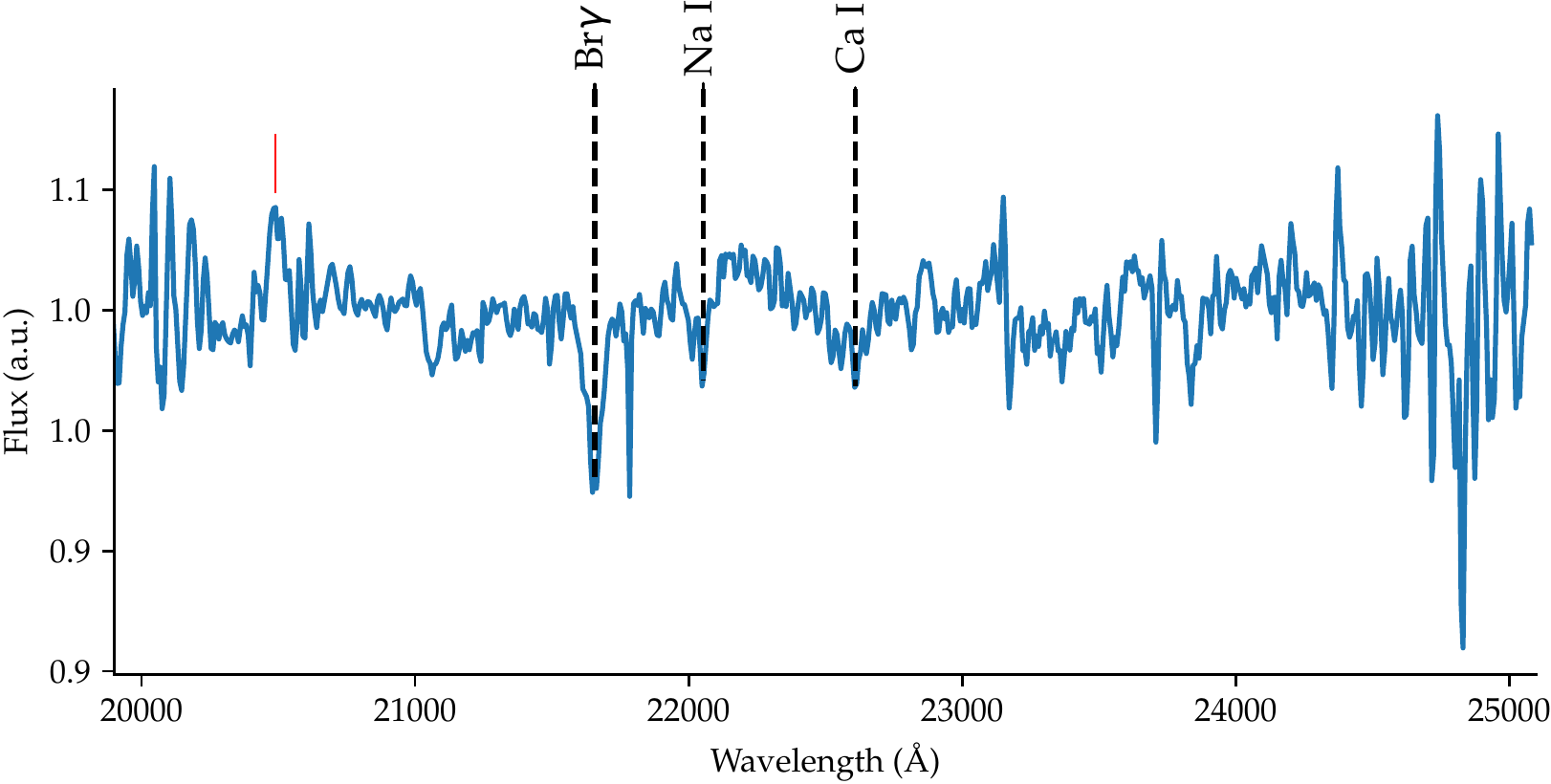}
\caption{IGR J19308$+$0530\label{fig:spec:19308}}
\end{minipage}
\end{subfigure}

\end{figure*}

\appendix
\onecolumn
\section{Finding charts}

\begin{figure}[ht!]
\caption{K$_s$ finding charts of IGR sources in this study. Each frame is 33\arcsec$\times$33\arcsec; north is up and east is left. Red circles indicate the high-energy position at 90\% confidence.\label{fig:findingchart1}}

\begin{subfigure}{.24\textwidth}
\includegraphics[width=\textwidth]{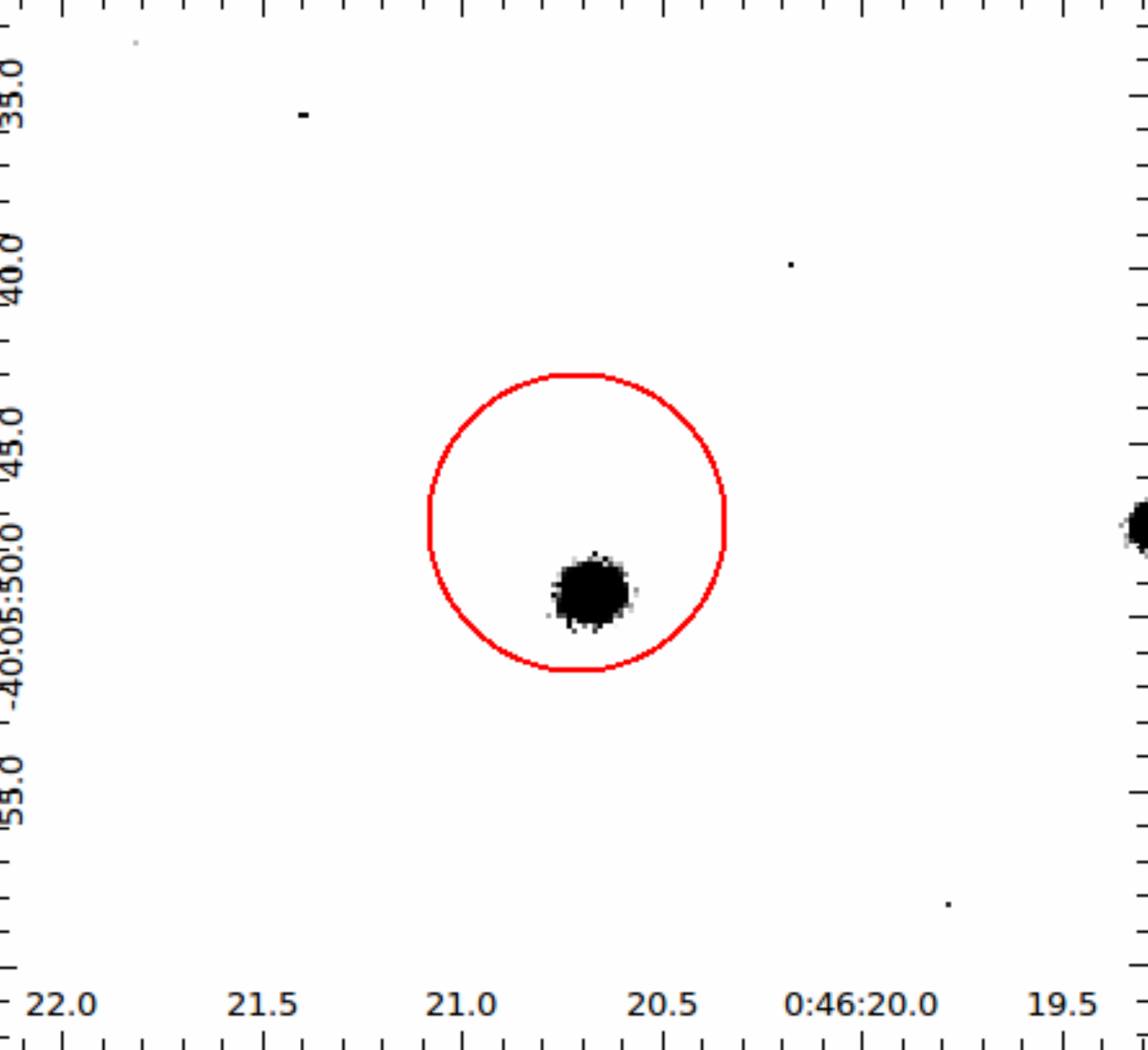}
\caption{IGR J00465$-$4005\label{fig:fc:00465}}
\end{subfigure}
\hfill
\begin{subfigure}{.24\textwidth}
\includegraphics[width=\textwidth]{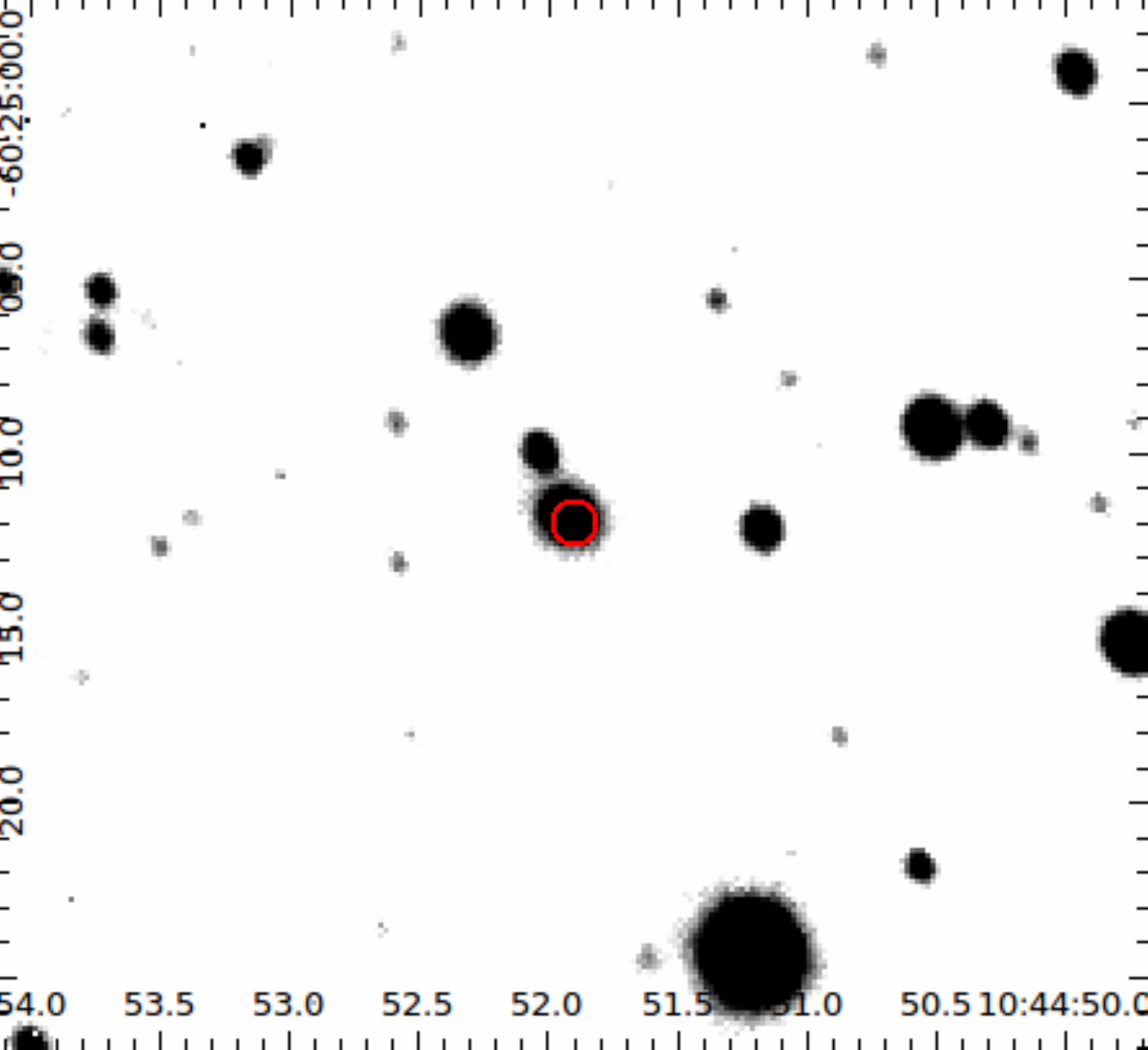}
\caption{IGR J10447$-$6027\label{fig:fc:10447}}
\end{subfigure}
\hfill
\begin{subfigure}{.24\textwidth}
\includegraphics[width=\textwidth]{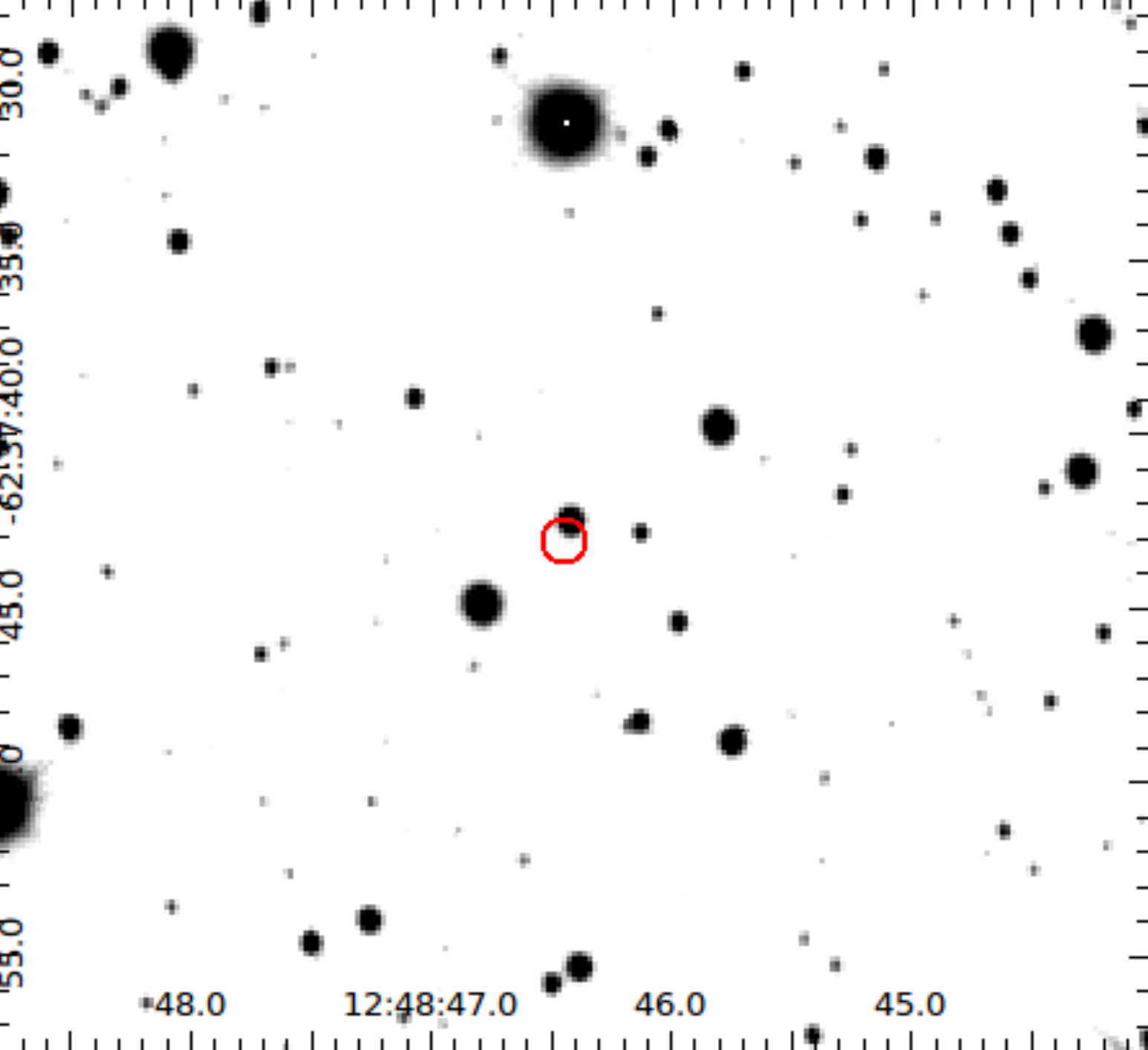}
\caption{IGR J12489$-$6243\label{fig:fc:12489}}
\end{subfigure}
\hfill
\begin{subfigure}{.24\textwidth}
\includegraphics[width=\textwidth]{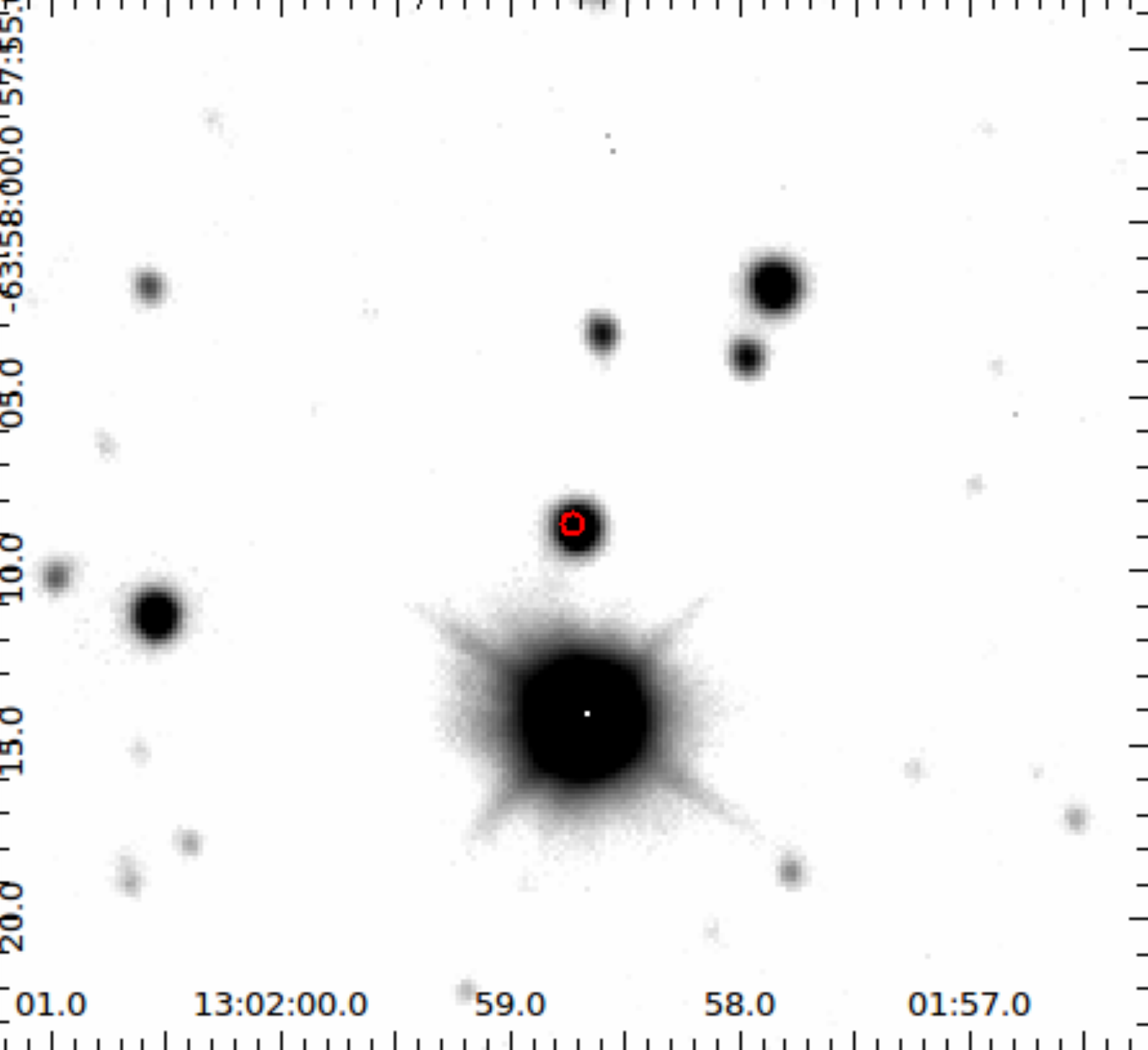}
\caption{IGR J13020$-$6359\label{fig:fc:13020}}
\end{subfigure}

\begin{subfigure}{.24\textwidth}
\includegraphics[width=\textwidth]{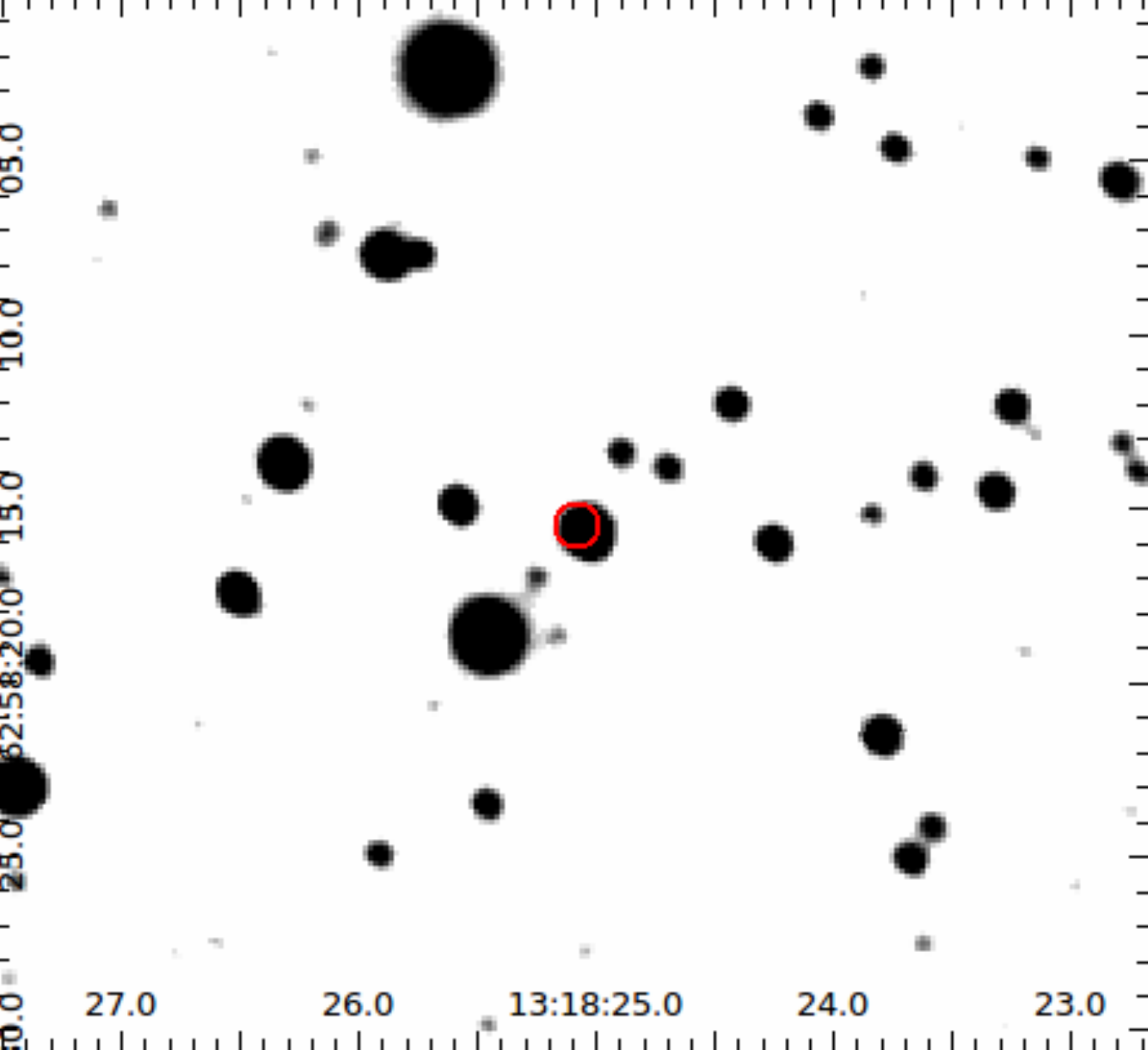}
\caption{IGR J13186$-$6257\label{fig:fc:13186}}
\end{subfigure}
\hfill
\begin{subfigure}{.24\textwidth}
\includegraphics[width=\textwidth]{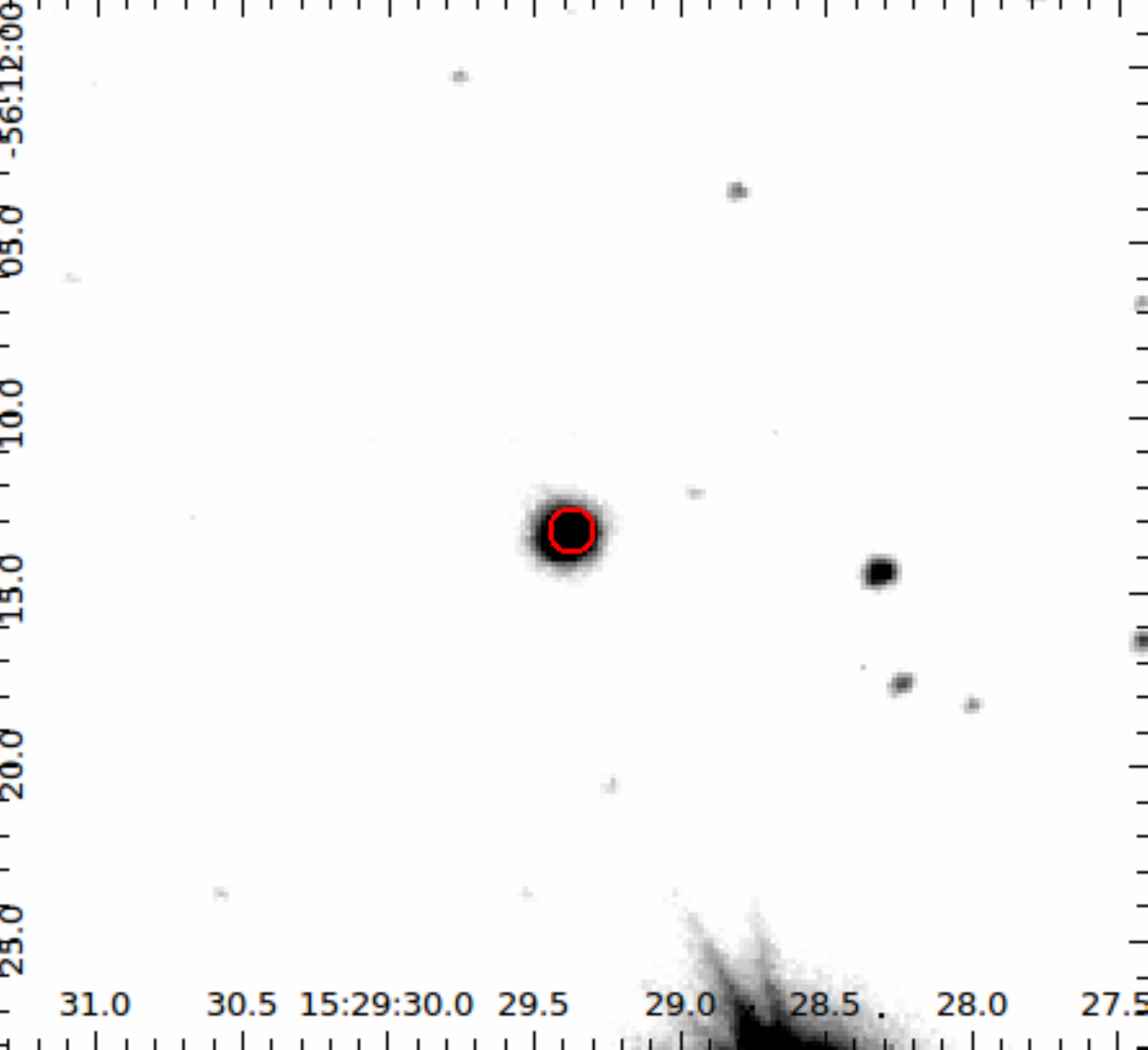}
\caption{IGR J15293$-$5609\label{fig:fc:15293}}
\end{subfigure}
\hfill
\begin{subfigure}{.24\textwidth}
\includegraphics[width=\textwidth]{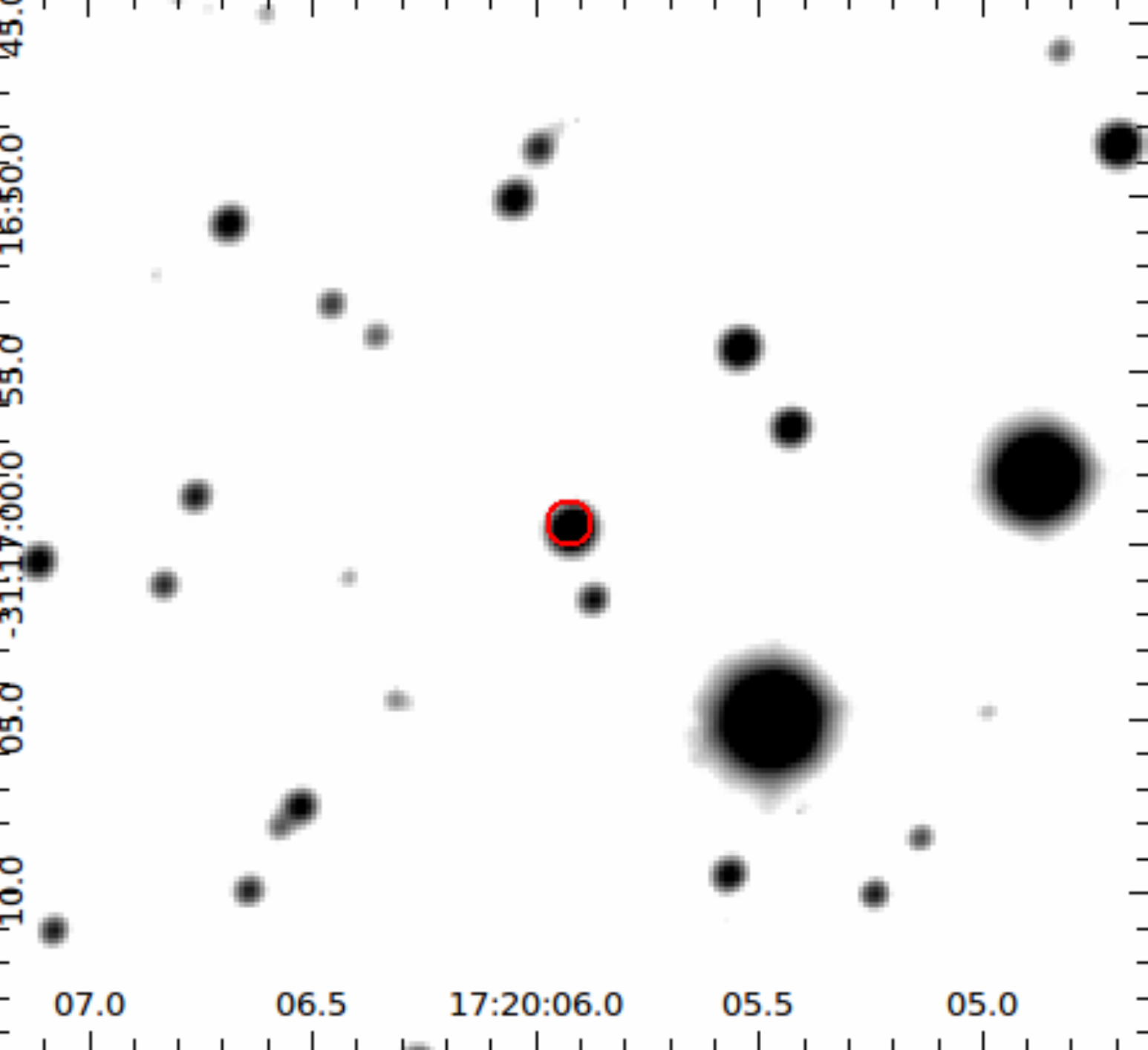}
\caption{IGR J17200$-$3116\label{fig:fc:17200}}
\end{subfigure}
\hfill
\begin{subfigure}{.24\textwidth}
\includegraphics[width=\textwidth]{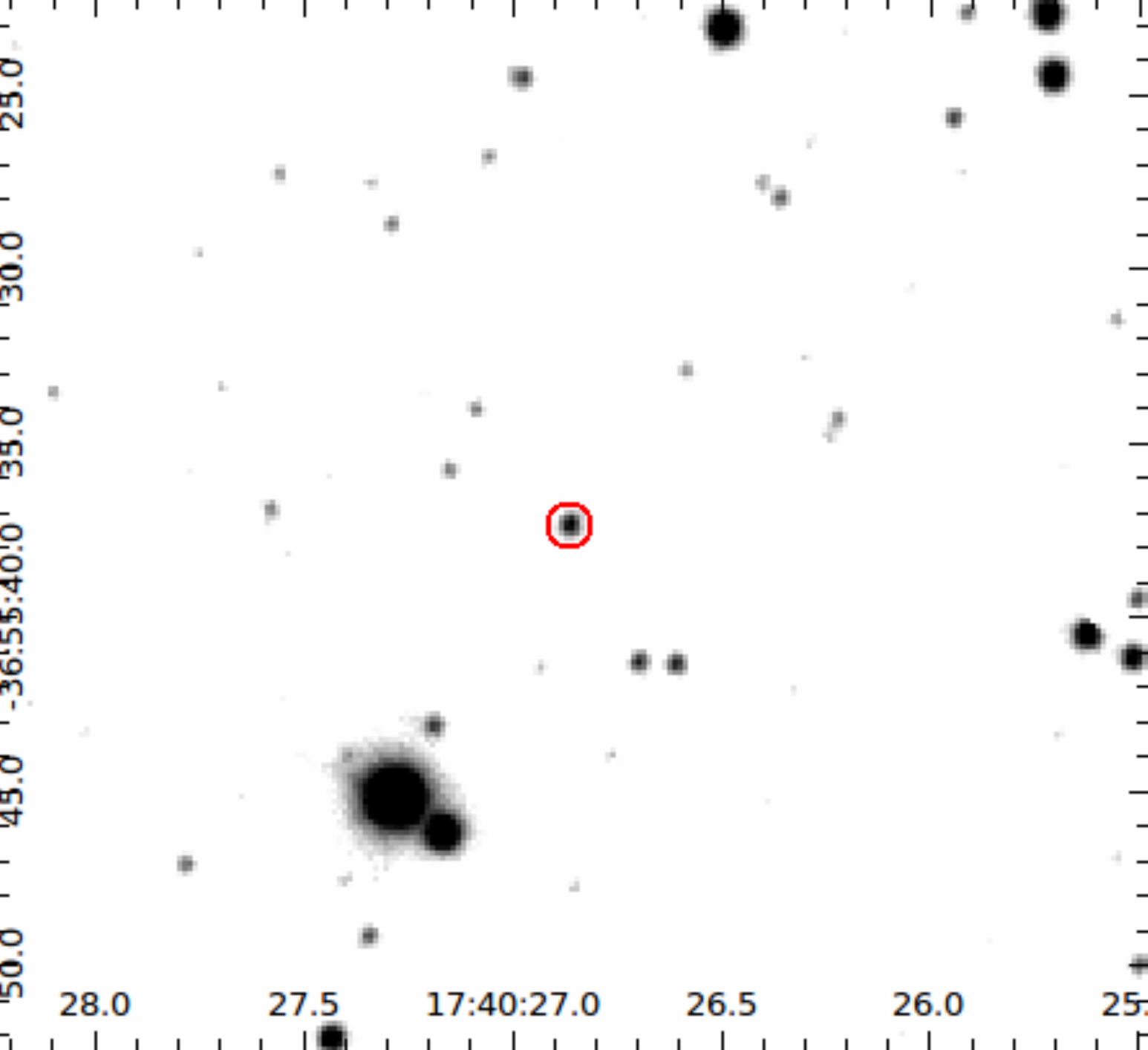}
\caption{IGR J17404$-$3655\label{fig:fc:17404}}
\end{subfigure}

\begin{subfigure}{.24\textwidth}
\includegraphics[width=\textwidth]{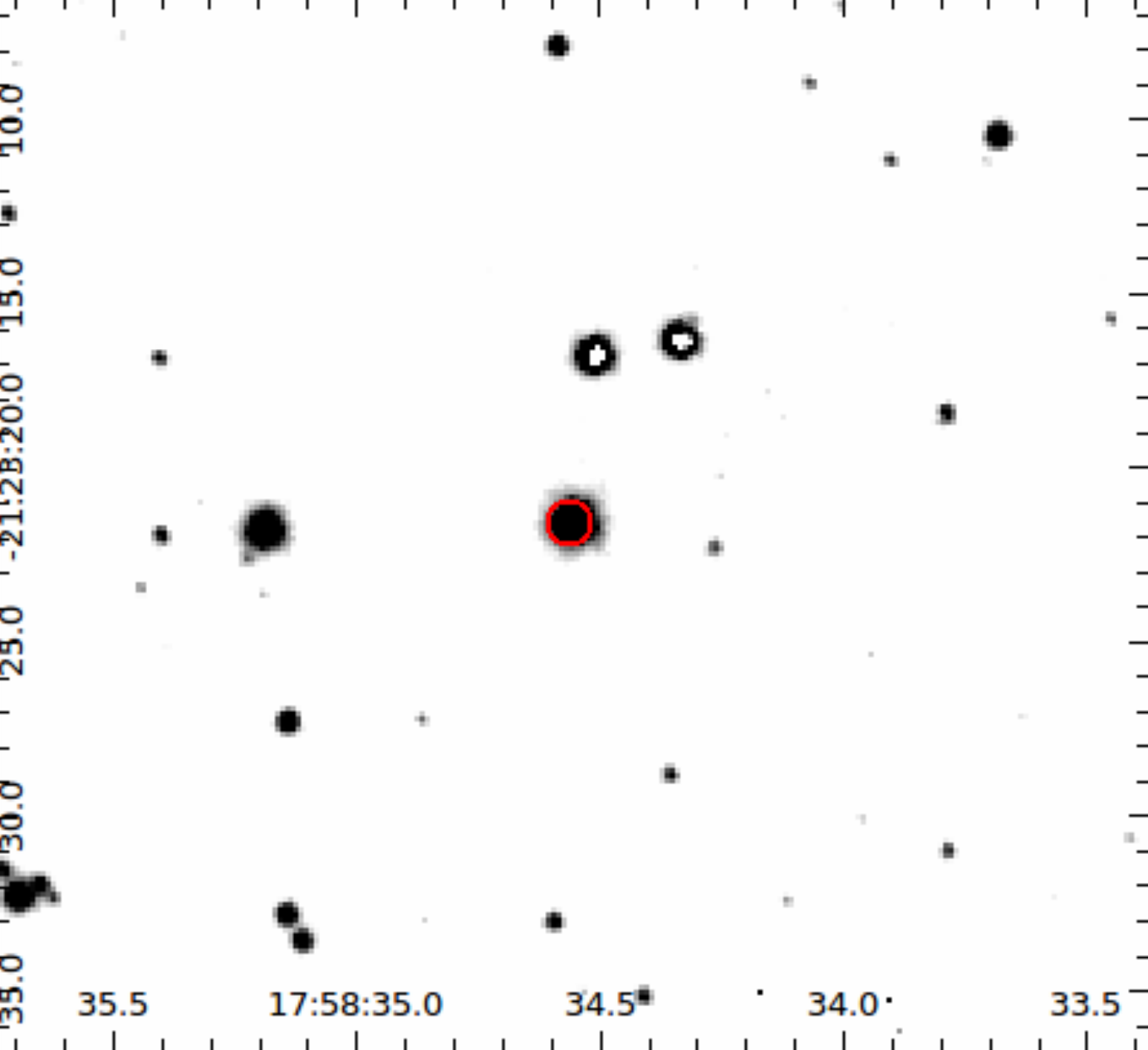}
\caption{IGR J17586$-$2129\label{fig:fc:17586}}
\end{subfigure}
\hfill
\begin{subfigure}{.24\textwidth}
\includegraphics[width=\textwidth]{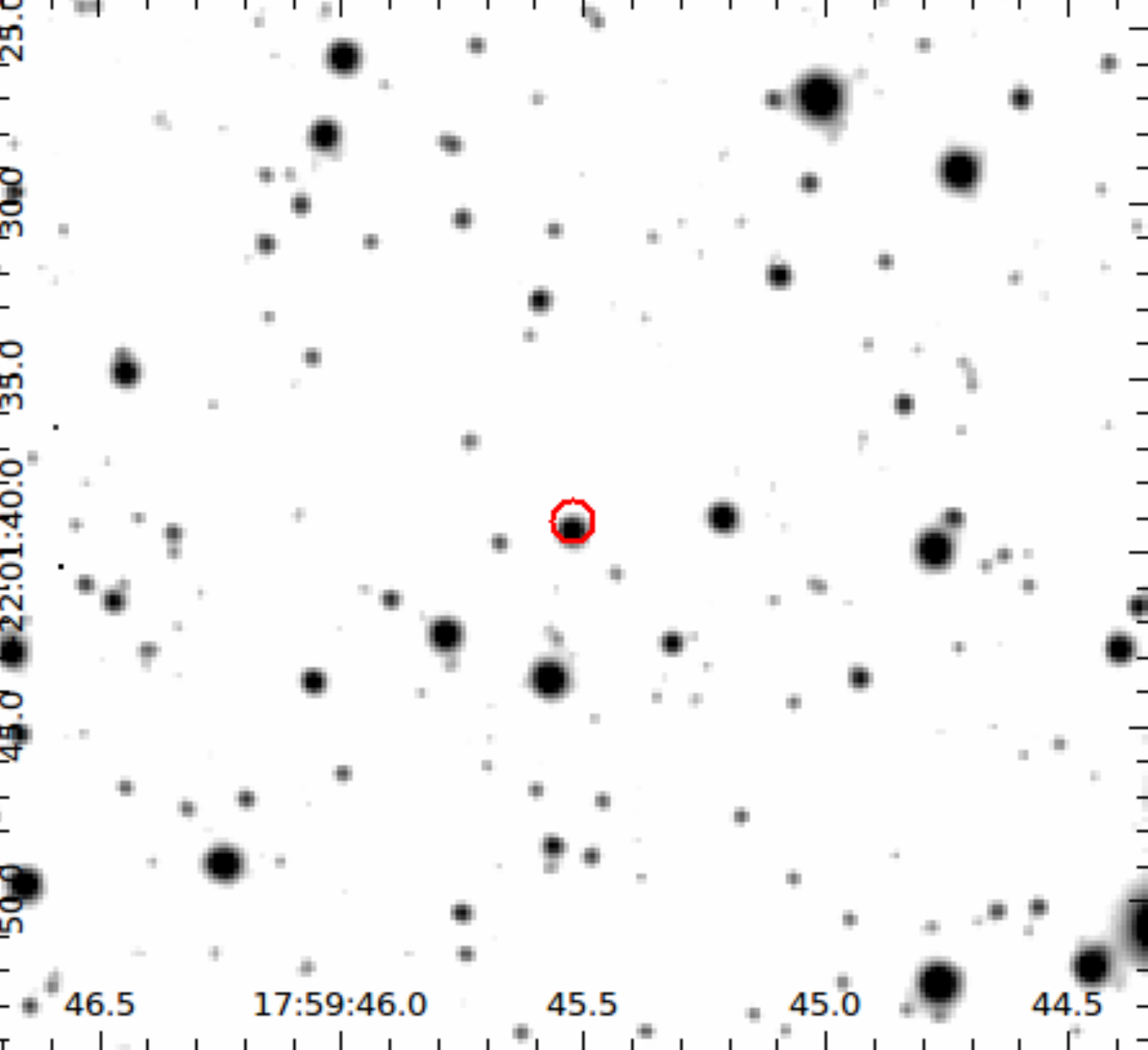}
\caption{IGR J17597$-$2201\label{fig:fc:17597}}
\end{subfigure}
\hfill
\begin{subfigure}{.24\textwidth}
\includegraphics[width=\textwidth]{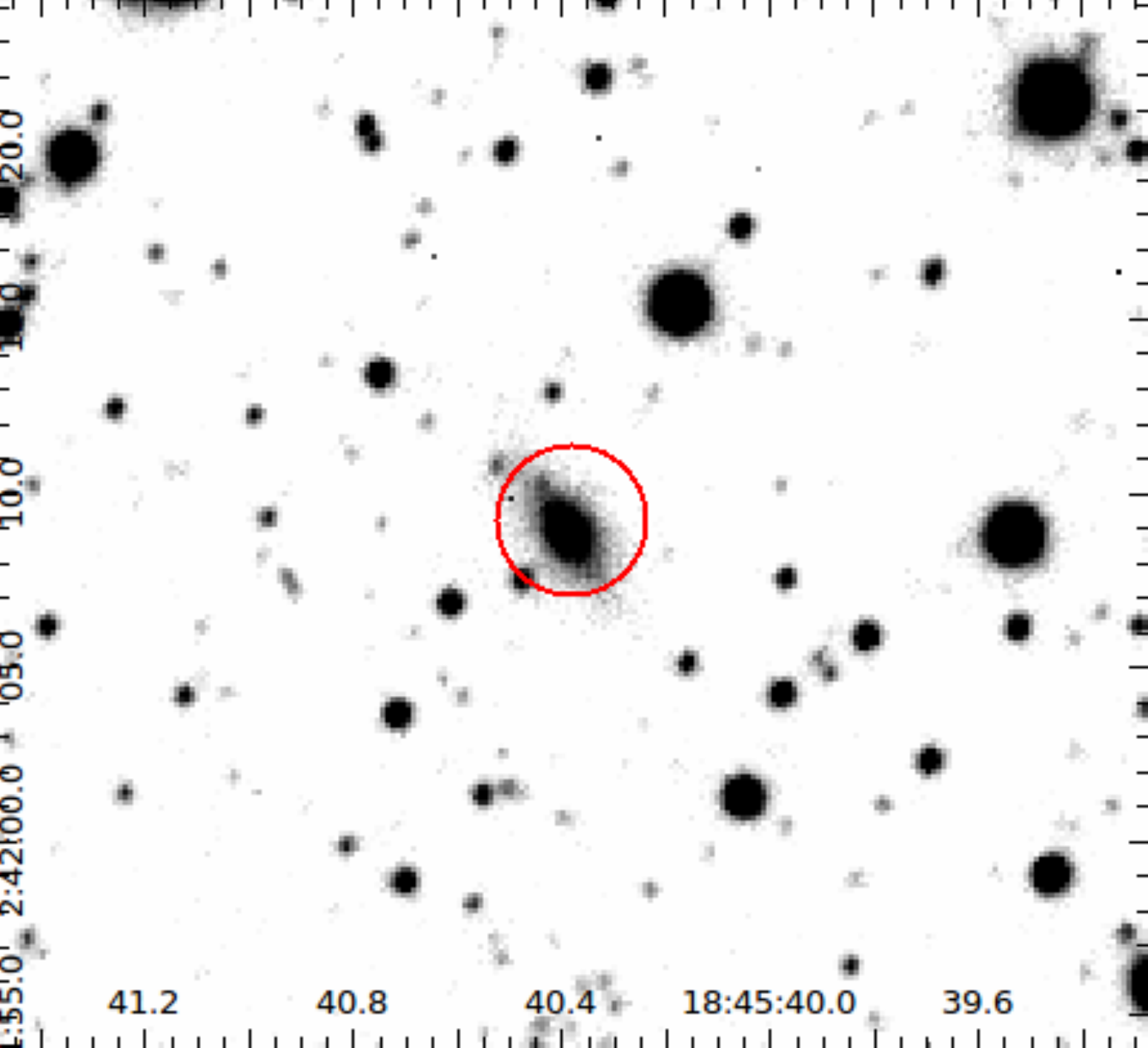}
\caption{IGR J18457$+$0244\label{fig:fc:18457}}
\end{subfigure}
\hfill
\begin{subfigure}{.24\textwidth}
\includegraphics[width=\textwidth]{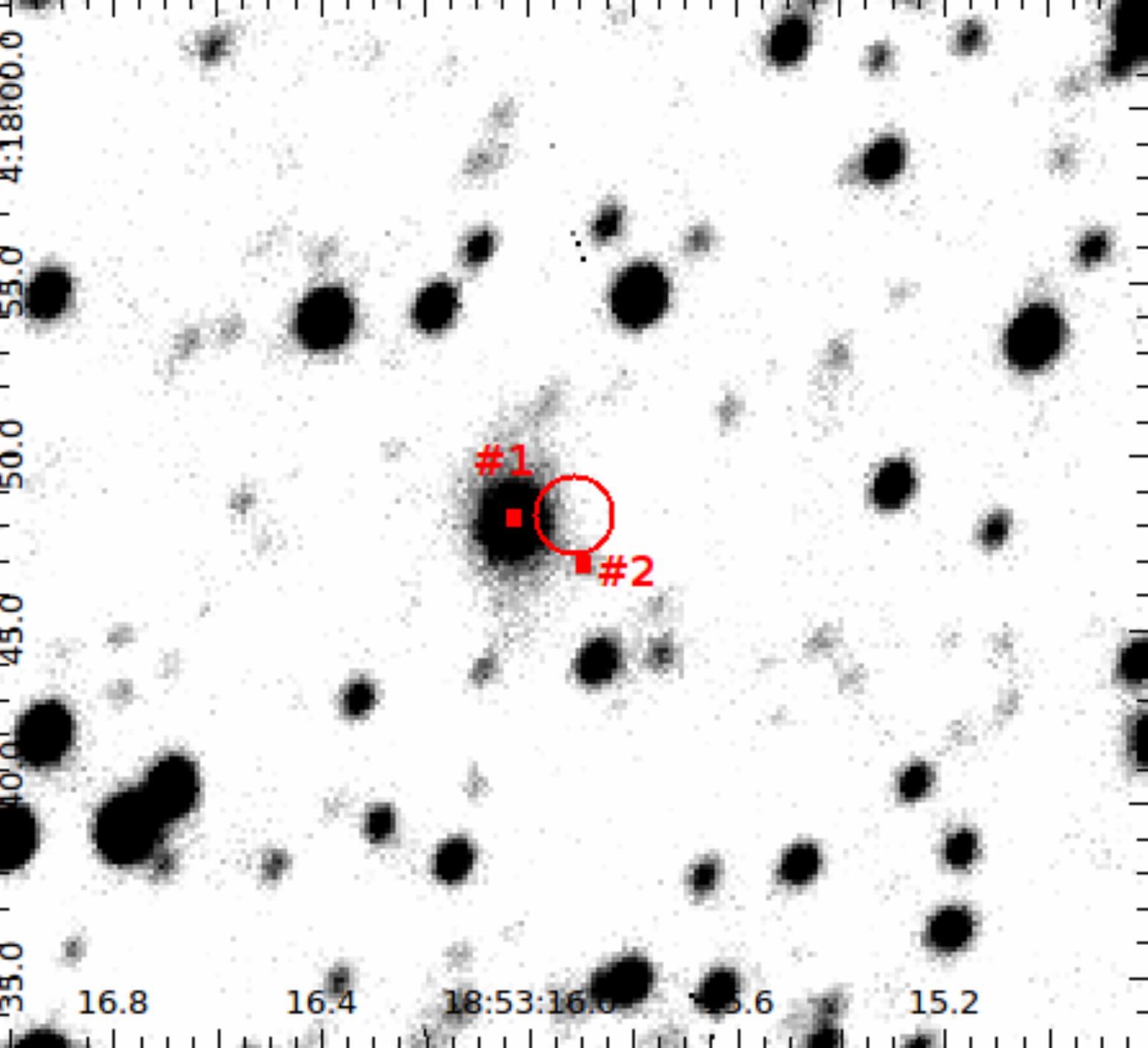}
\caption{IGR J18532$+$0416\label{fig:fc:18532}}
\end{subfigure}

\begin{subfigure}{.24\textwidth}
\includegraphics[width=\textwidth]{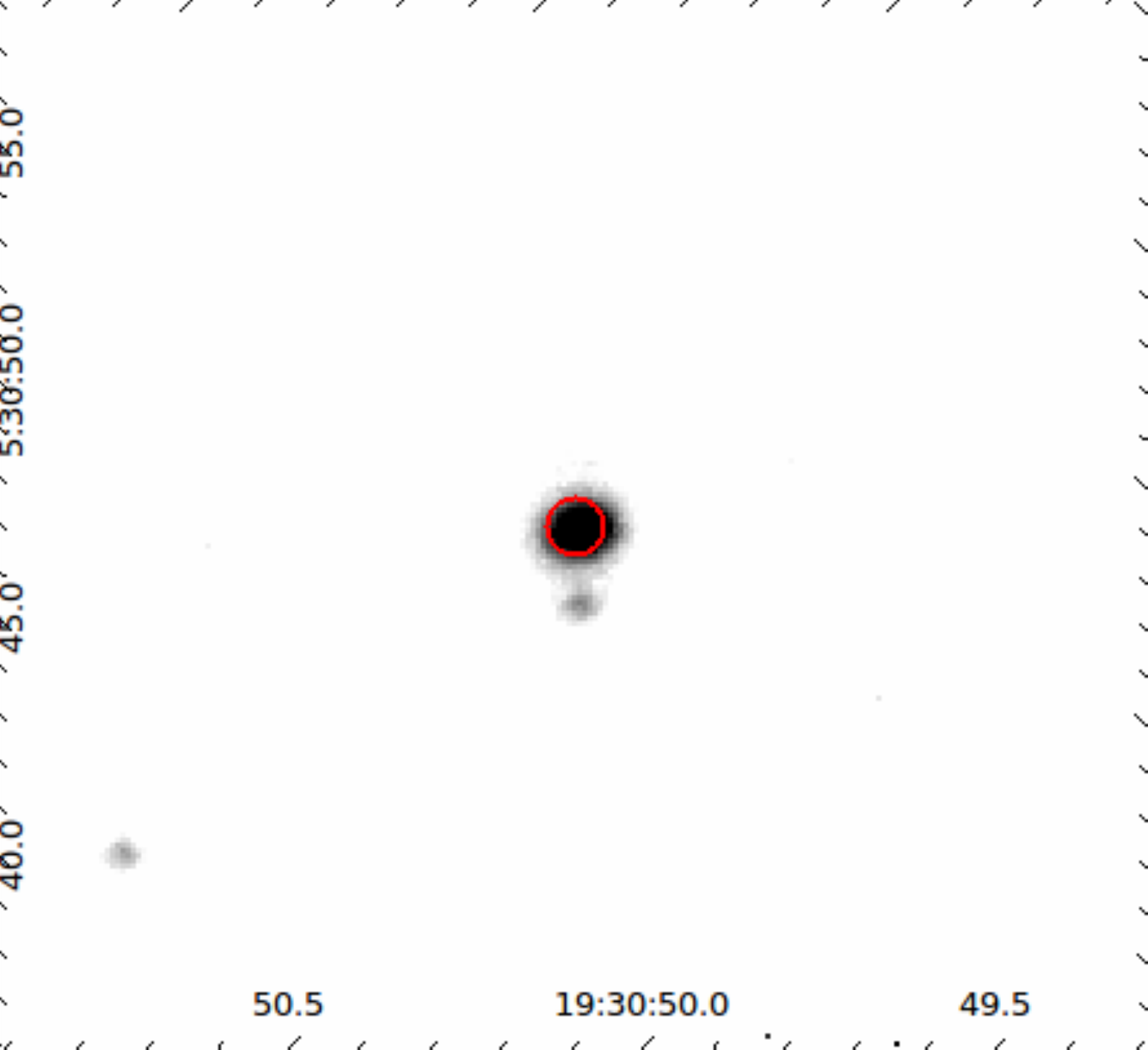}
\caption{IGR J19308$+$0530\label{fig:fc:19308}}
\end{subfigure}
\hfill
\begin{subfigure}{.24\textwidth}
\includegraphics[width=\textwidth]{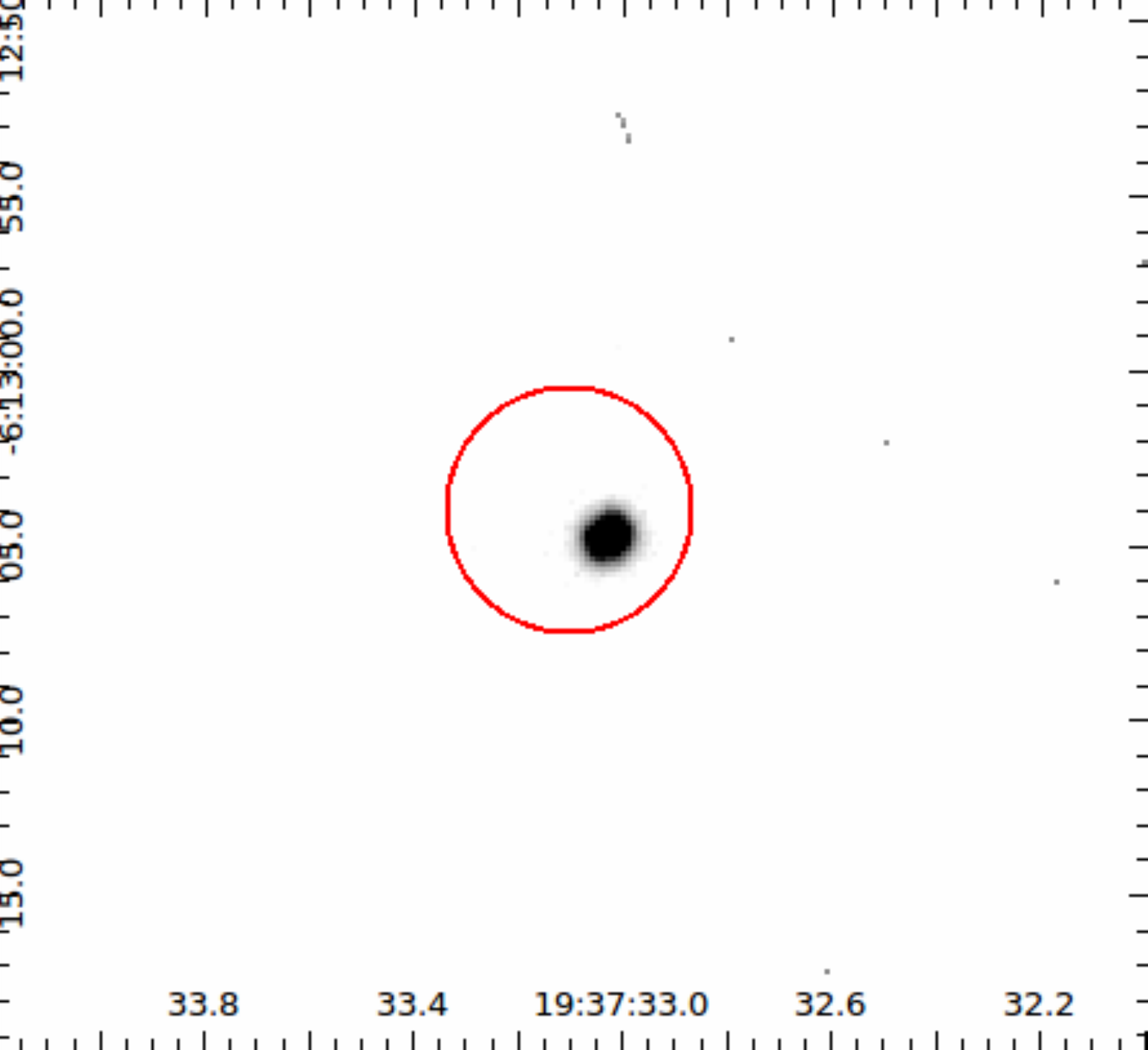}
\caption{IGR J19378$-$0617\label{fig:fc:19378}}
\end{subfigure}
\begin{minipage}{.24\textwidth}
~~~~~~
\end{minipage}
\hfill
\begin{minipage}{.24\textwidth}
~~~~~~
\end{minipage}

\end{figure}

\end{document}